\documentclass[superscriptaddress,secnumarabic,amssymb,amsmath,nobibnotes,aps,prd,showkeys,nofootinbib,onecolumn,12pt]{revtex4}

\usepackage{graphicx}
\usepackage{subcaption}

\usepackage{multirow}
\usepackage[utf8]{inputenc} 
\usepackage[T1]{fontenc}    
\usepackage{hyperref}       
\usepackage{url}            
\usepackage{booktabs}       
\usepackage{amsfonts}       
\usepackage{nicefrac}       
\usepackage{microtype}      
\usepackage{lipsum}		
\usepackage{natbib}
\usepackage{doi}
\usepackage{float}
\usepackage{epsf}
\usepackage{bm}
\usepackage{amsmath}
\usepackage{amsfonts}
\usepackage{amssymb}
\usepackage{graphicx}
\usepackage{tabularx}
\usepackage{color}%
\providecommand{\U}[1]{\protect\rule{.1in}{.1in} }

\newcommand{\be}{\begin{equation}}
\newcommand{\ee}{\end{equation}}

\newcommand{\mincir}{\raise
-3.truept\hbox{\rlap{\hbox{$\sim$}}\raise4.truept\hbox{$<$}\ }}
\newcommand{\magcir}{\raise
-3.truept\hbox{\rlap{\hbox{$\sim$}}\raise4.truept\hbox{$>$}\ }}

\providecommand{\U}[1]{\protect\rule{.1in}{.1in}}

\usepackage{tikz,xcolor,hyperref}

\definecolor{lime}{HTML}{A6CE39}
\DeclareRobustCommand{\orcidicon}{%
	\begin{tikzpicture}
	\draw[lime, fill=lime] (0,0) 
	circle [radius=0.16] 
	node[white] {{\fontfamily{qag}\selectfont \tiny ID}};
	\draw[white, fill=white] (-0.0625,0.095) 
	circle [radius=0.007];
	\end{tikzpicture}
	\hspace{-2mm}
}

\foreach \x in {A, ..., Z}{%
	\expandafter\xdef\csname orcid\x\endcsname{\noexpand\href{https//orcid.org/\csname orcidauthor\x\endcsname}{\noexpand\orcidicon}}
}

\begin{document}

\title{Fractional Einstein-Gauss-Bonnet scalar field cosmology}
\author{Bayron Micolta-Riascos\orcidA{}}
\email{bayron.micolta@alumnos.ucn.cl}
\affiliation{Departamento de Física, Universidad Católica del Norte, Avenida Angamos 0610, Casilla 1280, Antofagasta, Chile}
\author{Alfredo D.  Millano\orcidB{}}
\email{alfredo.millano@alumnos.ucn.cl}
\affiliation{Departamento de Matem\'{a}ticas, Universidad Cat\'{o}lica del Norte, Avenida
Angamos 0610, Casilla 1280 Antofagasta, Chile}
\author{Genly Leon\orcidC{}}
\email{genly.leon@ucn.cl}
\affiliation{Departamento de Matem\'{a}ticas, Universidad Cat\'{o}lica del Norte, Avenida
Angamos 0610, Casilla 1280 Antofagasta, Chile}
\affiliation{Institute of Systems Science, Durban University of Technology, P.O. Box 1334, \mbox{Durban 4000,  Republic of South Africa}}
\author{Byron Droguett\orcidD{}}
\email{byron.droguett@uantof.cl}
\affiliation{Department of Physics, Universidad de Antofagasta, 1240000 Antofagasta, Chile}
\author{Esteban Gonz\'alez\orcidE{}} 
\email{esteban.gonzalez@ucn.cl}
\affiliation{Departamento de Física, Universidad Católica del Norte, Avenida Angamos 0610, Casilla 1280, Antofagasta, Chile}
\author{Juan Magaña\orcidF{}}
\email{juan.magana@ucentral.cl}
\affiliation{Escuela de Ingeniería, Universidad Central de Chile, Avenida
Francisco de Aguirre 0405, 171-0164, La Serena Coquimbo, Chile}

\begin{abstract}
Our paper introduces a new theoretical framework called the Fractional Einstein--Gauss--Bonnet scalar field cosmology, which has important physical implications. 
Using fractional calculus to modify the gravitational action integral, we derived a modified Friedmann equation and a modified Klein--Gordon equation.  Our research reveals non-trivial solutions associated with exponential potential, exponential couplings to the Gauss--Bonnet term, and a logarithmic scalar field, which are dependent on two cosmological parameters, $m$ and $\alpha_{0}=t_{0}H_{0}$ and the fractional derivative order $\mu$. By employing linear stability theory, we reveal the phase space structure and analyze the dynamic effects of the Gauss--Bonnet couplings.  The scaling behavior at some equilibrium points reveals that the geometric corrections in the coupling to the Gauss--Bonnet scalar can mimic the behavior of the dark sector in modified gravity.  Using data from cosmic chronometers, type Ia supernovae, supermassive Black Hole Shadows, and strong gravitational lensing, we estimated the values of $m$ and $\alpha_{0}$, indicating that the solution is consistent with an accelerated expansion at late times with the values $\alpha_0=1.38\pm 0.05$, $m=1.44\pm 0.05$, and  $\mu=1.48 \pm 0.17$ (consistent with $\Omega_{m,0}=0.311\pm 0.016$ and $h=0.712\pm 0.007$), resulting in an age of the Universe $t_{0}=19.0\pm 0.7$ [Gyr] at 1$\sigma$ CL. Ultimately, we obtained late-time accelerating power-law solutions supported by the most recent cosmological data, and we proposed an alternative explanation for the origin of cosmic acceleration other than $\Lambda$CDM. Our results generalize and significantly improve previous achievements in the literature, highlighting the practical implications of fractional calculus in cosmology.
\end{abstract}


\date{\today}
\maketitle

\section{Introduction}

In general relativity, the Cosmological Principle is satisfied by the homogeneous and isotropic Friedmann--Lemaître--Robertson--Walker (FRLW) metric. It is believed that cold dark matter (CDM), which is not visible and only interacts through gravity, exists, and dark energy (in the form of a cosmological constant $\Lambda$) explains the Universe's accelerated expansion. This model is named $\Lambda$CDM and successfully explains the observed late-time acceleration of the Universe, as initially suggested by Type Ia supernovae (SNe Ia) observations~\cite{Riess:1998} and later confirmed by   Cosmic Microwave Background (CMB) measurements~\cite{Planck:2018}. Moreover, it accurately describes the formation of the Universe's large-scale structure. However, the model faces the long-standing Cosmological Constant problem \cite{Zeldovich, Weinberg}, and the true mechanism driving the Universe's accelerated expansion remains unknown \cite{Carroll:2000}. 
Various alternative theories have been proposed to address these issues, including
noncommutative theories, quantum cosmology, quantum deformed phase space models, and noncommutative minisuperspace approaches, in \cite{Rasouli:2013sda, Jalalzadeh:2014jea, Rasouli:2014dba, Rasouli:2016syh, Jalalzadeh:2017jdo} and modified Brans–Dicke theory in \cite{Rasouli:2014dxa,Rasouli:2016xdo}. 
 Phantom fields introduce exotic physics through their negative kinetic energy and inherent quantum instabilities, though they are still supported by some observational evidence \cite{Melchiorri:2002ux, Vikman:2004dc, Nesseris:2006er}. In previous studies, the cosmological evolution of phantom and quintom dark energy and their connection to late-time accelerated expansion were reviewed \cite{Elizalde:2004mq, Guo:2004fq, Zhang:2005eg}. Constraints based on cosmic age and SNae I data and theoretical implications and observations can be found in \cite{Feng:2004ad,Cai:2009zp}. The cosmological evolution of related hessence dark energy models was presented in \cite{Wei:2005fq,Wei:2005nw}. The examination of crossing the phantom divide in the hybrid dark energy model context can be found in \cite{Wei:2005si}. The reference \cite{Zhao:2005vj} discusses cosmological perturbations of the quintom models of dark energy and their effects on observations. References~\cite{Feng:2004ff,Xia:2004rw} investigated the oscillatory behavior of quintom models. Additionally, ref.
 \cite{Zhang:2005kj} studied interacting two-fluid scenarios for quintom dark energy, while ref.  \cite{Zhang:2005yz} presented a statefinder diagnostic for holographic dark energy model, and refs.  \cite{Zhang:2005hs,Zhang:2006qu} studied some holographic quintom dark energy models. Interestingly, ref.  \cite{Lazkoz:2006pa} proved that quintom models admit either tracking or phantom attractors, and refs.  \cite{Alimohammadi:2006tw, Lazkoz:2007mx, Leon:2012vt} analyzed the use of arbitrary potentials and classes of potential beyond the exponential potential. Furthermore, ref.  \cite{MohseniSadjadi:2006hb} examined the transition from quintessence to phantom phase in the quintom model, and ref.  \cite{Elizalde:2008yf} traced reconstructing attempts to relate these models with the Universe's history, from inflation to acceleration, with phantom and canonical scalar fields. Additionally, refs.  \cite{Leon:2018lnd, Tot:2022dpr,Leon:2023ywb} extended the examination in \cite{Lazkoz:2007mx,Leon:2012vt}. The crossing of the phantom divide is also possible within the framework of scalar-tensor theories \cite{Apostolopoulos:2006si, Bamba:2008xa, Bamba:2008hq, Setare:2008mb}, as well as in various modified gravity theories
~\cite{Nojiri:2003ft, Nojiri:2006gh, Cognola:2007zu,Capozziello:2006dj, Nojiri:2010wj, Nojiri:2017ncd}. 
 Ho\v{r}ava--Lifshitz cosmology was investigated using phase-space analysis in \cite{Leon:2009rc} obtaining late-time bouncing oscillatory solutions under detailed balance. Recently, it was proven that the nonprojectable Ho\v rava version is renormalizable \cite{Bellorin:2023dwk, Bellorin:2024qyy}. In \cite{Leon:2008de}, the author investigated scaling solutions in the regime where the scalar field diverges, with an interest in describing the past asymptotic dynamics in nonminimally coupled dark energy. 

Traditional mathematical models often need adjustments when dealing with power-law phenomena. Fractional calculus provides a suitable framework, accurately capturing non-local, frequency-dependent, and history-dependent properties inherent in power-law behavior. Fractional calculus is interdisciplinary and crucial for ongoing research. Recent studies show that fractional calculus is valuable in modeling fractional derivatives in quantum fields \cite{Lim:2006hp, LimEab+2019+237+256}, as well as in exploring quantum gravity and cosmology \cite{El-Nabulsi:2013hsa, El-Nabulsi:2013mwa, Moniz:2020emn, Rasouli:2021lgy, VargasMoniz:2020hve}, black holes \cite{Vacaru:2010wn, Jalalzadeh:2021gtq}, and fractional action cosmology \cite{El-Nabulsi:2007wgc, El-Nabulsi:2009bup, EL-Nabulsi:2009ejs, Jamil:2011uj, Shchigolev:2015rei}. We can develop fractional versions of Newtonian mechanics and Friedman--Robertson--Walker cosmology by substituting partial and fractional derivatives into familiar equations. For example, the study~\cite{ELNABULSI201765} investigated the emergence of a non-local-in-time fractional higher-order Newton's second law of motion and the non-local-in-time fractional dynamics exhibiting disordered motions. Additionally, fractional variational problems have been presented in \cite{RAMI20119492}. Fractional derivative methods have been developed through two primary approaches. The first, the last-step modification method, involves substituting the original cosmological field equations with fractional field equations tailored for a specific model. An illustrative example of this method is presented in \cite{Barrientos:2020kfp}. The second approach, the first-step modification method, is more fundamental. In this method, fractional derivative geometry is established initially, followed by the application of the Fractional Action-like Variational Approach (FALVA)~\cite{El-Nabulsi-Torres-2008, Baleanu-Trujillo-2010, Odzijewicz-Malinowska-Torres-2013c}. 

Using dynamical system methods, alongside observational data testing, provides a robust framework for analyzing the physical behavior of models, such as in fractional cosmology. This approach yields a cosmology with late-time acceleration without needing dark energy \cite{Garcia-Aspeitia:2022uxz, Gonzalez:2023who, LeonTorres:2023ehd}. In the study by \cite{Garcia-Aspeitia:2022uxz}, researchers conducted a joint analysis using cosmic chronometers (CCs) and SNe Ia data to determine the best-fit values for the fractional order of the derivative. The observational tests for fractional cosmology in the study by~\cite{Garcia-Aspeitia:2022uxz} were improved in subsequent studies by \cite{Gonzalez:2023who, LeonTorres:2023ehd}. In reference \cite{Micolta-Riascos:2023mqo}, the equation of state for a matter component was deduced based on compatibility conditions \cite{Micolta-Riascos:2023mqo}, which had not been previously analyzed in the study by \cite{Garcia-Aspeitia:2022uxz}.  

In Gauss--Bonnet (GB) theory, the addition of the GB term changes the Einstein--Hilbert action as referenced in \cite{Nojiri:2005vv, Nojiri:2006je, Cognola:2006sp, Nojiri:2007te}. Studies such as \cite{Chakraborty:2018scm, Fomin:2018typ} have demonstrated that scalar-coupled Gauss--Bonnet gravity in four dimensions can have significant effects on the early inflationary stage of our Universe. Various scenarios, including GB theory, $F(R)$, and Lorentz non-invariant models, are discussed in the reviews \cite{Nojiri:2010wj, Nojiri:2017ncd}, providing unified descriptions of early-time inflation, bounce, and late-time cosmic acceleration. The mass of a scalar field is influenced when it interacts with the GB scalar through a non-constant coupling function, leading to non-zero effects on gravitational theory, as evidenced in~\cite{Kanti:2015pda, Hikmawan:2015rze, Motaharfar:2016dqt, Rashidi:2020wwg}. In four dimensions, equilibrium points correspond to solutions with zero acceleration or de Sitter solutions in various coupling configurations, as found in \cite{Millano:2023czt, Millano:2023gkt}. A five-dimensional Einstein--Scalar--Gauss--Bonnet model was examined by \cite{Millano:2024vju}, exploring different coupling functions and scalar field potentials while analyzing the dynamics and stability of equilibrium points. These models can explain the Universe's early and late-time acceleration phases, making them suitable for studying inflation and dark energy. Reference \cite{ODINTSOV2020115135} introduces a new theoretical framework for Einstein--Gauss--Bonnet theories, based on the requirement that the theory be compatible with the GW170817 event, ensuring that the gravitational wave speed is close to one in natural units.

In this paper, we introduce the Fractional Einstein--Gauss--Bonnet scalar field gravity framework, a novel model with significant cosmological implications. We present an evolution equation for an effective scalar field related to the coupling with the GB term. The research uncovers non-trivial solutions associated with exponential potentials, exponential couplings to the GB term, and logarithmic scalar fields. Stability conditions for the exact solutions are established using linear stability theory, and the scaling behavior of these solutions is investigated. Notably, the geometric corrections in the coupling to the GB scalar mimic the behavior of the dark sector in modified gravity. We estimate the free parameters using observational data from the SNe Ia, CC, gravitational lensing (GL), and Black Hole Shadow (BHS) data. Our solutions are consistent with current observational constraints and result in a late-time accelerating power-law solution for the scale factor. Additionally, we discuss the physical interpretation of the cosmological solutions, emphasizing the role of the fractional derivative order within our gravitational framework.

The manuscript is organized as follows. In Section \ref{Einstein-G}, we will apply fractional differential calculus to compute specific physical quantities, highlighting their applications in cosmology. Rather than using covariant fractional derivatives as replacements for the standard covariant derivatives, we will adopt a point-like Lagrangian formulation of cosmology based on the flat FLRW metric and use FALVA to derive a modified cosmological model. Due to its relevance to cosmology, Section \ref{Scaling-Solutions} explores scaling solutions and outlines a reconstruction procedure for both the potential and the coupling function. The stability analysis of the exact solution is conducted in Section \ref{sect4.1}. In Section \ref{sec:fitting}, we will constrain the free parameters of the derived exact scaling solution. We will compute the best-fit parameters at the  $1\sigma$ of the confidence level for SNe Ia, CC, GL, and BHS data to achieve this. This approach offers the main advantages of combining dynamical systems analysis and observational testing, as it allows for using priors from the former analysis to inform the latter, aligning with the anticipated late-time behavior in a cosmological model. We present our conclusions in Section \ref{CONclusion}. For completeness, Appendix \ref{app0} contains our variational equations, while Appendix \ref{app1} outlines the formalism for the stability analysis of power-law solutions \(\psi_{s}(t)\) of a generic ordinary differential equation given by $
\mathcal{F}\left(t, \psi(t), \dot{\psi}(t), \ddot{\psi}(t), \ldots \right) \equiv 0,
$ where \(t > 0\) is the independent variable and \(\psi(t)\) is the dependent variable \cite{Ratra:1987rm, Liddle:1998xm, Uzan:1999ch}. Our proposal aims to construct regular differential equations on a bounded state space to describe dynamics globally, including the early- and late-time behavior of the intermediate stage of evolution, and alleviate cosmological problems like the $H_0$ tension and the Cosmic Coincidence problem.

\section{Einstein--Gauss--Bonnet Scalar Field Gravity}
\label{Einstein-G}
We begin with an action with a scalar field with a non-zero coupling to the {GB term,} 
\begin{equation}\label{action}
    S=\int d^4x\sqrt{-g}\left[\frac{R}{2\kappa^2}-\frac{1}{2}g^{\mu\nu}\partial_\mu \phi \partial_\nu \phi-V(\phi)-f(\phi)\mathcal{G}\right],
 \end{equation}
where $R$ is the curvature  scalar and we have assumed  units where $\kappa^2= 8\pi G=1$, and $\mathcal{G}$ is the GB term:
\begin{equation}
    \mathcal{G}=R^2-4R_{\alpha \beta}R^{\alpha \beta}+R_{\alpha\beta\gamma\delta}R^{\alpha\beta\gamma\delta}.
\end{equation}
{Here,} 
 \(\phi\) represents a scalar field that is minimally coupled to gravity and possesses a self-interacting potential \(V(\phi)\). Additionally, it is nonminimally coupled to the GB invariant \(\mathcal{G}\) through the coupling function \(f(\phi)\).
We are considering the FLRW
\begin{equation}
    ds^2=-N^2(t)\,dt^2+a^2(t)\left(dx^2+\,dy^2+\,dz^2\right).
\end{equation}
{Defining} $\displaystyle H={\Dot{a}}/{(N a)}$ 
we obtain that the curvature scalar and the GB term are given by 
\begin{equation}
    R=12H^2+{6\Dot{H}}/{N}, \quad
    \mathcal{G}=24 H^2 \left(H^2 + {\dot{H}}/{N}\right).
  \end{equation}
{By substituting} $R$ and $\mathcal{G}$ into the action, we can find the point-like Lagrangian:
\begin{align}\label{L1}
    \mathcal{L} & =\dot{N}(\theta ) \left[\frac{24 \dot{a}(\theta )^3 f(\phi (\theta ))}{N(\theta )^4}-\frac {3 a(\theta )^2 \dot{a}(\theta )}{N(\theta )^2}\right]+\frac{3 a(\theta ) \dot{a}(\theta
   )^2}{N(\theta )} \nonumber \\
   & +\ddot{a}(\theta ) \left[\frac{3 a(\theta )^2}{N(\theta )}-\frac{24 \dot{a}(\theta )^2 f(\phi (\theta ))}{N(\theta )^3}\right]+\frac{a(\theta )^3 \dot{\phi
   }(\theta )^2}{2 N(\theta )}-a(\theta )^3 N(\theta ) V(\phi (\theta )).
\end{align}
{Extending} this model using fractional calculus, we consider that the action is defined by
\begin{equation}\label{actionkernel}
    S(\tau)=\frac{1}{\Gamma(\mu)}\int_0^\tau \mathcal{L}(\theta,q_i(\theta),\Dot{q}_i(\theta), \Ddot{q}_i(\theta))(\tau-\theta)^{\mu-1}\,d\theta.
\end{equation}
{We follow} the procedures of reference \cite{10.5555/1466940.1466942}, which start with the action \eqref{actionkernel}. We make the steps in Appendix \ref{app0} to obtain the equations of motion of the fields. 
Implementing the derivatives in \eqref{Variational-Eqs}, we have 
\begin{align}
    &\frac{\partial \mathcal{L}(\theta,q_i(\theta),\Dot{q}_i(\theta),\Ddot{q}_i(\theta))}{\partial q_i}-\frac{d}{d\theta}\frac{\partial \mathcal{L}(\theta,q_i(\theta),\Dot{q}_i(\theta),\Ddot{q}_i(\theta))}{\partial \Dot{q}_i}+\frac{d^2}{d\theta^2}\frac{\partial \mathcal{L}(\theta,q_i(\theta),\Dot{q}_i(\theta),\Ddot{q}_i(\theta))}{\partial \Ddot{q}_1}\nonumber\\
    &=\frac{1-\mu}{\tau-\theta}\left[\frac{\partial \mathcal{L}(\theta,q_i(\theta),\Dot{q}_i(\theta),\Ddot{q}_i(\theta))}{\partial \Dot{q}_i}-2\frac{d}{d\theta}\frac{\partial \mathcal{L}(\theta,q_i(\theta),\Dot{q}_i(\theta),\Ddot{q}_i(\theta))}{\Ddot{q}_i}\right]\nonumber\\
    &-\frac{(1-\mu)(2-\mu)}{(\tau-\theta)^2}\frac{\partial \mathcal{L}(\theta,q_i(\theta),\Dot{q}_i(\theta),\Ddot{q}_i(\theta))}{\Ddot{q}_i}. \label{EP}
\end{align}

Then, by applying Hamilton's principle, we obtain the Euler--Poisson equations modified by the fractional parameter $\mu$ given by \eqref{EP} by taking the variations with respect to $q_i \in \{N, a, \phi\}$, and making $N=1$ after the variation. Then, using the parameterization $(\tau,\theta)=(2 t, t)$, we obtain respectively the Friedmann, Raychaudhuri and Klein--Gordon equations:

\begin{align}
    \frac{24 \dot{a}^3 \dot{\phi} f'(\phi)}{a^3}-\frac{24 (\mu -1) \dot{a}^3 f(\phi)}{t a^3}-\frac{3 \dot{a} \left(t \dot{a}-\mu  a+a\right)}{t a^2}+V(\phi)+\frac{1}{2} \dot{\phi}^2&=0, \label{eq1} \\
    -\frac{2 \ddot{a}}{a}+\frac{8 \dot{a}^2 \dot{\phi}^2 f''(\phi)}{a^2}+\frac{\dot{a} \left[2 (\mu -1) a-t \dot{a}\right]}{t a^2} +f'(\phi) \left[-\frac{16 (\mu -1) \dot{a}^2 \dot{\phi}}{t a^2}+\frac{8 \dot{a}^2 \ddot{\phi}}{a^2}+\right.&\left.\frac{16 \dot{a} \ddot{a} \dot{\phi}}{a^2}\right] \nonumber \\
    +f(\phi) \left[\frac{8 (\mu -2) (\mu -1) \dot{a}^2}{t^2 a^2}-\frac{16 (\mu -1) \dot{a} \ddot{a}}{ta^2}\right]-\frac{(\mu -2) (\mu -1)}{t^2}+V(\phi)-\frac{1}{2} \dot{\phi}^2&=0, \label{eq2} \\
    -\frac{3 \dot{a} \dot{\phi}}{a}-\frac{24 \dot{a}^2 \ddot{a} f'(\phi)}{a^3}+\frac{(\mu -1) \dot{\phi}}{t}-V'(\phi)-\ddot{\phi}&=0. \label{eq3}
\end{align}
Substituting $H= \dot{a}/a$ (with $N\equiv 1$) in \eqref{eq1}--\eqref{eq3}, we have
\begin{align}
    24 H^3 \dot{\phi} f'(\phi)-\frac{24 (\mu -1) H^3 f(\phi)}{t}+\frac{3 (\mu -1) H}{t}-3 H^2+V(\phi)+\frac{1}{2} \dot{\phi}^2&=0,\label{FL1}\\
    8 H^2 \dot{\phi}^2 f''(\phi)+f'(\phi) \left[16 H \dot{H} \dot{\phi}+H^2 \left(8 \ddot{\phi}-\frac{16 (\mu -1) \dot{\phi}}{t}\right)+16 H^3 \dot{\phi}\right] \nonumber\\
    +f(\phi) \left[-\frac{16 (\mu -1) H \dot{H}}{t}+\frac{8 (\mu -2) (\mu -1)H^2}{t^2}-\frac{16 (\mu -1) H^3}{t}\right] \nonumber\\
    -2 \dot{H}+\frac{2 (\mu -1) H}{t}-3 H^2-\frac{(\mu -2) (\mu -1)}{t^2}+V(\phi)-\frac{1}{2} \dot{\phi}^2&=0, \label{RL1}\\
    -24 H^2 \dot{H} f'(\phi)-24 H^4 f'(\phi)+\frac{(-3 t H+\mu -1) \dot{\phi}}{t}-V'(\phi)-\ddot{\phi}&=0.\label{KGL1}
\end{align}
By solving the system \eqref{FL1}--\eqref{KGL1} for $V(\phi)$, $\Ddot{\phi}$ and $\Dot{H}$, we obtain the following equations:
\vspace{-6pt}

\begin{align}
    &V(\phi)=\frac{(3-3 \mu ) H}{t}+3 H^2+\frac{24 (-1+\mu ) f(\phi) H^3}{t}-24 H^3 f'(\phi) \dot{\phi}-\frac{1}{2} \dot{\phi}^2,\label{V}
\\
    &  \Ddot{\phi}=\nonumber\\
    & \left\{12 t (-1+\mu ) H^3 f'(\phi)+t \left[-t V'(\phi)+(-1+\mu ) \dot{\phi}\right]+288 t H^5 f'(\phi) \left[(1-\mu) f(\phi)\hspace{-1mm}+\hspace{-1mm}t f'(\phi) \dot{\phi}\right]\right.\nonumber\\
    & +12 H^2 \left[-2 t (-1+\mu ) f(\phi) \dot{\phi}+f'(\phi) \left(2-3 \mu +\mu ^2+3 t^2 \dot{\phi}^2\right)\right]+\nonumber\\
    & H \left[-8 (-1+\mu ) f(\phi) \left(t V'(\phi)-(-1+\mu ) \dot{\phi}\right)+t \dot{\phi} \left(-3 t+8 f'(\phi) \left(t V'(\phi)-(-1+\mu) \dot{\phi}\right)\right)\right]\nonumber\\
    & \left.\left.-24 H^4 f'(\phi) \left[4 (-2+\mu ) (-1+\mu ) f(\phi)+t \left(t+4 \dot{\phi} \left(-2 (-1+\mu ) f'(\phi)+t \dot{\phi}f''(\phi)\right)\right)\right]\right\}\right/\nonumber\\
    & \left\{t \left[t+8 H \left((-1+\mu ) f(\phi)+t f'(\phi) \left(12 H^3 f'(\phi)-\dot{\phi}\right)\right)\right]\right\},\label{phipp}
\end{align}
\begin{align}
    & \Dot{H} = \nonumber\\
    & -\left\{\left[2-3 \mu +\mu ^2+t (-1+\mu ) H+192 t^2 H^6 f'(\phi)^2+t^2 \dot{\phi}^2+8 t H^3 \left((1-\mu ) f(\phi)\hspace{-1mm}+\hspace{-1mm}4 t f'(\phi) \dot{\phi}\right)\right.\right.\nonumber\\
    & \left.\left.-8 H^2 \left((-2+\mu ) (-1+\mu ) f(\phi)+t \left(f'(\phi) \left(-t V'(\phi)-(-1+\mu ) \dot{\phi}\right)+t \dot{\phi}^2f''(\phi)\right)\right)\right]\right/\nonumber\\
    & \left.\left[2 t \left(t+8 H \left((-1+\mu ) f(\phi)+t f'(\phi) \left(12 H^3 f'(\phi)-\dot{\phi}\right)\right)\right)\right]\right\}.\label{Hp}
\end{align}
We now impose that the time derivative of Friedmann Equation \eqref{FL1} is zero under the hypothesis that $\mu\neq 1$ (for $\mu=1$ is trivial); 
that is,
\begin{align}
\frac{d}{d t}\left[24 H^3 \dot{\phi} f'(\phi)-\frac{24 (\mu -1) H^3 f(\phi)}{t}+\frac{3 (\mu -1) H}{t}-3 H^2+V(\phi)+\frac{1}{2} \dot{\phi}^2\right]=0. \label{cons1}
\end{align}
Computing the time derivative and replacing $V(\phi)$, $\Ddot{\phi}$ and $\Dot{H}$ given by \eqref{V}--\eqref{Hp} into \eqref{cons1}, we obtain
\vspace{-12pt}
\begin{align}
    0&= \Bigg\{3 (-2+\mu ) (-1+\mu )+t^2 \dot{\phi}^2-4608 t H^7 f'(\phi)^2 \left[-((-3+\mu ) f(\phi))+t f'(\phi)\dot{\phi}\right]  \nonumber\\
    & +48 t H^3 \left[-((1+\mu ) f(\phi))+2 t f'(\phi) \dot{\phi}\right]+t H \left[15-3 \mu -16 (-1+\mu ) f(\phi) \dot{\phi}^2+16 t f'(\phi) \dot{\phi}^3\right]  \nonumber\\
    & -576 t H^5 \left[-\left((-1+\mu ) f(\phi)^2\right)+(-3+\mu) f'(\phi)^2+2 t f(\phi) f'(\phi) \dot{\phi}\right]  \nonumber\\
    & -48 H^4 \Big[-4 (-1+\mu ) (-8+3 \mu ) f(\phi)^2-4 t^2 f'(\phi)^2 \dot{\phi}^2  \nonumber\\
    & +t f(\phi) \left(-t+4 f'(\phi) \left(tV'(\phi)+(-9+5 \mu ) \dot{\phi}\right)-4 t \dot{\phi}^2 f''(\phi)\right)\Big]  \nonumber\\
    & -6 H^2 \Big(4 f(\phi) \left(10\hspace{-1mm}-\hspace{-1mm}14 \mu +4 \mu ^2+t^2\dot{\phi}^2\right) \hspace{-1mm} +t \hspace{-0.5mm}\left(t\hspace{-1mm}-\hspace{-1mm}4 f'(\phi) \left(t V'(\phi)+(-7+3 \mu ) \dot{\phi}\right)\hspace{-1mm}+\hspace{-1mm}4 t \dot{\phi}^2 f''(\phi)\right)\Big)\Bigg\}\Big/\nonumber \\
    & \Bigg\{2t^2 \left(t+8 H \left((-1+\mu ) f(\phi)+t f'(\phi) \left(12 H^3 f'(\phi)-\dot{\phi}\right)\right)\right)\Bigg\}.\label{ConsPsi}
\end{align}
We introduce the auxiliary field  
\begin{equation}
 \psi=f(\phi).
\end{equation}
Moreover, we calculate the successive derivatives using the chain rule:
\begin{equation}
    \Dot{\psi}=\frac{d}{dt}f(\phi)=f^{\prime}(\phi)\Dot{\phi}, \quad
\Ddot{\psi}=f^{\prime\prime}(\phi)\Dot{\phi}^2+f^{\prime}(\phi)\Ddot{\phi}=f^{\prime\prime}(\phi)\Dot{\phi}^2+\frac{\Dot{\psi}}{\Dot{\phi}}\Ddot{\phi}.
\end{equation}
In this way, we obtain
\begin{equation}\label{fp-fpp-psi}
    f^{\prime}(\phi)=\frac{\Dot{\psi}}{\Dot{\phi}},\quad f^{\prime\prime} (\phi)=\frac{1}{\Dot{\phi}^2}\left(\Ddot{\psi}-\frac{\Dot{\psi}}{\Dot{\phi}}\Ddot {\phi}\right).
\end{equation}
Replacing the expression \eqref{fp-fpp-psi} into \eqref{V}--\eqref{Hp} and \eqref{ConsPsi}, then solving the resulting system for $\Ddot{\phi}$, $\Ddot{\psi}$, $\Dot {H}$ and  $V(\phi)$, it turns out that 
\begin{align}
    \Ddot{\phi}&=\frac{\frac{24 (\mu -3) H^3 \dot{\psi}}{\dot{\phi}}+(\mu -1) \dot{\phi}}{t} +\frac{192 H^5 \dot{\psi}^2}{\left(1-8 H^2 \psi \right) \dot{\phi}}+\dot{\phi} \left(\frac{8 H^2 \dot{\psi}}{1-8 H^2 \psi }-3 H\right)-V'(\phi ),\label{phippf}
\\
    \Ddot{\psi}&=\frac{(\mu -1) \left[\frac{\mu -2}{H^2}+8 (8-3 \mu )
   \psi \right]}{8 t^2}\nonumber\\ 
   & +\frac{2 \dot{\psi} \left(\frac{\mu -1}{8 H^2 \psi -1}+3 \mu -5\right)-3 (\mu -1) H
   \psi +\frac{2 (\mu -1) \psi  \dot{\phi}^2}{3 H \left(8 H^2 \psi -1\right)}-\frac{\mu -5}{8
   H}}{t} \nonumber\\
   & +\frac{16 H^2 \dot{\psi}^2}{1-8 H^2 \psi }+H \left(\frac{2}{8 H^2 \psi -1}+3\right) \dot{\psi}  +\dot{\phi}^2 \left(\frac{2 \dot{\psi}}{3 H-24 H^3 \psi }+\frac{\frac{2}{8 H^2 \psi -1}+3}{24 H^2}\right)-\frac{1}{4}, \label{psippf}
\\
    \Dot{H}&=-\frac{(\mu -3) H}{t}+\frac{8 H^3 \dot{\psi}}{8 H^2 \psi-1}+\frac{\dot{\phi}^2}{24 H^2 \psi -3}-H^2,\label{Hpf}
\\
    V(\phi)&= \frac{3 (\mu -1) H \left(8 H^2 \psi -1\right)}{t}-24 H^3 \dot{\psi}+3 H^2-\frac{1}{2} \dot{\phi}^2.\label{Vf}
\end{align}
From Equation \eqref{Vf}, we define the effective energy densities as follows:
\vspace{-6pt}
\begin{align}
    \rho_{\phi} & = \frac{1}{2}\dot{\phi}^2 + V(\phi),\quad  
    \rho_{\text{fracc}}  = \frac{3(\mu -1)}{t} H,\quad 
    \rho_{\text{GB}}  = 24 H^3 \dot{\psi},\quad 
    \rho_{\text{GB,\,fracc}}  =- \frac{24(\mu -1)}{t} H^3 \psi, 
\end{align}
such that \eqref{FL1} is reduced to
\begin{align}
    3 H^2 &= \rho_{\phi} +  \rho_{\text{fracc}} + \rho_{\text{GB}} + \rho_{\text{GB,\,fracc}}.
    \end{align}

\section{Exact Scaling Solutions}
\label{Scaling-Solutions}

We can find an analytical solution to the system by considering
\begin{align}
    & H=\frac{2}{m}t^{-p}, \\ 
    &\psi=F_0 t^q, \\
    & \phi  =\frac{2 \ln (t)}{\lambda }, \quad \lambda\neq 0, \label{exactphi}
\end{align}
where $m$, $F_0$, $p$, $q$ and $\lambda$ are constants. 

Then, Equations \eqref{phippf}--\eqref{Vf} are written as
\begin{align}
& \frac{64 F_0 (q-\mu ) t^q+2 \mu  m^2 t^{2 p}}{\lambda  t^2 \left(m^2 t^{2 p}-32 F_0
   t^q\right)}-\frac{96 F_0 \lambda  q (-\mu +q+3) t^{q-3 p-1}}{m^3}+\frac{3 \lambda  m q^2 t^{p-1}}{m^2 t^{2
   p}-32 F_0 t^q}\nonumber\\
   & -\frac{3 t^{-p-1} \left(\lambda ^2 q^2+4\right)}{\lambda  m}-V'\left(\ln \left(t^{2/\lambda}\right)\right)=0, \label{(47)}
\\
   & \frac{m^4 \left[3 \lambda ^2 (\mu -2) (\mu -1)+4\right] t^{4 p-2}}{96 \lambda ^2 \left(m^2 t^{2
   p}-32 F_0 t^q\right)}-\frac{1}{4} \nonumber \\
   & +  \frac{32 F_0^2 \left[-11 \mu +3 \left(\mu ^2-2 \mu  q+q (q+3)\right)+8\right] t^{2 q-2}}{m^2
   t^{2 p}-32 F_0 t^q}  \nonumber\\
   & -\frac{F_0 m^2 t^{2 p+q-2} \left[\lambda ^2 (-2 \mu +q+2) (-2 \mu+q+5)+4\right]}{\lambda ^2 \left(m^2 t^{2 p}-32 F_0 t^q\right)} +\frac{4 F_0 m^3 (-\mu +q+1) t^{3 p+q-3}}{3 \lambda ^2 \left(m^2 t^{2 p}-32 F_0 t^q\right)}  \nonumber\\
   & +\frac{1}{16} m t^{p-1} \left(-\frac{64
   F_0 q t^q}{m^2 t^{2 p}-32 F_0 t^q}-\mu +5\right)+\frac{6 F_0 (-\mu +q+1)
   t^{q-p-1}}{m}  =0, \label{(48)}
\\
   & \frac{4}{3 \lambda ^2 t^2 \left(\frac{32 F_0 t^{q-2
   p}}{m^2}-1\right)}+\frac{2 t^{-p-1} (-\mu +p+q+3)}{m}-\frac{2 m q t^{p-1}}{m^2 t^{2 p}-32 F_0 t^q}-\frac{4 t^{-2 p}}{m^2}  =0, \label{(49)}\\
  & \frac{6 t^{-3 p-1} \left[32 F_0 (\mu -q-1) t^q-(\mu -1) m^2 t^{2
   p}\right]}{m^3}+\frac{12 t^{-2 p}}{m^2}-V\left(\ln \left(t^{2/\lambda }\right)\right)-\frac{2}{\lambda ^2 t^2}  =0.
\end{align}
By comparing the coefficients of $t$ in the last equation, we have, by dimensional analysis, that
\begin{equation}
 q=2p,\quad  -2p=-2 \quad\implies\quad  p=1,\quad  q=2.
\end{equation}
Then, for $p=1$ and $q=2$, and assuming $m^2-32 F_0\neq 0, m\neq 0$ and $\lambda\neq 0$ we have
\begin{align}
    6144 F_0^2 \lambda ^2 (\mu -5)+64 F_0 m^2 \left[-3 \lambda ^2 (\mu -3)+(\mu -2) m-6\right] \nonumber\\
    +\lambda m^3 t^2 \left(m^2-32 F_0\right) V'\left(\ln \left(t^{2/\lambda }\right)\right)-2 m^4 (\mu  m-6)&=0, \label{eq-pq-1}\\
    3072 F_0^2 \lambda ^2 \left[6 (\mu -3)+(\mu  (3 \mu -23)+38) m\right] \nonumber\\
    -64 F_0 m \left[-12 \lambda ^2+2 (\mu -3) m^3+m^2\left(3 \lambda ^2 (\mu -2) (2 \mu -7)+6\right)+6 \lambda ^2 \mu  m\right] \nonumber\\
    +m^3 \left[-24 \lambda ^2+m^2 \left(3\lambda ^2 (\mu -2) (\mu -1)+4\right)-6 \lambda ^2 (\mu -5) m\right]&=0,\label{eq-pq-2}\\
    192 F_0 \lambda ^2+96 F_0\lambda ^2 (\mu -6) m-2 m^4-3 \lambda ^2 (\mu -4) m^3-6 \lambda ^2 m^2&=0,\label{eq-pq-3}\\
    -\frac{2 \left[-96 F_0 \lambda^2 (\mu -3)+m^3+3 \lambda ^2 (\mu -1) m^2-6 \lambda ^2 m\right]}{\lambda ^2 m^3 t^2}-V\left(\ln
   \left(t^{2/\lambda }\right)\right)&=0. \label{eq-pq-4}
\end{align}
 Equation \eqref{eq-pq-4} allows us to reconstruct the GB coupling and scalar field potentials, which replicate scaling behaviors. 

In our scenario, say we acquire the exponential potential 
\begin{equation}\label{Potential-pq}
    V\left(\phi\right) =\frac{2 e^{-\lambda  \phi } \left[96 F_0 \lambda ^2 (\mu -3)-m^3-3 \lambda ^2 (\mu -1) m^2+6 \lambda ^2 m\right]}{\lambda ^2 m^3}.
\end{equation}
Replacing the derivative of potential \eqref{Potential-pq} in Equation \eqref{eq-pq-1}, and assuming $(\mu -6) m+2\neq 0$ and $m\neq 0$, results in two independent equations: 
\begin{align}
    F_0&= \frac{1}{96} m^2 \left[\frac{2 m \left(3 \lambda ^2+m\right)}{\lambda ^2 ((\mu -6)m+2)}+3\right],\\
    0&=\left(3 \lambda ^2+m\right) \left[m \left(3 \lambda ^2 (\mu +2)-4 \mu +((\mu -5) \mu +2) m+30\right)-6 \left(3 \lambda ^2+2\right)\right]. \label{eq43}
\end{align}
The first condition is the definition of $F_0$. Hence, we have the two-parametric family of solutions (recall $\lambda=\lambda(m,\mu)$):
\begin{align}
    R(t) & = -\frac{12 (m-4)}{m^2 t^2} \underbrace{\implies}_{\text{Using}\; \eqref{exactphi}}  R(\phi)   = -\frac{12 (m-4) e^{-\lambda  \phi }}{m^2},\\
    \mathcal{G}(t) & = -\frac{192 (m-2)}{m^4 t^4} \implies   \mathcal{G}(\phi)  =-\frac{192 (m-2) e^{-2 \lambda  \phi }}{m^4}, 
 \\
    H(t) & = \frac{2}{m t}
    \implies  H(\phi)  = \frac{2}{m}e^{-\frac{\lambda\phi }{2}}, \label{exactH}
\\
    \psi(t) & = \frac{1}{96} m^2 t^2 \left[\frac{2 m\left(3 \lambda ^2+m\right)}{\lambda ^2 \left[(\mu -6) m+2\right]}+3\right] \nonumber \\
     &   \implies f(\phi) = \frac{1}{96} m^2 e^{\lambda  \phi } \left[\frac{2 m \left(3 \lambda^2+m\right)}{\lambda ^2 \left[(\mu -6) m+2\right]}+3\right], \label{exactpsi}
  \\
    V(t) & = \frac{2 \left[12\lambda ^2+\mu  m^3+2 \left(9 \lambda ^2-1\right) m^2+6 \lambda ^2 (\mu -8) m\right]}{\lambda ^2 m^2 t^2 \left[(\mu -6)m+2\right]} \label{exactV} \\
    & \implies  V(\phi)  =\frac{2 e^{-\lambda  \phi } \left[12 \lambda ^2+\mu  m^3+2\left(9 \lambda ^2-1\right) m^2+6 \lambda ^2 (\mu -8) m\right]}{\lambda ^2 m^2 \left[(\mu -6) m+2\right]}.\label{exp}
\end{align}
Replacing the derivative of potential \eqref{Potential-pq} in Equation \eqref{eq-pq-1}, we obtain the conditions 
\begin{align}
    6144 F_0^2 \lambda ^2 (\mu -4)+32 F_0 m \left[6 \lambda ^2+m \left(3 \lambda ^2 (7-3 \mu)+(\mu -3) m-6\right)\right]\nonumber\\
    +m^3 \left[m \left(3 \lambda ^2 (\mu -1)-\mu  m+m+6\right)-6 \lambda^2\right]&=0,
\end{align}
together with \eqref{eq-pq-2} and \eqref{eq-pq-3}. 
Reducing the conditions and assuming $(\mu -6) m+2\neq 0$ and $m\neq 0$, we obtain the cases
\begin{enumerate}
\item For $0<\mu \leq\frac{1}{2} \left(5-\sqrt{17}\right)$ and $\frac{2 \mu -15}{\mu ^2-5 \mu +2}+\sqrt{\frac{16 \mu ^2-120
\mu +249}{\left(\mu ^2-5 \mu +2\right)^2}}<m<\frac{6}{\mu +2}$, \newline $\lambda =\frac{1}{3} \sqrt{\frac{36-3 m \left[-4 \mu
   +((\mu -5) \mu +2) m+30\right]}{(\mu +2) m-6}}$ and $F_0= \frac{m^2 (m (-4 \mu +(\mu -2) (\mu -1)m+18)-12)}{32 (m (-4 \mu +((\mu -5) \mu +2)
   m+30)-12)}$. For $\mu=\frac{1}{2} \left(5-\sqrt{17}\right)$ and $\frac{1}{83} \left(60-6 \sqrt{17}\right)<m<\frac{1}{16}
   \left(27+3 \sqrt{17}\right)$,  $\lambda =\frac{2}{3} \sqrt{\frac{3 \left(10+\sqrt{17}\right)
   m-18}{\left(\sqrt{17}-9\right) m+12}}$ and $F_0= -\frac{m^2 \left(\left(\sqrt{17}-5\right)
   m^2-2 \left(4+\sqrt{17}\right)
   m+12\right)}{64\left(\left(10+\sqrt{17}\right) m-6\right)}$. 
\item For $\frac{1}{2}
   \left(5-\sqrt{17}\right)<\mu \leq 4$ and $m>\frac{2 \mu -15}{\mu ^2-5 \mu +2}-\sqrt{\frac{16 \mu ^2-120 \mu
   +249}{\left(\mu ^2-5 \mu +2\right)^2}}$ we have 
 \newline $\lambda =\frac{\sqrt{36-3 m \left[-4 \mu +((\mu -5) \mu +2) m+30\right]}}{3\sqrt{(\mu +2) m-6}}$  {and} 
 $F_0= \frac{m^2 (m (-4 \mu +(\mu -2) (\mu -1)m+18)-12)}{32 (m (-4 \mu +((\mu -5) \mu +2)
   m+30)-12)}$.  
\item For $4<\mu \leq\frac{1}{2}
   \left(5+\sqrt{17}\right)$ and  $\frac{6}{\mu +2}<m<\frac{2 \mu -15}{\mu ^2-5 \mu +2}+\sqrt{\frac{16 \mu ^2-120 \mu
   +249}{\left(\mu ^2-5 \mu +2\right)^2}}$ we  {have} 
  $\lambda =\frac{\sqrt{36-3 m \left[-4 \mu +((\mu -5) \mu +2) m+30\right]}}{3
   \sqrt{(\mu +2) m-6}}$   {and} 
 $F_0= \frac{m^2 (m (-4 \mu +(\mu -2) (\mu -1)m+18)-12)}{32 (m (-4 \mu +((\mu -5) \mu +2) m+30)-12)}$. For $\mu =\frac{1}{2} \left(5+\sqrt{17}\right)$ and $\frac{1}{16} \left(27-3
   \sqrt{17}\right)<m<\frac{1}{83} \left(60+6 \sqrt{17}\right)$ we have \newline $\lambda =2 \sqrt{\frac{\left(\sqrt{17}-10\right) m+6}{3 \left(9+\sqrt{17}\right) m-36}}$ and  $F_0=\frac{m^2 \left(12-m \left(\left(5+\sqrt{17}\right) m-2 \sqrt{17}+8\right)\right)}{64\left(\left(\sqrt{17}-10\right) m+6\right)}$. 
   
\item For $4<\mu <\frac{1}{2} \left(5+\sqrt{17}\right)$ and $m>\frac{2 \mu -15}{\mu ^2-5
   \mu +2}-\sqrt{\frac{16 \mu ^2-120 \mu +249}{\left(\mu ^2-5 \mu +2\right)^2}}$ we have \newline $\lambda =\frac{\sqrt{36-3 m \left[-4\mu +((\mu -5) \mu +2) m+30\right]}}{3 \sqrt{(\mu +2) m-6}}$ and $F_0= \frac{m^2 (m (-4 \mu +(\mu -2) (\mu -1)m+18)-12)}{32 (m (-4 \mu +((\mu -5) \mu +2) m+30)-12)}$.  
\item For $\mu >\frac{1}{2} \left(5+\sqrt{17}\right)$ and
   $\frac{6}{\mu +2}<m<\frac{2 \mu -15}{\mu ^2-5 \mu +2}+\sqrt{\frac{16 \mu ^2-120 \mu +249}{\left(\mu ^2-5 \mu +2\right)^2}}$ we have 
   $\lambda =\frac{1}{3} \sqrt{\frac{36-3 m \left[-4 \mu +((\mu -5) \mu +2) m+30\right]}{(\mu +2) m-6}}$  and  $F_0=\frac{m^2 (m (-4 \mu +(\mu -2) (\mu -1)m+18)-12)}{32 (m (-4 \mu +((\mu -5) \mu +2) m+30)-12)}$. 
\end{enumerate}

\begin{figure}[H]
    \includegraphics[scale=0.7]{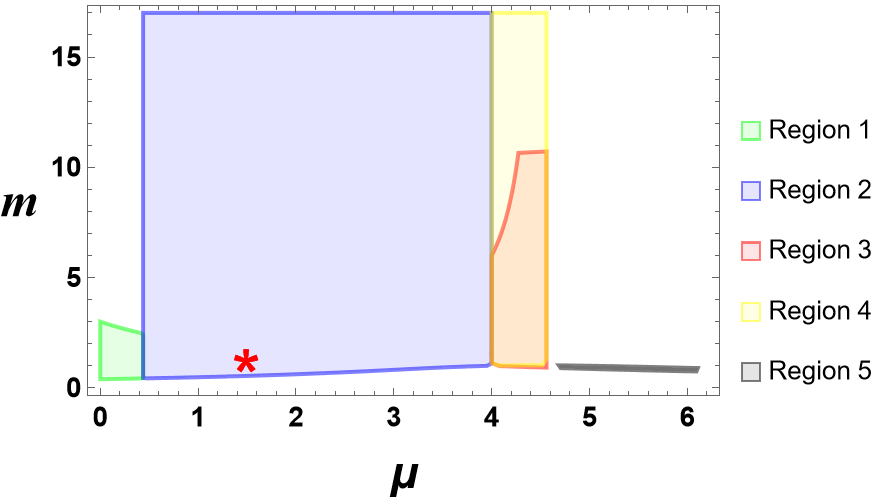}
        \caption{Allowed regions for the parameters $\mu$ and $m$. The red star represents the point in the parameter space with the best-fit values of $\mu$ and $m$ according to the analysis in Section \ref{sec:fitting}.}
    \label{allowed-regions-3}
\end{figure}

In Figure \ref{allowed-regions-3}, we see the allowed regions for the parameters $\mu$ and $m$.

Replacing the ansatz
\begin{equation}
    H=\frac{2}{m t},\quad  m \neq \frac{6}{\mu},
\end{equation}
in Equations \eqref{phippf}--\eqref{Vf}, we obtain
\begin{align}
 \ddot{\phi} & = -\frac{6 (m-2) \left[(\mu -4) m+2\right] \left(m^2 t^2-32 \psi\right)}{m^5 t^5
   \dot{\phi}}+\frac{\left[(\mu -2) m-4\right] \dot{\phi}}{m t}-V'(\phi (t)),
\\
 \ddot{\psi} &= \frac{\left[m (\mu ((\mu -7) m+4)-6)+4\right] \psi}{m^2 t^2}\nonumber\\
 & +\frac{1}{96} \left[m^2 t^2 (2-\mu  m) \dot{\phi}^2+3
   m (-4 \mu +(2-(\mu -5) \mu ) m+2)-12\right],
   \end{align}
\begin{align}
 \dot{\psi} & = \frac{\left[(\mu -4) m+2\right] \psi}{m t}-\frac{1}{192} m t \left[m^2 t^2 \dot{\phi }^2+6 (\mu -4) m+12\right],
 \\
 V(\phi) &  = \frac{6 (4-3 m) m^2 t^2+192 (3 m-2) \psi}{m^4 t^4}+\frac{1}{2} \dot{\phi}^2.
\end{align}
To investigate the stability of the solution \eqref{exactphi} and \eqref{exactH}--\eqref{exactV}, we specify the potential \eqref{exp}, obtaining  the equations
\begin{align}
    \ddot{\phi} &= \frac{192 \lambda  (3 m-2) \psi}{m^4 t^4}+\frac{6 \lambda  (4-3 m)}{m^2t^2}+\frac{\frac{192 (m-2) \left[(\mu -4) m+2\right] \psi}{m^5 t^5}-\frac{6 (m-2) \left[(\mu -4)m+2\right]}{m^3 t^3}}{\dot\phi} \nonumber\\
    & +\frac{\left[(\mu -2) m-4\right] \dot\phi}{m t}+\frac{1}{2} \lambda \dot{\phi}^2,\\
    \ddot{\psi}&= \frac{\left[m (\mu  ((\mu -7) m+4)-6)+4\right] \psi}{m^2t^2}\nonumber\\
    & +\frac{1}{96}\left[m^2 t^2 (2-\mu  m) \dot\phi^2+3 m (-4 \mu +(2-(\mu -5) \mu ) m+2)-12\right], \\
    \dot\psi &= \frac{\left[(\mu -4) m+2\right]\psi}{m t}-\frac{1}{192} m t\left[m^2 t^2 \dot\phi^2+6(\mu -4) m+12\right].
\end{align}
After reducing the quadratic term $\dot\phi^2/2$ using the last equation, we obtain 
\begin{align}
    \ddot{\phi}&= -\frac{96 \lambda  \dot{\psi}}{m^3 t^3}-\frac{6 (m-2) \left[(\mu -4) m+2\right]}{m^3 t^3 \dot{\phi}}-\frac{3 \lambda  \left[(\mu +2) m-6\right]}{m^2 t^2}\nonumber \\
    & + \psi \left[\frac{192 (m-2) \left[(\mu -4)m+2\right]}{m^5 t^5 \dot{\phi}}+\frac{96 \lambda  \left[(\mu +2) m-2\right]}{m^4 t^4}\right]+\frac{\left[(\mu -2)m-4\right] \dot{\phi}}{m t}, \label{syst0}\\
    \ddot{\psi}&=\frac{1}{32} \left[m (-4\mu +(\mu -2) (\mu -1) m+18)-12\right] \nonumber\\
    & + \frac{\left[m (\mu  (-\mu  m+m+4)-22)+12\right] \psi}{m^2 t^2}+\frac{2 \left(\mu -\frac{2}{m}\right) \dot{\psi}}{t},\label{system2}\\
    \dot{\phi}^2&=\frac{192 \left[(\mu -4) m+2\right] \psi}{m^4 t^4}-\frac{192 \dot{\psi}}{m^3 t^3}-\frac{6
   \left[(\mu -4) m+2\right]}{m^2 t^2}. \label{system1}
\end{align}
Equation \eqref{system1} defines $\dot\phi$, and we analyze \eqref{system2}, our equation of interest. The general solution of \eqref{system2} is 
\begin{align}
    \psi (t) & = \frac{m^2 t^2 \left[m (-4 \mu +(\mu -2) (\mu -1) m+18)-12\right]}{32 \left[m (-4 \mu +((\mu-5) \mu +2) m+30)-12\right]} \nonumber\\
    & +c_1 t^{\frac{2 \mu  m +m-4-\sqrt{m (8 \mu  m+m-96)+64}}{2 m}}+c_2t^{\frac{2 \mu  m+m-4 +\sqrt{m (8 \mu  m+m-96)+64}}{2 m}}. \label{psi-sol}
\end{align}
Taking $\psi(t)$ from \eqref{psi-sol}, calculating its time derivative, $\dot{\psi}(t)$, and substituting in \eqref{system1},  $\phi(t)$ is given through the quadrature
\begin{equation}
    \phi (t)= \pm  \int
   _1^t\frac{\sqrt{-6 \left[m (\mu -4)+2\right] \left[m^2 \eta^2-32 \psi (\eta)\right]-192 m \eta \psi
   '(\eta)}}{m^2 \eta^2}d\eta+c_3.
\end{equation}

\section{Stability Analysis of the Exact Solution}
\label{sect4.1}
Denoting 
\begin{align}
    \psi_c(t) & =\frac{1}{96} m^2 t^2 \left(\frac{2 m\left(3 \lambda ^2+m\right)}{\lambda ^2 \left[(\mu -6) m+2\right]}+3\right), \label{Nexactpsi}\\
    \phi_c(t)  & =\frac{2 \ln (t)}{\lambda}. \label{Nexactphi}
\end{align}
In this case, the evolution equation for $\psi$ is decoupled.

Specializing the  procedure in Appendix \ref{app1} to the system \eqref{system2} and \eqref{system1}, we define 
\begin{equation}
    \varepsilon(\tau)= \frac{\psi(\tau)}{\psi_c(\tau)}-1.
\end{equation}
Hence,
\begin{align}
    \ddot{\psi}(t)& = \frac{m^2 e^{2 \tau } \left[6 \lambda ^2+2 m^2+3 \lambda ^2 (\mu-4) m\right] \left[\varepsilon''(\tau )+3 \varepsilon'(\tau )+2 \varepsilon(\tau )+2\right]}{96 \lambda ^2 t^2 \left[(\mu -6)m+2\right]},\\
    \dot{\psi}(t) & = \frac{m^2 e^{2 \tau } \left[6 \lambda ^2+2 m^2+3 \lambda ^2 (\mu -4)m\right] \left[\varepsilon'(\tau )+2 \varepsilon(\tau )+2\right]}{96 \lambda ^2 t \left[(\mu -6) m+2\right]},\\
    \psi (t) & =\frac{m^2 e^{2 \tau } \left[\varepsilon(\tau )+1\right] \left[6 \lambda ^2+2 m^2+3 \lambda ^2 (\mu -4)m\right]}{96 \lambda ^2 \left[(\mu -6) m+2\right]}.
\end{align}
Combining with Equation \eqref{system2}, we obtain 
\begin{align}
    \varepsilon^{\prime \prime}(\tau )&= \underbrace{\frac{36 \lambda ^2-2 m \left[3 \lambda ^2 (\mu +2)-4 \mu +((\mu -5) \mu +2)m+30\right]+24}{6 \lambda ^2+2 m^2+3 \lambda ^2 (\mu -4) m}}_{= 0 \;\text{by}\; \text{(42)} } \nonumber\\
    & +\frac{\left[12-m (-4 \mu +((\mu -5)\mu +2) m+30)\right]}{m^2} \varepsilon(\tau )+\left(2 \mu -\frac{4}{m}-3\right) \varepsilon'(\tau),
\end{align}
where using the relation  \eqref{eq43}, 
the first term is zero. Defining $v(\tau)= \varepsilon^{\prime}(\tau)$, 
we obtain the linear dynamical system
\begin{align}
    \varepsilon^{\prime}&=v,\label{equ}\\
    v^{\prime}&=\frac{\left[12-m (-4 \mu +((\mu -5)\mu +2) m+30)\right]}{m^2} \varepsilon+\left(2 \mu -\frac{4}{m}-3\right) v. \label{eqv}
\end{align}

\subsection{Stability of the Scaling Solution}

The scaling solution corresponds to the fixed point $(\varepsilon,v)=(0,0)$.
The matrix of the linear system is defined by 
\begin{equation}
    J(\varepsilon, v)= \left(
    \begin{array}{cc}
        0 & 1 \\
        \frac{12-m \left[-4 \mu +((\mu -5) \mu +2) m+30\right]}{m^2} & 2 \mu -\frac{4}{m}-3 \\
    \end{array}
    \right).
\end{equation}
\textls[-15]{Evaluating the linear matrix around the fixed point $(\varepsilon=0, v=0)$, we obtain the eigenvalues}
\begin{align}
    \lambda_{1,2}= \frac{1}{2} \left(2 \mu -\frac{4}{m}-3 \pm\sqrt{\frac{m (8 \mu m+m-96)+64}{m^2}}\right). \label{lambda1-2}
\end{align}
The origin is a sink for 
\begin{enumerate}
    \item $0<\mu <\frac{1}{2} \left(5-\sqrt{17}\right),  \frac{2 \mu -15}{\mu ^2-5 \mu+2}+\sqrt{\frac{16 \mu ^2-120 \mu +249}{\left(\mu ^2-5 \mu +2\right)^2}}<m\leq \frac{48}{8\mu +1}-8 \sqrt{-\frac{8 \mu -35}{(8 \mu +1)^2}}$, or
    \item $ 0<\mu <\frac{1}{2}\left(5-\sqrt{17}\right),  m\geq 8 \sqrt{-\frac{8 \mu -35}{(8 \mu +1)^2}}+\frac{48}{8 \mu+1}$, or
    \item $ \mu =\frac{1}{2} \left(5-\sqrt{17}\right), -\frac{6}{-10-\sqrt{17}}<m\leq \frac{48}{1+4 \left(5-\sqrt{17}\right)}-\frac{8 \sqrt{35-4\left(5-\sqrt{17}\right)}}{1+4 \left(5-\sqrt{17}\right)}$, or
    \item $ \mu =\frac{1}{2}\left(5-\sqrt{17}\right),  m\geq \frac{48}{1+4 \left(5-\sqrt{17}\right)}+\frac{8\sqrt{35-4 \left(5-\sqrt{17}\right)}}{1+4 \left(5-\sqrt{17}\right)}$, or
    \item \textls[-15]{$\frac{1}{2} \left(5-\sqrt{17}\right)<\mu <\frac{287+146 \sqrt{3}}{148+80\sqrt{3}},  \frac{2 \mu -15}{\mu ^2-5 \mu +2}-\sqrt{\frac{16 \mu ^2-120 \mu+249}{\left(\mu ^2-5 \mu +2\right)^2}}<m\leq \frac{48}{8 \mu +1}-8 \sqrt{-\frac{8 \mu-35}{(8 \mu +1)^2}}$, or}
    \item \textls[-15]{$\frac{1}{2} \left(5-\sqrt{17}\right)<\mu <\frac{287+146 \sqrt{3}}{148+80\sqrt{3}},  8 \sqrt{-\frac{8 \mu -35}{(8 \mu +1)^2}}+\frac{48}{8 \mu +1}\leq m<\frac{2 \mu -15}{\mu ^2-5 \mu +2}+\sqrt{\frac{16 \mu ^2-120 \mu +249}{\left(\mu ^2-5 \mu+2\right)^2}}$, or}
    \item $ \frac{287+146 \sqrt{3}}{148+80 \sqrt{3}}\leq \mu<\frac{146 \sqrt{3}-287}{80 \sqrt{3}-148},  \frac{2 \mu -15}{\mu ^2-5 \mu+2}-\sqrt{\frac{16 \mu ^2-120 \mu +249}{\left(\mu ^2-5 \mu +2\right)^2}}<m\leq \frac{48}{8\mu +1}-8 \sqrt{-\frac{8 \mu -35}{(8 \mu +1)^2}}$.
\end{enumerate}
The origin is a source for 
\begin{enumerate}
    \item $\frac{287+146 \sqrt{3}}{148+80 \sqrt{3}}<\mu \leq \frac{146 \sqrt{3}-287}{80\sqrt{3}-148},  8 \sqrt{-\frac{8 \mu -35}{(8 \mu +1)^2}}+\frac{48}{8 \mu +1}\leq m<\frac{2 \mu -15}{\mu^2-5 \mu +2}+\sqrt{\frac{16 \mu ^2-120 \mu +249}{\left(\mu ^2-5 \mu+2\right)^2}}$, or
    \item $\frac{146 \sqrt{3}-287}{80 \sqrt{3}-148}<\mu<\frac{35}{8},  \frac{2 \mu -15}{\mu ^2-5 \mu +2}-\sqrt{\frac{16 \mu ^2-120 \mu+249}{\left(\mu ^2-5 \mu +2\right)^2}}<m\leq \frac{48}{8 \mu +1}-8 \sqrt{-\frac{8 \mu-35}{(8 \mu +1)^2}}$, or
    \item $\frac{146 \sqrt{3}-287}{80 \sqrt{3}-148}<\mu<\frac{35}{8},    8 \sqrt{-\frac{8 \mu -35}{(8 \mu +1)^2}}+\frac{48}{8 \mu +1}\leq m<\frac{2 \mu -15}{\mu ^2-5 \mu +2}+\sqrt{\frac{16 \mu ^2-120 \mu +249}{\left(\mu ^2-5 \mu+2\right)^2}}$, or
    \item $ \frac{35}{8}\leq \mu <\frac{1}{2}\left(5+\sqrt{17}\right),  \frac{2 \mu -15}{\mu ^2-5 \mu +2}-\sqrt{\frac{16 \mu ^2-120\mu +249}{\left(\mu ^2-5 \mu +2\right)^2}}<m<\frac{2 \mu -15}{\mu ^2-5 \mu+2}+\sqrt{\frac{16 \mu ^2-120 \mu +249}{\left(\mu ^2-5 \mu +2\right)^2}}$, or
    \item $\mu =\frac{1}{2} \left(5+\sqrt{17}\right),  m>-\frac{6}{\sqrt{17}-10}$, or
    \item $\mu >\frac{1}{2} \left(5+\sqrt{17}\right), m>\frac{2 \mu -15}{\mu ^2-5 \mu  +2}+\sqrt{\frac{16 \mu ^2-120 \mu +249}{\left(\mu ^2-5 \mu +2\right)^2}}$. 
\end{enumerate}
The origin is a saddle for
\begin{enumerate}
    \item $0<\mu <\frac{1}{2} \left(5-\sqrt{17}\right),  0<m<\frac{2 \mu -15}{\mu ^2-5 \mu+2}+\sqrt{\frac{16 \mu ^2-120 \mu +249}{\left(\mu ^2-5 \mu +2\right)^2}}$, or
    \item $\mu =\frac{1}{2} \left(5-\sqrt{17}\right),  0<m<-\frac{6}{-10-\sqrt{17}}$, or
    \item $\frac{1}{2} \left(5-\sqrt{17}\right)<\mu <\frac{1}{2} \left(5+\sqrt{17}\right), 0<m<\frac{2 \mu -15}{\mu ^2-5 \mu +2}-\sqrt{\frac{16 \mu ^2-120 \mu +249}{\left(\mu ^2-5 \mu
   +2\right)^2}}$, or
   \item $\frac{1}{2}\left(5-\sqrt{17}\right)<\mu<\frac{1}{2}\left(5+\sqrt{17}\right), m>\frac{2 \mu -15}{\mu ^2-5 \mu +2}+\sqrt{\frac{16 \mu ^2-120 \mu +249}{\left(\mu ^2-5 \mu
   +2\right)^2}}$
    \item $\mu =\frac{1}{2} \left(5+\sqrt{17}\right), 
   0<m<-\frac{6}{\sqrt{17}-10}$, or
   \item $\mu >\frac{1}{2} \left(5+\sqrt{17}\right), 
   0<m<\frac{2 \mu -15}{\mu ^2-5 \mu +2}+\sqrt{\frac{16 \mu ^2-120 \mu +249}{\left(\mu ^2-5 \mu
   +2\right)^2}}$.
\end{enumerate}

\begin{figure}[H]
    \includegraphics[scale=0.7]{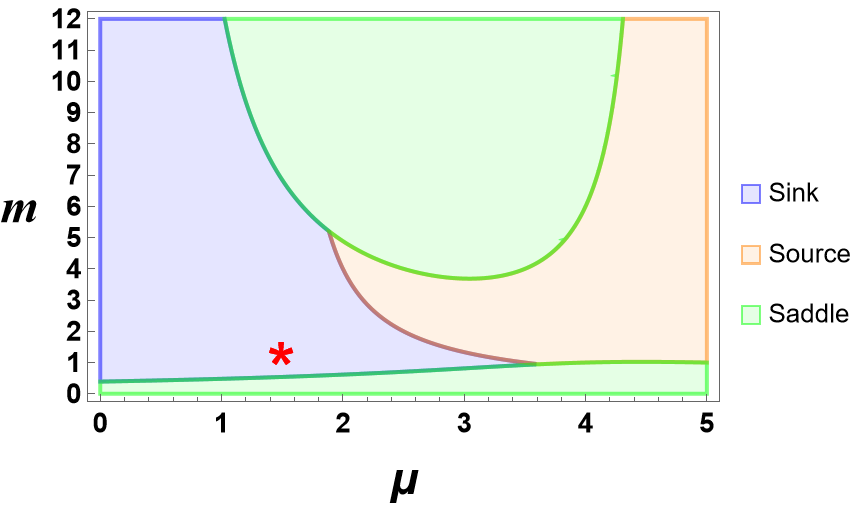}
    \caption{Allowed regions by the parameters $\mu$ and $m$, for the sink, source, and saddle cases. The red star represents the point in the parameter space with the best-fit values of $\mu$ and $m$ according to the analysis in Section \ref{sec:fitting}.}
    \label{allowed-regions-1}
\end{figure}

In Figure \ref{allowed-regions-1}, we can see the allowed regions for the parameters $\mu$ and $m$ for the sink, source, and saddle cases.

\begin{figure}[H]
   \begin{subfigure}{0.45\textwidth}
    \includegraphics[scale=0.4]{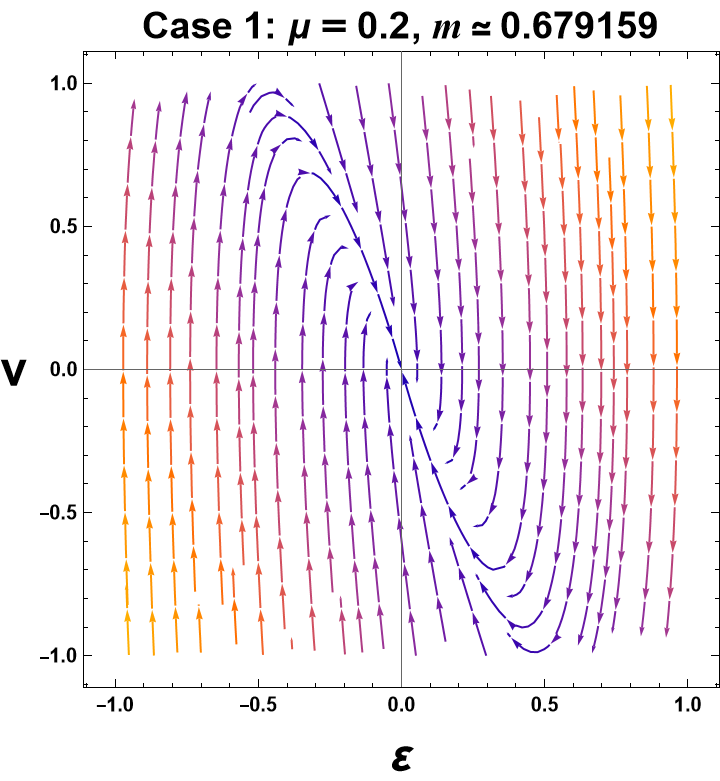} 
\end{subfigure}
\begin{subfigure}{0.45\textwidth}
   \includegraphics[scale=0.4]{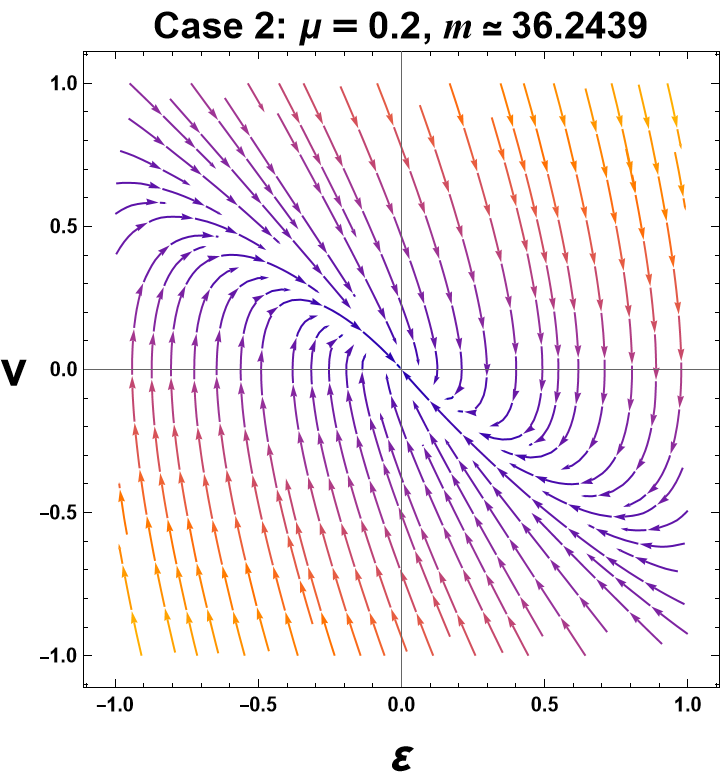} 
\end{subfigure}\\
   \begin{subfigure}{0.45\textwidth}
    \includegraphics[scale=0.4]{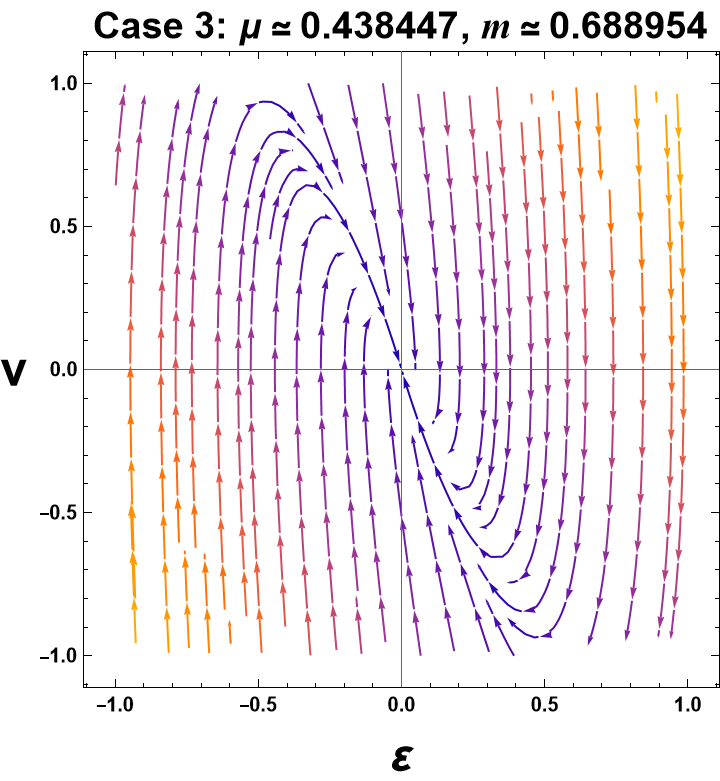} 
\end{subfigure}
\begin{subfigure}{0.45\textwidth}
    \includegraphics[scale=0.4]{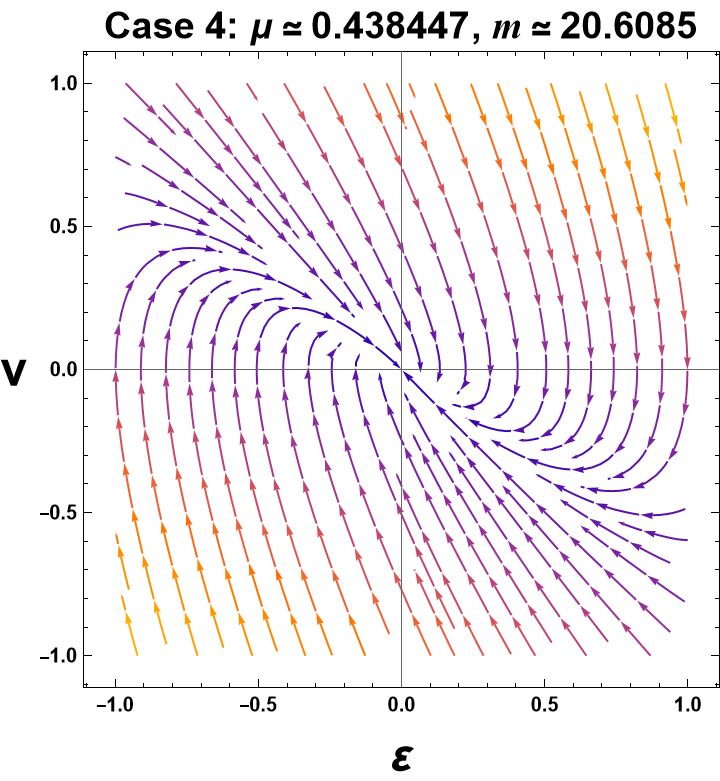} 
\end{subfigure}\\
   \begin{subfigure}{0.45\textwidth}
   \includegraphics[scale=0.4]{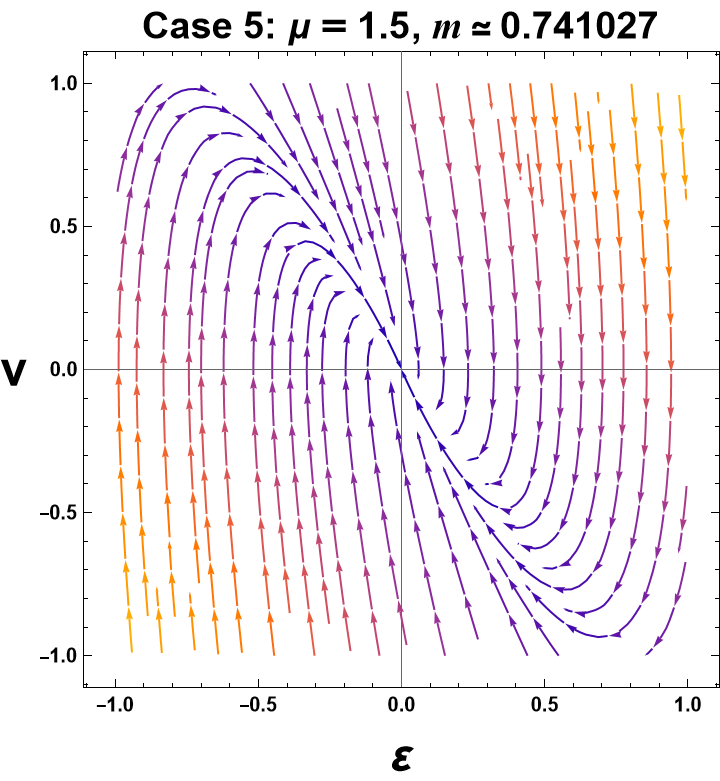} 
\end{subfigure}
\begin{subfigure}{0.45\textwidth}
  \includegraphics[scale=0.4]{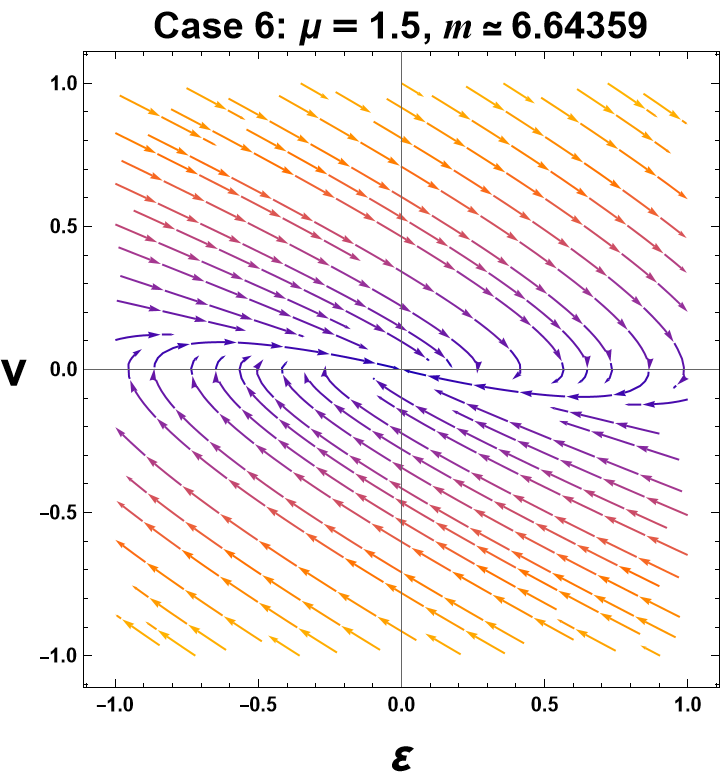}
\end{subfigure}\\
\begin{subfigure}{0.4\textwidth}
\hspace{85pt} \includegraphics[scale=0.45]{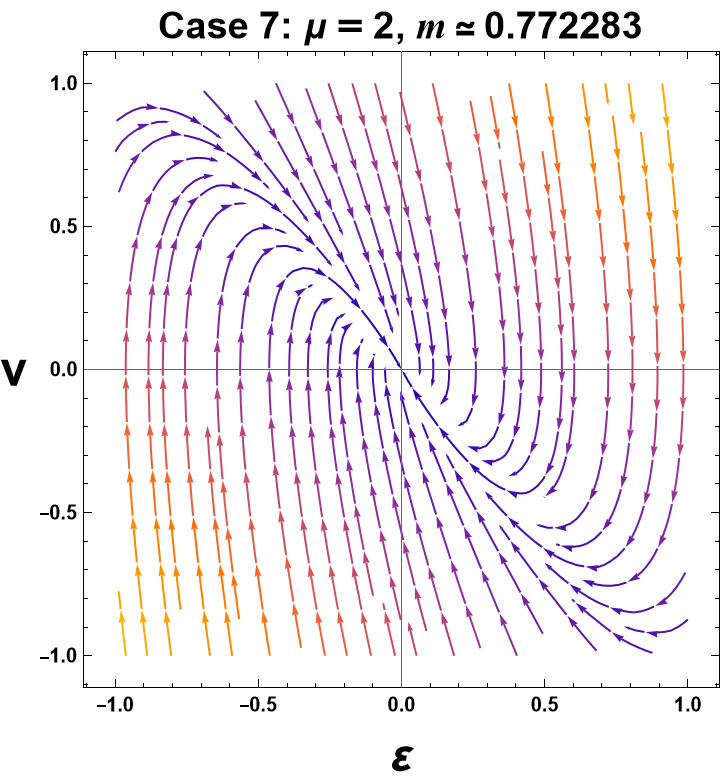} 
\end{subfigure}
\caption{Flow of system \eqref{equ} and \eqref{eqv} for the sink cases. The streamlines are colored by default according to the magnitude of the vector field, with the arrow pointing in the time-increasing direction.} \label{fig:sinkcase}
\end{figure}

\begin{figure}[H]
   \begin{subfigure}{0.45\textwidth}
    \includegraphics[scale=0.45]{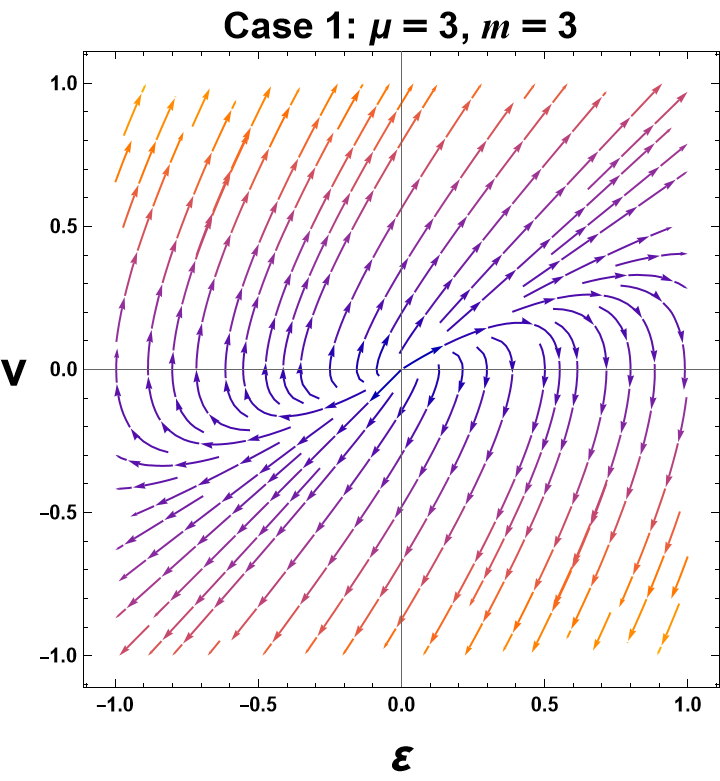}
\end{subfigure}
\hspace{6pt}
\begin{subfigure}{0.45\textwidth}
    \includegraphics[scale=0.45]{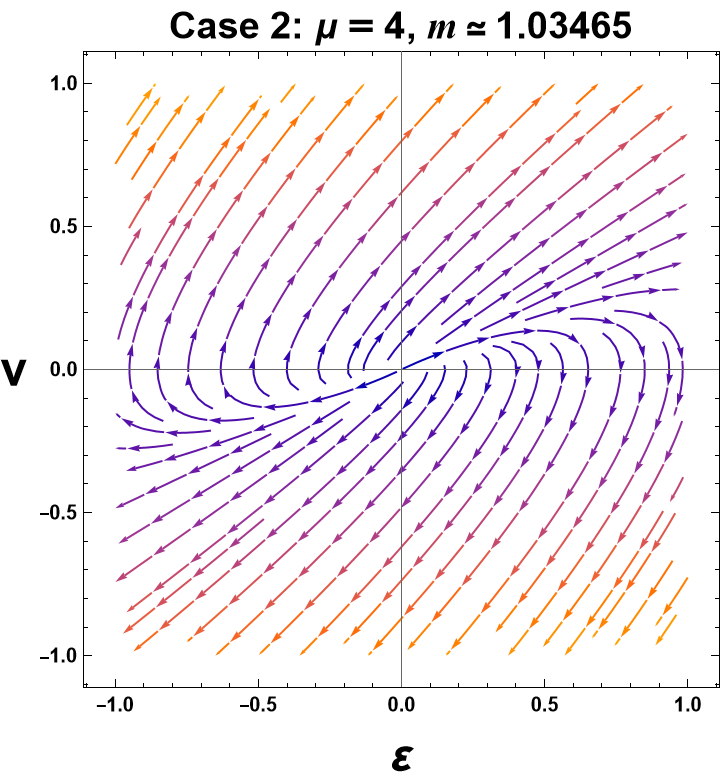}
\end{subfigure}\\
   \begin{subfigure}{0.45\textwidth}
    \includegraphics[scale=0.45]{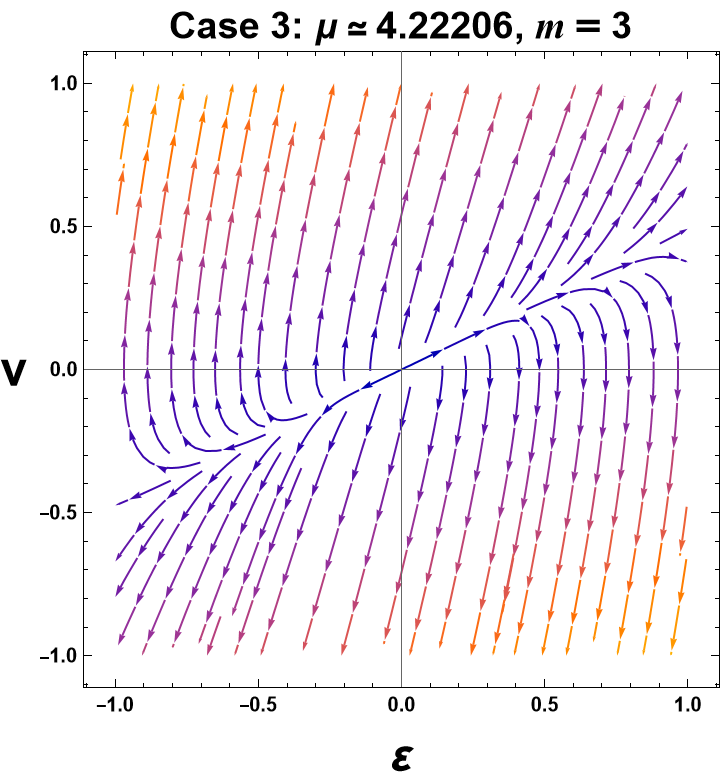}
\end{subfigure}\hspace{6pt}
\begin{subfigure}{0.45\textwidth}
    \includegraphics[scale=0.45]{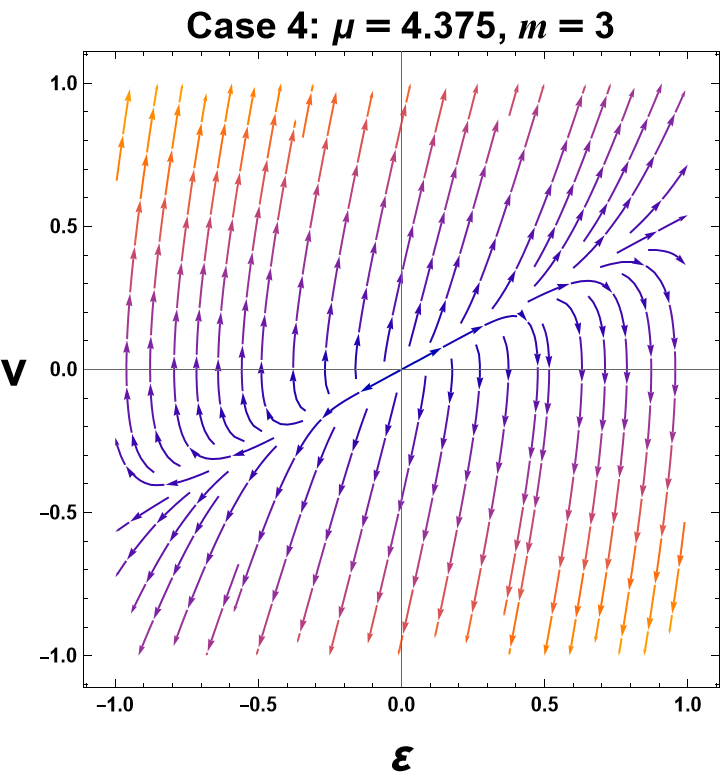}
\end{subfigure}\\
   \begin{subfigure}{0.45\textwidth}
    \includegraphics[scale=0.45]{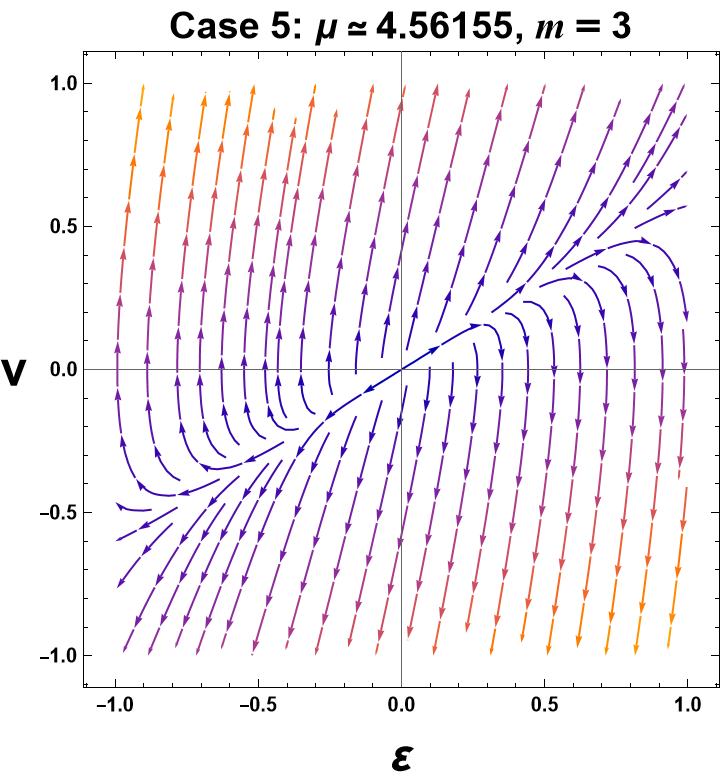}
\end{subfigure}\hspace{6pt}
\begin{subfigure}{0.45\textwidth}
    \includegraphics[scale=0.45]{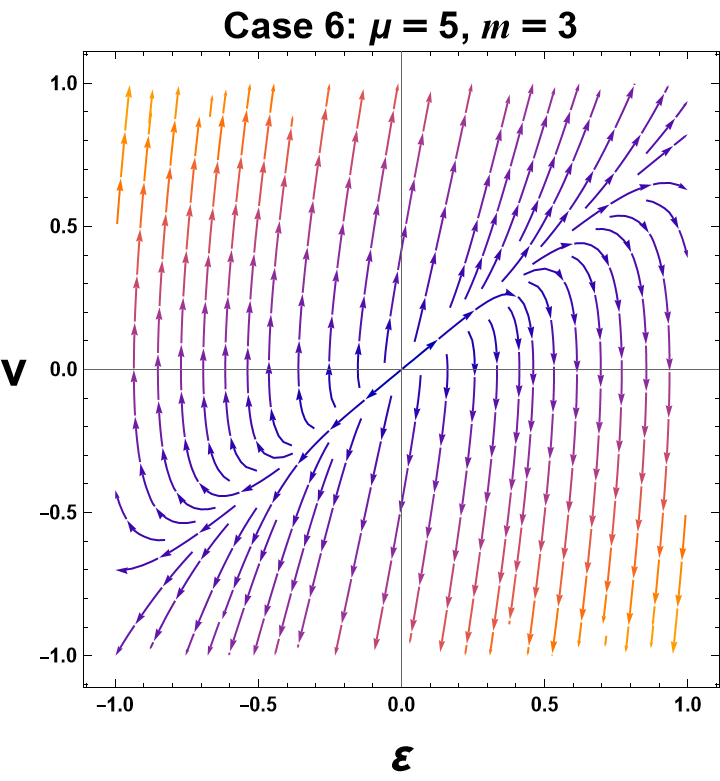}
\end{subfigure}

    \caption{Flow of system \eqref{equ} and \eqref{eqv} for the source cases. The streamlines are colored by default according to the magnitude of the vector field, with the arrow pointing in the time-increasing direction.} \label{fig:sourcecase}
\end{figure}

\begin{figure}[H]
   \begin{subfigure}{0.45\textwidth}
    \includegraphics[scale=0.4]{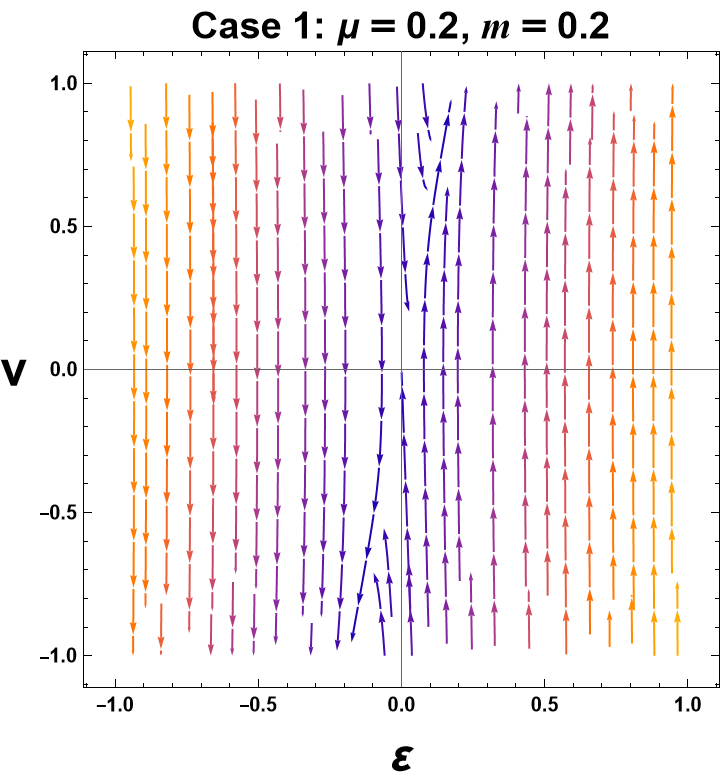}
\end{subfigure}
\begin{subfigure}{0.45\textwidth}
    \includegraphics[scale=0.4]{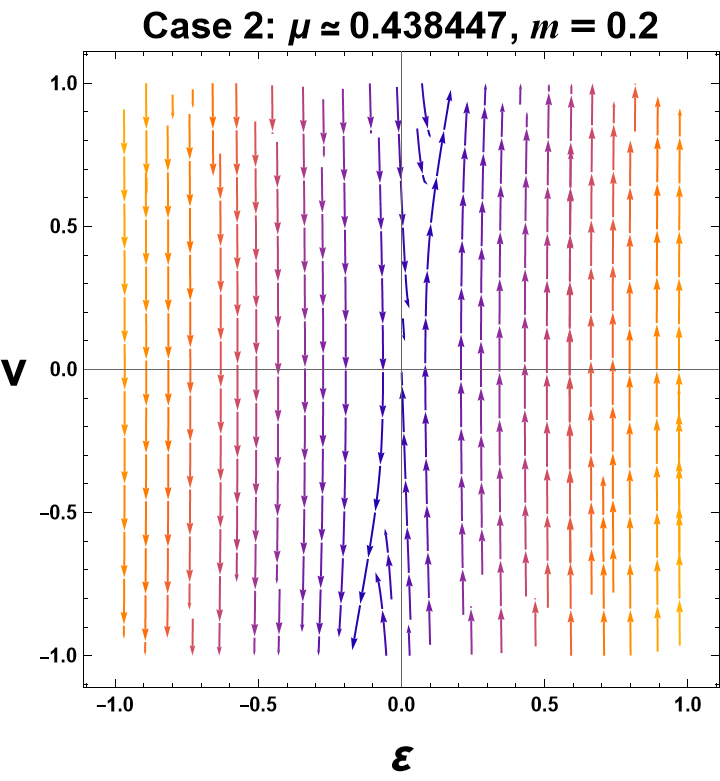}
\end{subfigure}\\
   \begin{subfigure}{0.45\textwidth}
    \includegraphics[scale=0.4]{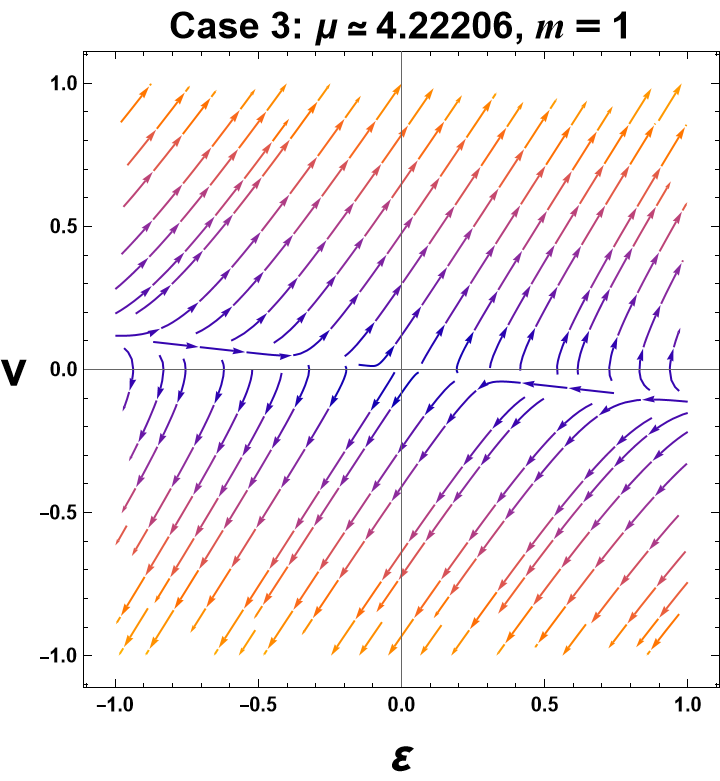}
\end{subfigure}
\begin{subfigure}{0.45\textwidth}
    \includegraphics[scale=0.4]{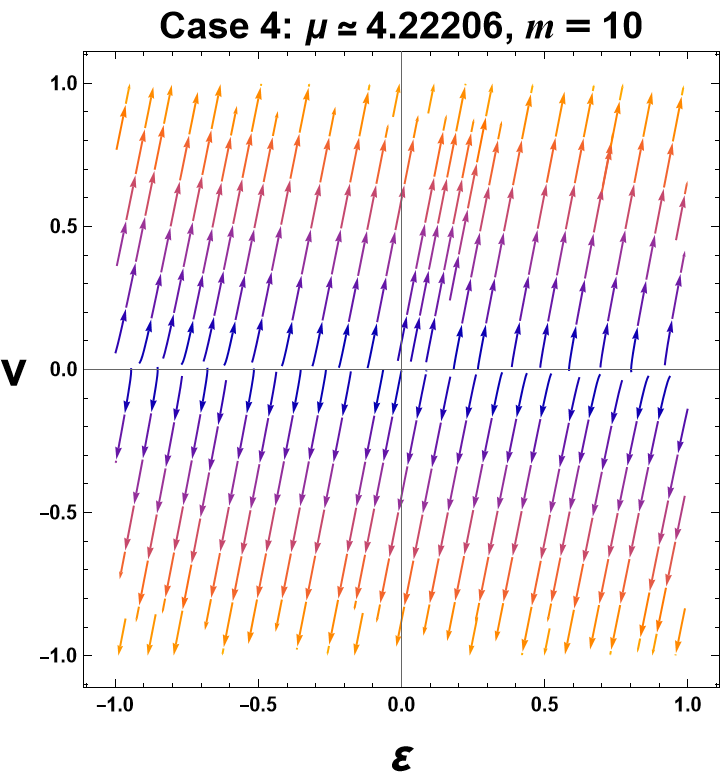}
\end{subfigure}\\
   \begin{subfigure}{0.45\textwidth}
    \includegraphics[scale=0.4]{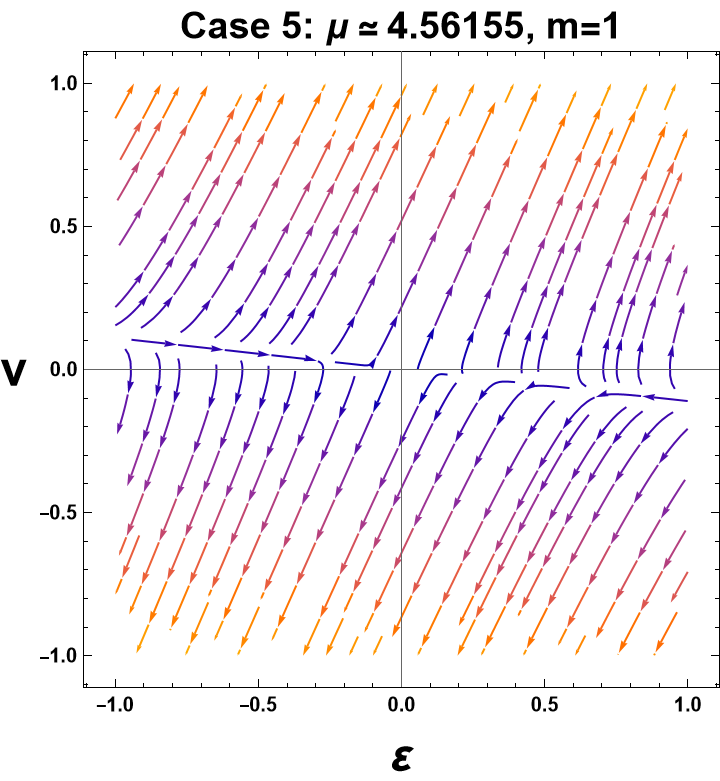}
\end{subfigure}
\begin{subfigure}{0.45\textwidth}
    \includegraphics[scale=0.4]{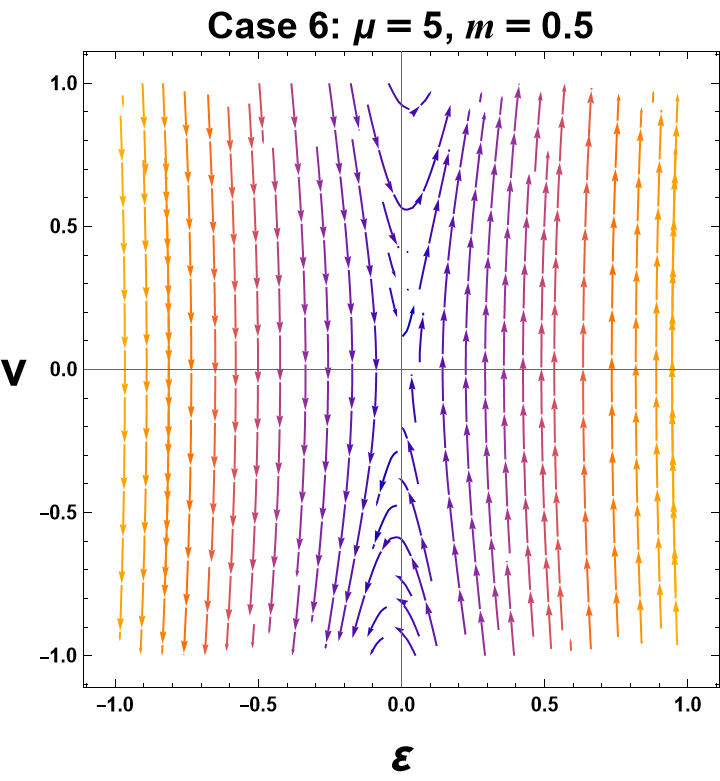}
\end{subfigure}
    \caption{Flow of system \eqref{equ} and \eqref{eqv} for the saddle cases. The streamlines are colored by default according to the magnitude of the vector field, with the arrow pointing in the time-increasing direction.} \label{fig:saddlecase}
\end{figure}

Now, we will show the flow of system \eqref{equ} and \eqref{eqv}. In Figure \ref{fig:sinkcase}, we show some phase space diagrams for sink cases, in Figure \ref{fig:sourcecase} the source cases, and in Figure \ref{fig:saddlecase}, the saddle cases. 
We used specific numerical values of $m$ and $\mu$ representing the regions depicted in Figure \ref{allowed-regions-1}. These values are chosen to meet the specified constraints. These regions are then utilized to determine the priors for the Bayesian analysis carried out in Section \ref{sec:fitting}, where we calculate the best-fit parameters at the \(1\sigma\) confidence level (CL) using the affine-invariant Markov chain Monte Carlo (MCMC) method \cite{Goodman_Ensemble_2010}, which is implemented in the Python code emcee version 3.1.6 
 \cite{Foreman-Mackey:2012any}. 

This approach offers the main advantages of combining dynamical systems analysis and observational testing, as it allows for using priors from the former analysis to inform the latter, aligning with the anticipated late-time behavior in a cosmological model.

\subsection{Dynamics at Infinity}
To investigate the late-time dynamics of system \eqref{equ} and \eqref{eqv} at infinity, we define the following variables:
\begin{equation}
    X=\frac{\varepsilon}{\sqrt{1+\varepsilon^2+v^2}},\quad Y=\frac{v}{\sqrt{1+\varepsilon^2+v^2}}.
\end{equation}
Hence, we obtain the system 
\begin{align}
\label{infty-eq-1}
    \frac{dX}{d\tau}&=\frac{Y}{m^2} \bigg[m \left(m X ((\mu -5) \mu  X+X+(3-2 \mu ) Y)+m+2 X ((15-2 \mu ) X+2 Y)\right)-12 X^2\bigg],\\
   \label{infty-eq-2} \frac{dY}{d\tau}&=-\frac{1}{m^2} \bigg[X \left(m^2 \left((\mu -5) \mu -\left(((\mu -5) \mu +1) Y^2\right)+2\right)\right.\nonumber\\
    & \left.+2 (2\mu -15) m \left(Y^2-1\right)+12 \left(Y^2-1\right)\right)+m Y \left(Y^2-1\right) ((2 \mu -3) m-4)\bigg],
\end{align}
defined in the compact set $\{(X,Y)\in \mathbb{R}^2: X^2+Y^2\leq 1\}$. 

There are five total equilibrium points for the system \eqref{infty-eq-1} and \eqref{infty-eq-2}. However, the existence conditions for each point and their respective eigenvalues depend on both $\mu$ and $m$; therefore, the analysis is not trivial. The stationary points in the coordinates $(X,Y)$ are

\begin{enumerate}
    \item $A=(0,0),$ with eigenvalues \eqref{lambda1-2}. 
The stability analysis is the same as in the previous finite regime.
   \item $B,C=\left( \pm \frac{k_2 m (k_1+(2 \mu -3)m-4)}{\sqrt{k} \left[2 m (-4 \mu +((\mu -5)\mu +2) m+30)-24\right]},\pm \frac{k_2}{\sqrt{k}}\right)$ are antipodal points, where 
   
 \begin{align}
        \label{def-of-k}
        k&=2 \left(\mu ^4-10 \mu ^3+31 \mu ^2-22 \mu +10\right) m^4\\ 
       \nonumber &-8 \left(2 \mu ^3-25\mu ^2+81 \mu -21\right) m^3-8 \left(2 \mu ^2+30 \mu -223\right) m^2+96 (2\mu -15) m+288,\\
        \label{def-of-k1}
        k_1&=\sqrt{m (8 \mu  m+m-96)+64},\\
        \label{def-of-k2-1}
        k_2&= \sqrt{\begin{array}{c}
        m^3 (3 (k_1+68)-2 \mu  (k_1+4 \mu  (2 \mu -25)+320))\\
        +4 m^2(k_1-4 \mu  (\mu +15)+436)+(2 (\mu -1) \mu  ((\mu -9) \mu +21)+13)m^4 \\
        +96 (2 \mu -15) m+288 \end{array}}.
   \end{align}
These points exist for the following intervals
\begin{enumerate}
\item[(1)] $\mu =\frac{1}{2} \left(5-\sqrt{17}\right),  0\leq m<\frac{1}{83}
   \left(60-6 \sqrt{17}\right)$, or 
   \item[(2)] $\mu =\frac{1}{2}
   \left(5-\sqrt{17}\right),  \frac{1}{83} \left(60-6
   \sqrt{17}\right)<m<\frac{8 \sqrt{15+4 \sqrt{17}}-48}{4
   \sqrt{17}-21}$, or 
   \item[(3)] $\mu =\frac{1}{2}
   \left(5-\sqrt{17}\right),  m>\frac{-48-8 \sqrt{15+4 \sqrt{17}}}{4
   \sqrt{17}-21}$, or 
   \item[(4)] $\frac{35}{8}<\mu <\frac{1}{2}
   \left(5+\sqrt{17}\right),  m>\frac{2 \mu-15}{\mu ^2-5 \mu +2}+\sqrt{\frac{16 \mu ^2-120\mu +249}{\left(\mu ^2-5 \mu +2\right)^2}}$, or 
   \item[(5)] $\frac{1}{4} \left(11-2 \sqrt{3}\right)<\mu<\frac{1}{4} \left(11+2 \sqrt{3}\right), 
   m>\frac{2 \mu -15}{\mu ^2-5 \mu
   +2}+\sqrt{\frac{16 \mu ^2-120 \mu
   +249}{\left(\mu ^2-5 \mu +2\right)^2}}$, or 
   \item[(6)] $\frac{1}{2} \left(5-\sqrt{17}\right)<\mu<\frac{1}{4} \left(11-2 \sqrt{3}\right), 
   m>\frac{2 \mu -15}{\mu ^2-5 \mu
   +2}+\sqrt{\frac{16 \mu ^2-120 \mu
   +249}{\left(\mu ^2-5 \mu +2\right)^2}}$, or 
   \item[(7)] $ \mu >\frac{1}{2} \left(5+\sqrt{17}\right), 
   m>\frac{2 \mu -15}{\mu ^2-5 \mu
   +2}+\sqrt{\frac{16 \mu ^2-120 \mu
   +249}{\left(\mu ^2-5 \mu +2\right)^2}}$, or 
   \item[(8)] $\frac{1}{4} \left(11+2 \sqrt{3}\right)<\mu \leq
   \frac{35}{8},  m>\frac{2 \mu -15}{\mu ^2-5\mu +2}+\sqrt{\frac{16 \mu ^2-120 \mu+249}{\left(\mu ^2-5 \mu +2\right)^2}}$, or 
   \item[(9)] $m=\frac{8 \left(\sqrt{13+4 \sqrt{3}}-6\right)}{4\sqrt{3}-23},  \mu =\frac{1}{4} \left(11-2
   \sqrt{3}\right)$, or 
   \item[(10)] $\mu =\frac{1}{4} \left(11-2 \sqrt{3}\right), 
   0\leq m<\frac{152+16 \sqrt{3}-32 \sqrt{13+4\sqrt{3}}}{55+4 \sqrt{3}}$, or
   \item[(11)] $\mu =\frac{1}{4} \left(11-2 \sqrt{3}\right), 
   \frac{152+16 \sqrt{3}-32 \sqrt{13+4
   \sqrt{3}}}{55+4 \sqrt{3}}<m<\frac{8 \sqrt{13+4\sqrt{3}}-48}{4 \sqrt{3}-23}$, or 
   \item[(12)] $\mu=\frac{1}{4} \left(11-2 \sqrt{3}\right), 
   m>\frac{-48-8 \sqrt{13+4 \sqrt{3}}}{4
   \sqrt{3}-23}$, or 
   \item[(13)] $m=\frac{8 \left(6+\sqrt{13-4
   \sqrt{3}}\right)}{23+4 \sqrt{3}},  \mu=\frac{1}{4} \left(11+2 \sqrt{3}\right)$, or 
   \item[(14)] $\mu =\frac{1}{4} \left(11+2 \sqrt{3}\right), 
   0\leq m<\frac{48-8 \sqrt{13-4 \sqrt{3}}}{23+4\sqrt{3}}$, or 
   \item[(15)] $\mu =\frac{1}{4} \left(11+2 \sqrt{3}\right), 
   \frac{48+8 \sqrt{13-4 \sqrt{3}}}{23+4
   \sqrt{3}}<m<\frac{-152+16 \sqrt{3}-32
   \sqrt{13-4 \sqrt{3}}}{4 \sqrt{3}-55}$, or 
   \item[(16)] $\mu =\frac{1}{4} \left(11+2 \sqrt{3}\right), 
   m>\frac{-152+16 \sqrt{3}-32 \sqrt{13-4
   \sqrt{3}}}{4 \sqrt{3}-55}$, or 
   \item[(17)] $\mu =\frac{1}{2} \left(5+\sqrt{17}\right), 
   0\leq m<-\frac{6}{\sqrt{17}-10}$, or 
   \item[(18)] $\mu =\frac{1}{2} \left(5+\sqrt{17}\right), 
   m>-\frac{6}{\sqrt{17}-10}$, or 
   \item[(19)] $\mu =\frac{35}{8},  m=\frac{4}{3}$, or 
   \item[(20)] $\mu >\frac{1}{2} \left(5+\sqrt{17}\right), 
   0\leq m<\frac{2 \mu -15}{\mu ^2-5 \mu
   +2}+\sqrt{\frac{16 \mu ^2-120 \mu
   +249}{\left(\mu ^2-5 \mu +2\right)^2}}$, or 
   \item[(21)] $ 0\leq \mu <\frac{1}{2}
   \left(5-\sqrt{17}\right),  0\leq m<\frac{2\mu -15}{\mu ^2-5 \mu +2}+\sqrt{\frac{16 \mu^2-120 \mu +249}{\left(\mu ^2-5 \mu+2\right)^2}}$, or \item[(22)] $\frac{1}{4} \left(11+2 \sqrt{3}\right)<\mu \leq\frac{35}{8},  0\leq m<\frac{2 \mu -15}{\mu
   ^2-5 \mu +2}-\sqrt{\frac{16 \mu ^2-120 \mu+249}{\left(\mu ^2-5 \mu +2\right)^2}}$, or 
   \item[(23)] $\frac{35}{8}<\mu <\frac{1}{2}
   \left(5+\sqrt{17}\right),  0\leq m<\frac{2\mu -15}{\mu ^2-5 \mu +2}-\sqrt{\frac{16 \mu^2-120 \mu +249}{\left(\mu ^2-5 \mu+2\right)^2}}$, or \item[(24)] $\frac{1}{2} \left(5-\sqrt{17}\right)<\mu<\frac{1}{4} \left(11-2 \sqrt{3}\right), 
   0\leq m<\frac{2 \mu -15}{\mu ^2-5 \mu
   +2}-\sqrt{\frac{16 \mu ^2-120 \mu
   +249}{\left(\mu ^2-5 \mu +2\right)^2}}$, or 
   \item[(25)] $\frac{1}{4} \left(11-2 \sqrt{3}\right)<\mu<\frac{1}{4} \left(11+2 \sqrt{3}\right), 
   0\leq m<\frac{2 \mu -15}{\mu ^2-5 \mu
   +2}-\sqrt{\frac{16 \mu ^2-120 \mu
   +249}{\left(\mu ^2-5 \mu +2\right)^2}}$, or 
   \item[(26)] $0\leq \mu <\frac{1}{4} \left(11-2\sqrt{3}\right),  m=\frac{-64\sqrt{\frac{35-8 \mu }{(8 \mu +1)^2}} \mu -8\sqrt{\frac{35-8 \mu }{(8 \mu +1)^2}}+48}{8 \mu
   +1}$, or 
   \item[(27)] $0\leq \mu <\frac{1}{4} \left(11-2\sqrt{3}\right),  m=\frac{-64\sqrt{\frac{35-8 \mu }{(8 \mu +1)^2}} \mu -8\sqrt{\frac{35-8 \mu }{(8 \mu +1)^2}}-48}{-8\mu -1}$, or 
   \item[(28)] $0\leq \mu <\frac{1}{2}
   \left(5-\sqrt{17}\right),  \frac{2 \mu-15}{\mu ^2-5 \mu +2}+\sqrt{\frac{16 \mu ^2-120\mu +249}{\left(\mu ^2-5 \mu
   +2\right)^2}}<m<\frac{48}{8 \mu +1}-8
   \sqrt{-\frac{8 \mu -35}{(8 \mu +1)^2}}$, or 
   \item[(29)] $0\leq \mu <\frac{1}{2}
   \left(5-\sqrt{17}\right),  m>8
   \sqrt{-\frac{8 \mu -35}{(8 \mu
   +1)^2}}+\frac{48}{8 \mu +1}$, or 
   \item[(30)] $\frac{1}{4} \left(11+2 \sqrt{3}\right)<\mu<\frac{35}{8},  m=\frac{-64 \sqrt{\frac{35-8\mu }{(8 \mu +1)^2}} \mu -8 \sqrt{\frac{35-8\mu }{(8 \mu +1)^2}}+48}{8 \mu +1}$, or 
   \item[(31)] $\frac{1}{4} \left(11+2 \sqrt{3}\right)<\mu<\frac{35}{8},  m=\frac{-64 \sqrt{\frac{35-8\mu }{(8 \mu +1)^2}} \mu -8 \sqrt{\frac{35-8\mu }{(8 \mu +1)^2}}-48}{-8 \mu -1}$, or 
   \item[(32)] $\frac{1}{4} \left(11+2 \sqrt{3}\right)<\mu \leq\frac{35}{8},  \frac{2 \mu -15}{\mu ^2-5 \mu
   +2}-\sqrt{\frac{16 \mu ^2-120 \mu
   +249}{\left(\mu ^2-5 \mu+2\right)^2}}<m<\frac{48}{8 \mu +1}-8\sqrt{-\frac{8 \mu -35}{(8 \mu +1)^2}}$, or 
   \item[(33)] $\frac{1}{4} \left(11+2 \sqrt{3}\right)<\mu
   <\frac{35}{8},  8 \sqrt{-\frac{8 \mu -35}{(8\mu +1)^2}}+\frac{48}{8 \mu +1}<m<\frac{2 \mu-15}{\mu ^2-5 \mu +2}+\sqrt{\frac{16 \mu ^2-120\mu +249}{\left(\mu ^2-5 \mu +2\right)^2}}$, or 
   \item[(34)] $\frac{1}{4} \left(11-2 \sqrt{3}\right)<\mu<\frac{1}{4} \left(11+2 \sqrt{3}\right),  m=\frac{-64 \sqrt{\frac{35-8 \mu }{(8 \mu
   +1)^2}} \mu -8 \sqrt{\frac{35-8 \mu }{(8 \mu+1)^2}}+48}{8 \mu +1}$, or 
   \item[(35)] $\frac{1}{4} \left(11-2 \sqrt{3}\right)<\mu<\frac{1}{4} \left(11+2 \sqrt{3}\right),  m=\frac{-64 \sqrt{\frac{35-8 \mu }{(8 \mu
   +1)^2}} \mu -8 \sqrt{\frac{35-8 \mu }{(8 \mu+1)^2}}-48}{-8 \mu -1}$, or 
   \item[(36)] $\frac{1}{4} \left(11-2 \sqrt{3}\right)<\mu<\frac{1}{4} \left(11+2 \sqrt{3}\right), 
   \frac{2 \mu -15}{\mu ^2-5 \mu
   +2}-\sqrt{\frac{16 \mu ^2-120 \mu
   +249}{\left(\mu ^2-5 \mu+2\right)^2}}<m<\frac{48}{8 \mu +1}-8\sqrt{-\frac{8 \mu -35}{(8 \mu +1)^2}}$, or 
   \item[(37)] $\frac{1}{4} \left(11-2 \sqrt{3}\right)<\mu<\frac{1}{4} \left(11+2 \sqrt{3}\right),  8
   \sqrt{-\frac{8 \mu -35}{(8 \mu+1)^2}}+\frac{48}{8 \mu +1}<m<\frac{2 \mu-15}{\mu ^2-5 \mu +2}+ \linebreak \sqrt{\frac{16 \mu ^2-120\mu +249}{\left(\mu ^2-5 \mu +2\right)^2}}$, or 
   
   \item[(38)] $\frac{1}{2} \left(5-\sqrt{17}\right)<\mu<\frac{1}{4} \left(11-2 \sqrt{3}\right), 
   \frac{2 \mu -15}{\mu ^2-5 \mu
   +2}-\sqrt{\frac{16 \mu ^2-120 \mu
   +249}{\left(\mu ^2-5 \mu+2\right)^2}}<m<\frac{48}{8 \mu +1}-\linebreak 8\sqrt{-\frac{8 \mu -35}{(8 \mu +1)^2}}$, or 
   \item[(39)] $\frac{1}{2} \left(5-\sqrt{17}\right)<\mu<\frac{1}{4} \left(11-2 \sqrt{3}\right),  8
   \sqrt{-\frac{8 \mu -35}{(8 \mu+1)^2}}+\frac{48}{8 \mu +1}<m<\frac{2 \mu-15}{\mu ^2-5 \mu +2}+\linebreak \sqrt{\frac{16 \mu ^2-120\mu +249}{\left(\mu ^2-5 \mu +2\right)^2}}$, or 
   \item[(40)] $\frac{35}{8}<\mu <\frac{1}{2}
   \left(5+\sqrt{17}\right),$ \newline  $\frac{2 \mu-15}{\mu ^2-5 \mu +2}-\sqrt{\frac{16 \mu ^2-120\mu +249}{\left(\mu ^2-5 \mu
   +2\right)^2}}<m<\frac{2 \mu -15}{\mu ^2-5 \mu+2} +\sqrt{\frac{16 \mu ^2-120 \mu
   +249}{\left(\mu ^2-5 \mu +2\right)^2}}.$
   \end{enumerate}
   
Recall that the flow in a neighborhood of antipodal points is topologically equivalent, and it may be reversed \cite{perko}, but in this case, both points share the same eigenvalues. Therefore, the stability is the same for both $B$ and $C.$ We will write $\lambda_1(B,C)=f_1(\mu,m)$ and $\lambda_2(B,C)=f_2(\mu,m)$ to represent their eigenvalues. Given that many existence conditions and the eigenvalues depend on both free parameters, we will only consider some cases to analyze the stability of $B$ and $C.$ In particular, if we only consider the existence conditions for which the points are hyperbolic (most), we verify by numerical inspection that the points can never be attractors. However, they can be sources or saddles. When one parameter is fixed and the other free, the behavior is as depicted in Figure \ref{fig:stability-of-BandC-2D-plots}. On the other hand, when both parameters are free, the behaviour is as shown in Figure \ref{fig:stability-of-BandC-3D-plots}.
   \begin{figure}[H]
   \begin{subfigure}{0.4\textwidth}
       \includegraphics[scale=0.45]{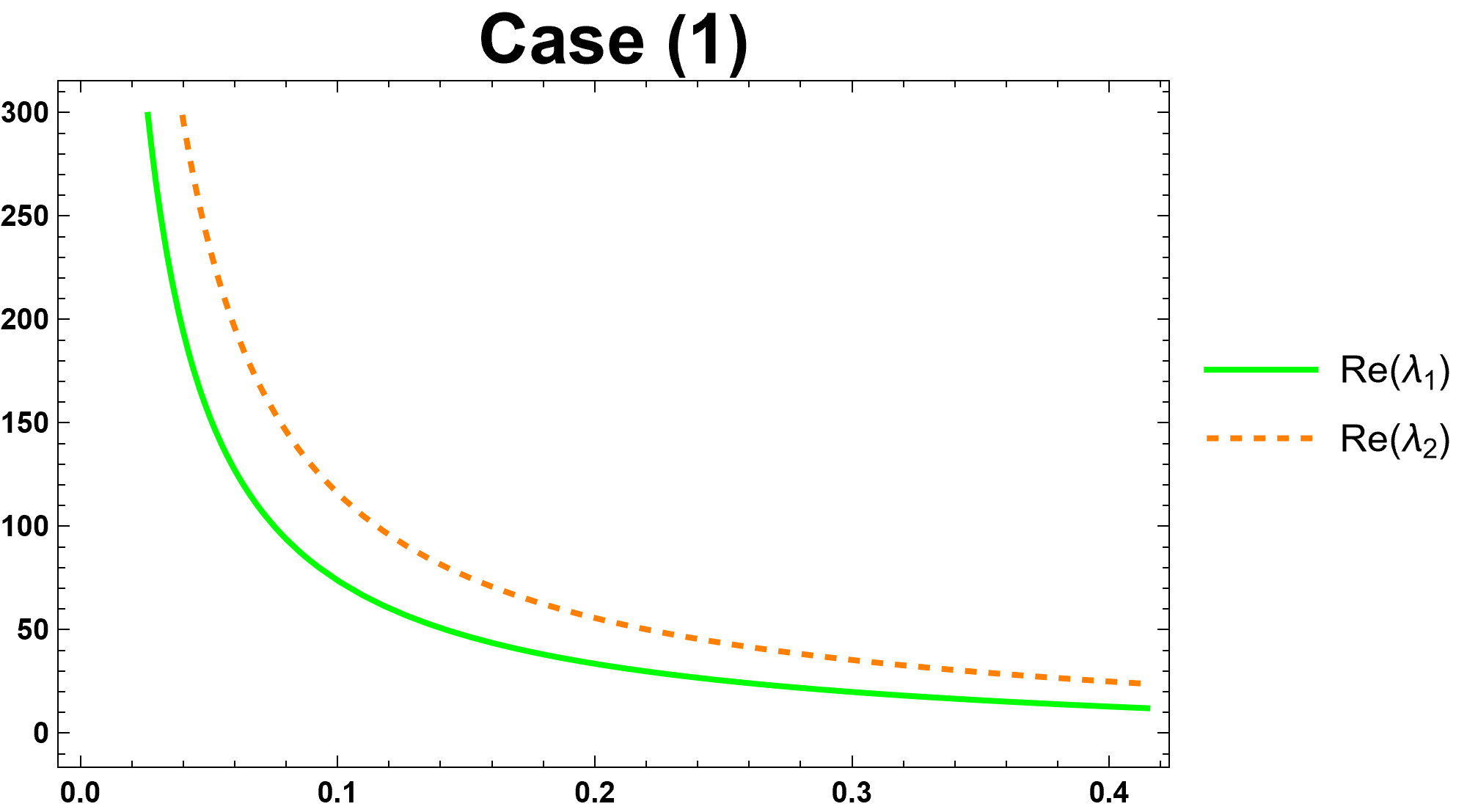}
\end{subfigure}\\
\begin{subfigure}{0.4\textwidth}
       \includegraphics[scale=0.45]{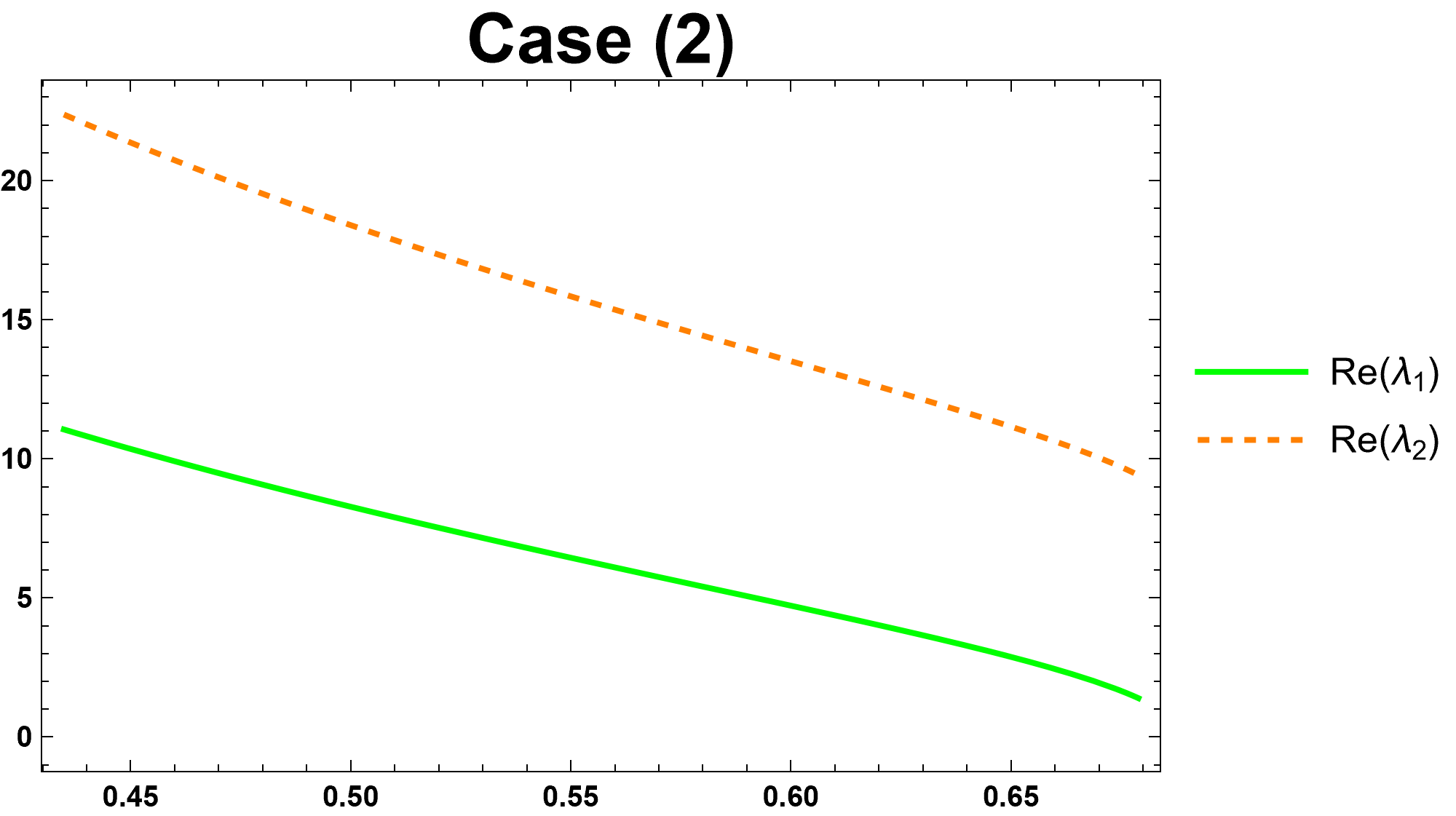}
\end{subfigure}\\
\begin{subfigure}{0.4\textwidth}
       \includegraphics[scale=0.45]{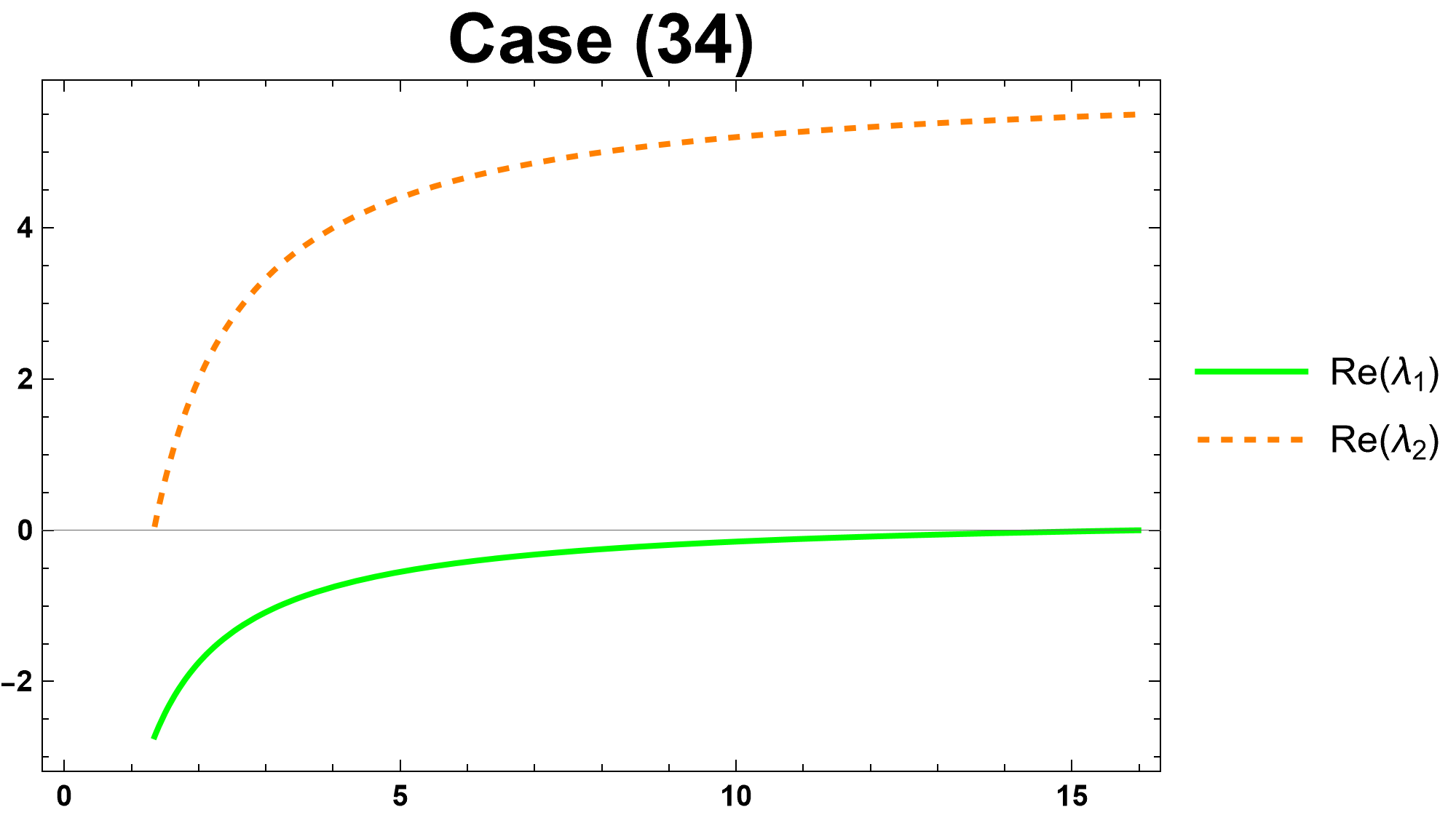}
\end{subfigure}
       \caption{Real part of the eigenvalues for $B$ and $C$ for a fixed value of one of the parameters according to the respective case. In cases {(1)} and (2) $\mu$ is fixed as $\mu=\frac{1}{2}\left(5-\sqrt{17}\right)$ while $m$ moves in two different intervals. In case {(34),} $\mu=\frac{35}{8}$ and $\frac{4}{5}<m<16.$ Both points behave as sources or saddles whenever they exist and are hyperbolic.}
       \label{fig:stability-of-BandC-2D-plots}
   \end{figure}
   \begin{figure}[H]
\begin{subfigure}{0.4\textwidth}
    \includegraphics[scale=0.45]{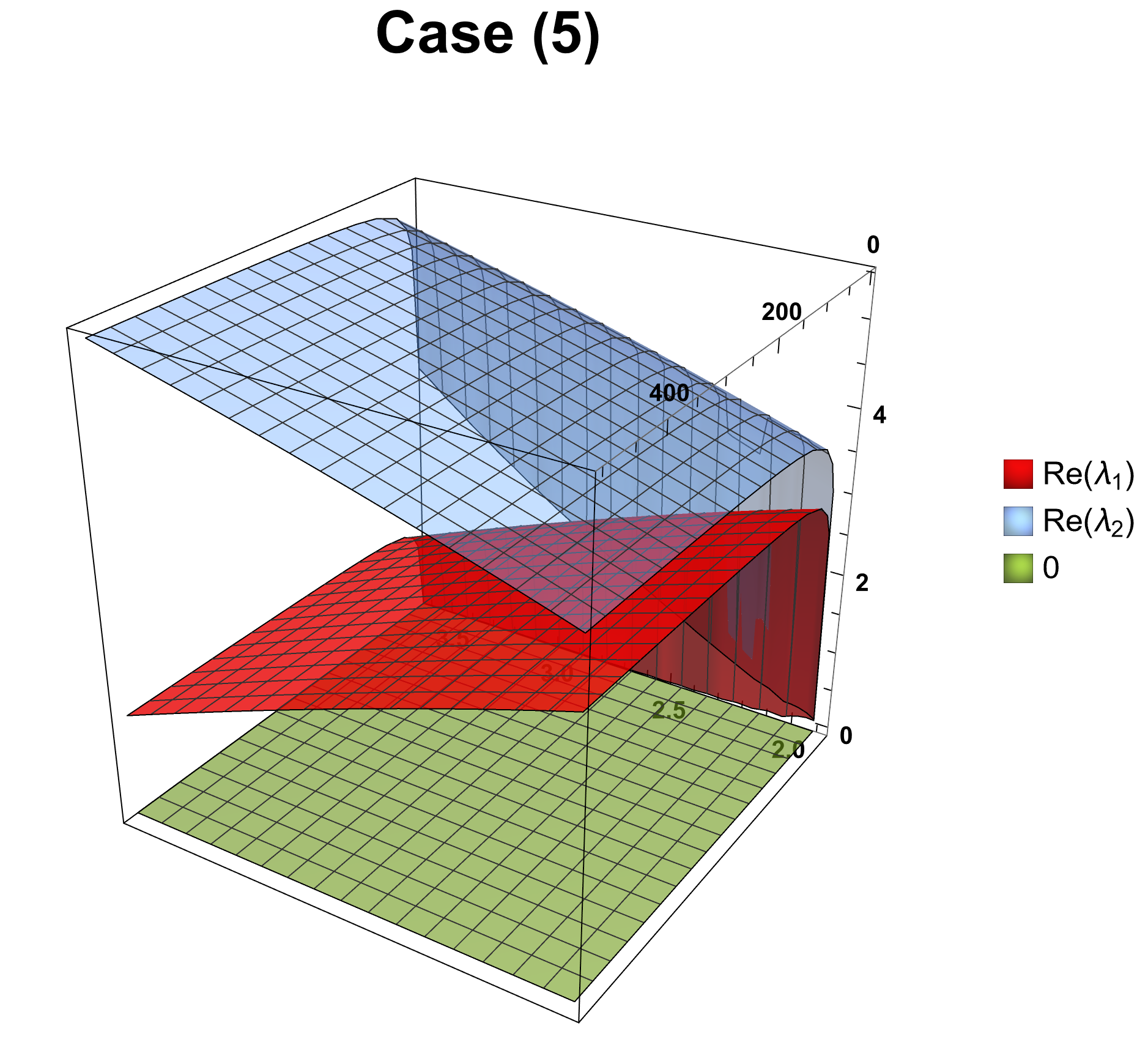}
\end{subfigure}\\
\begin{subfigure}{0.4\textwidth}
   \includegraphics[scale=0.45]{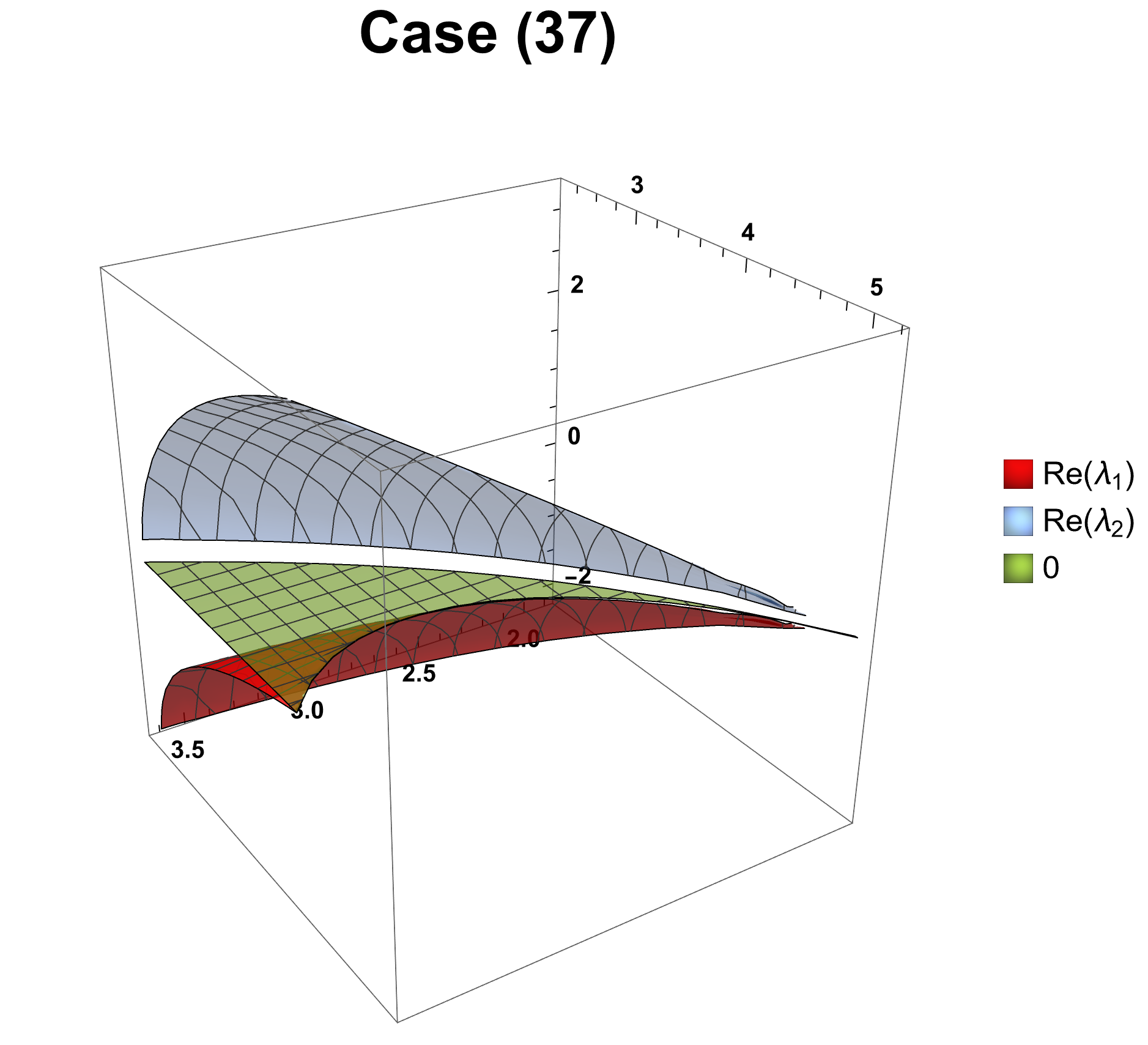}
\end{subfigure}\\
   \begin{subfigure}{0.4\textwidth}
        \includegraphics[scale=0.45]{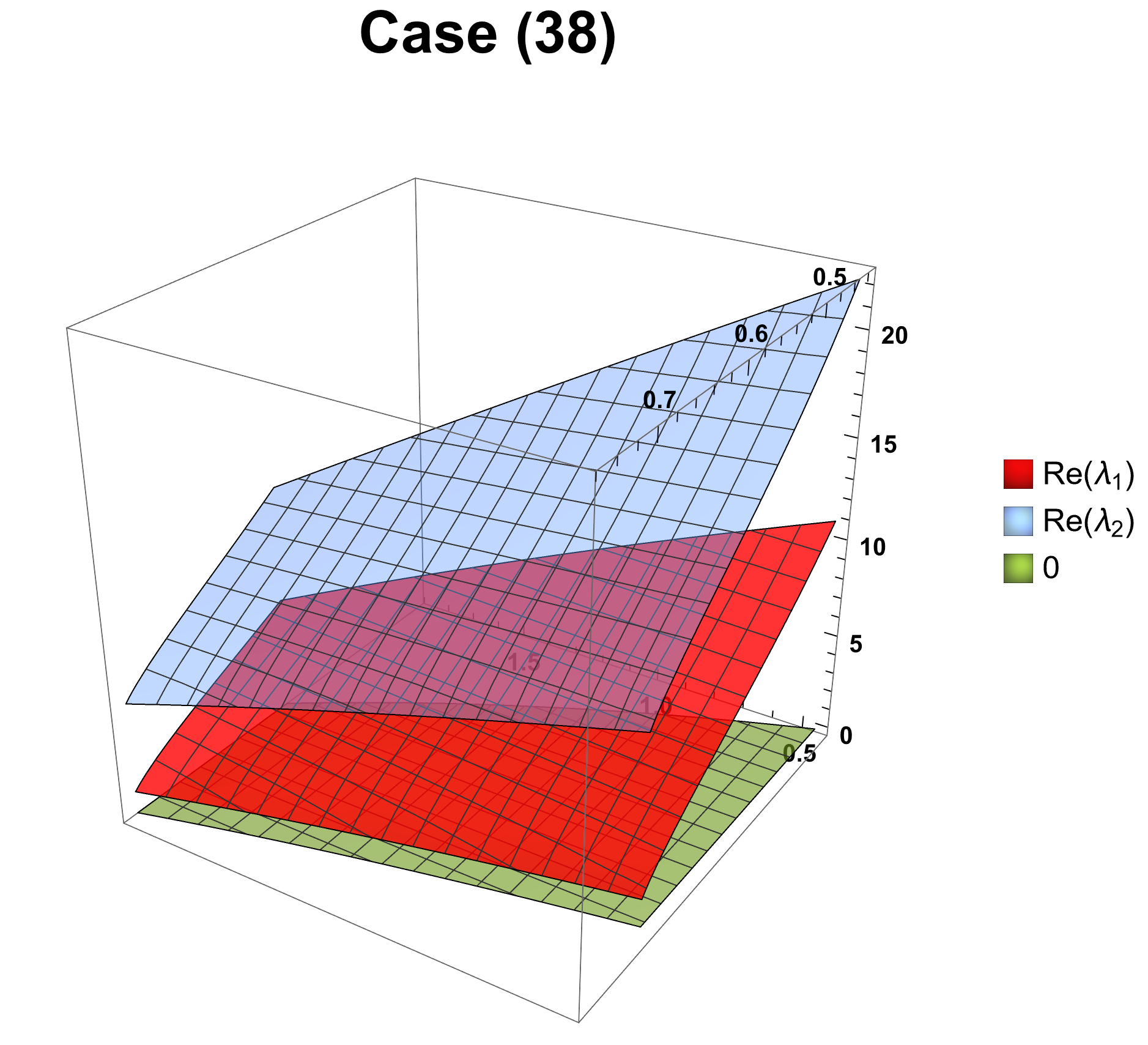}
\end{subfigure}    
        \caption{{Real part} 
 of the eigenvalues for $B$ and $C$ where both of the parameters $\mu, m$ remain free in some intervals. The cases depicted are cases (5), {(37)}
 and (38). Once again, as in  Figure \ref{fig:stability-of-BandC-2D-plots}, both points behave either as sources or saddles for every existence condition in which they are hyperbolic.}
       \label{fig:stability-of-BandC-3D-plots}
   \end{figure}
   \item $D,E=\left(\pm \frac{k_2 m \left[-k_1+(2 \mu -3)m-4\right]}{\sqrt{k} \left[2 m (-4 \mu +((\mu -5)
   \mu +2) m+30)-24\right]},\frac{k_2}{\sqrt{k}}\right),$ where $k$ and $k_1$ are defined the same as in \eqref{def-of-k} and \eqref{def-of-k1} but 
   \begin{equation}
   \label{def-of-k2-2}
        k_2=\sqrt{\begin{array}{cc}
        m^3 \left[k_1 (2 \mu -3)-8 \mu (\mu  (2 \mu -25)+80)+204\right]\\-4 m^2\left[k_1+4 \mu (\mu +15)-436\right]+\left[2(\mu -1) \mu  ((\mu -9) \mu +21)+13\right]m^4 \\ +96 (2 \mu -15) m+288
   \end{array}}.
   \end{equation}
\end{enumerate}
As before, $D$ and $E$ are antipodal points, and once again, they share the same eigenvalues. We will write them as $\gamma_1=g_1(\mu,m)$ and $\gamma_2=g_2(\mu,m)$. The conditions for $D$ and $E$ are the same as the forty cases. However, $D$ and $E$ can be attractors or saddles in those intervals. Additionally, numerical inspection in every interval shows that the points cannot be sources. See {Figures} \ref{fig:stability-of-DandE-2D-plots} and \ref{fig:stability-of-DandE-3D-plots}.

\begin{figure}[H]
   \begin{subfigure}{0.4\textwidth}
  \includegraphics[scale=0.45]{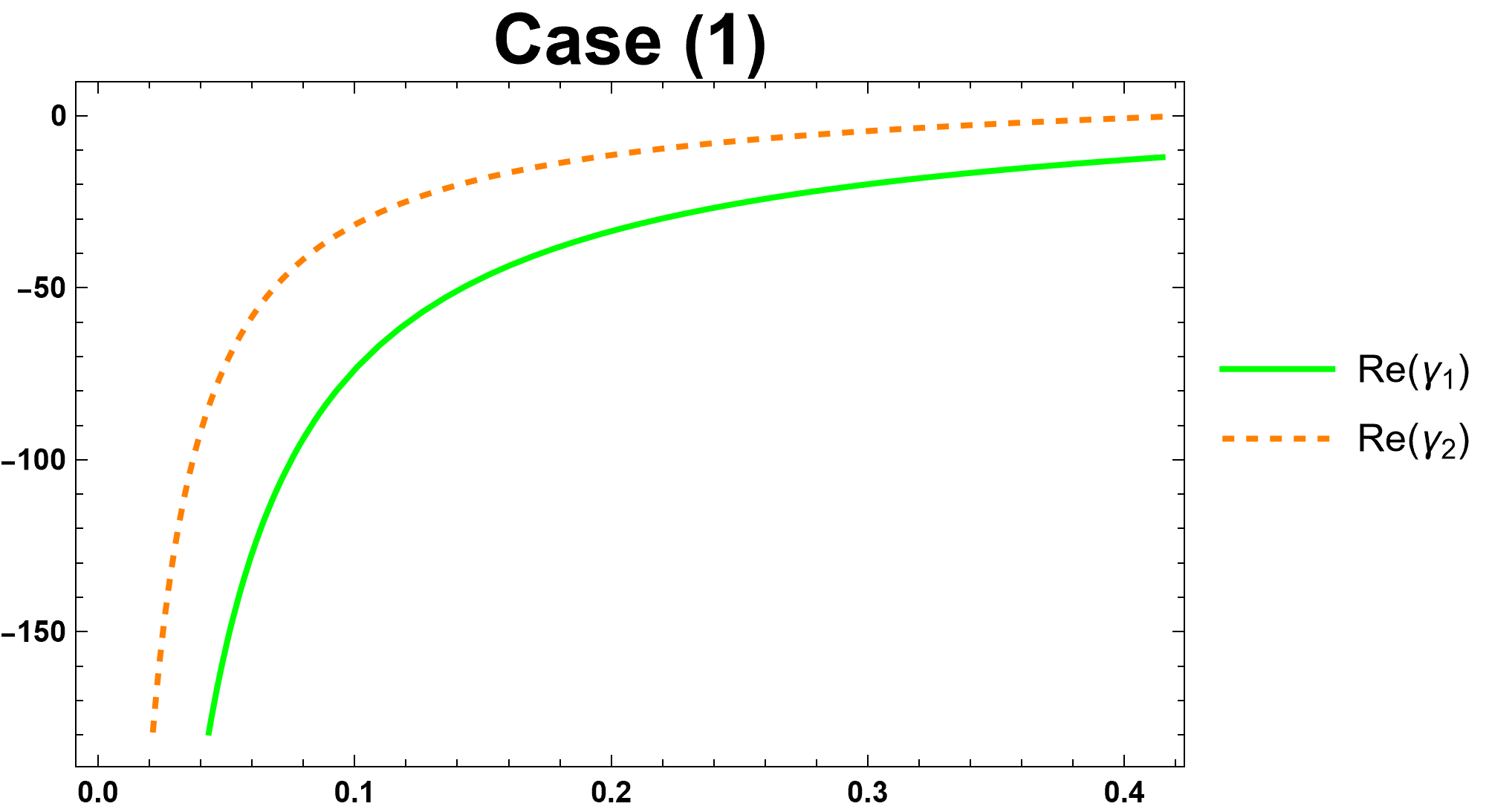}
\end{subfigure}\\
\begin{subfigure}{0.4\textwidth}
   \includegraphics[scale=0.45]{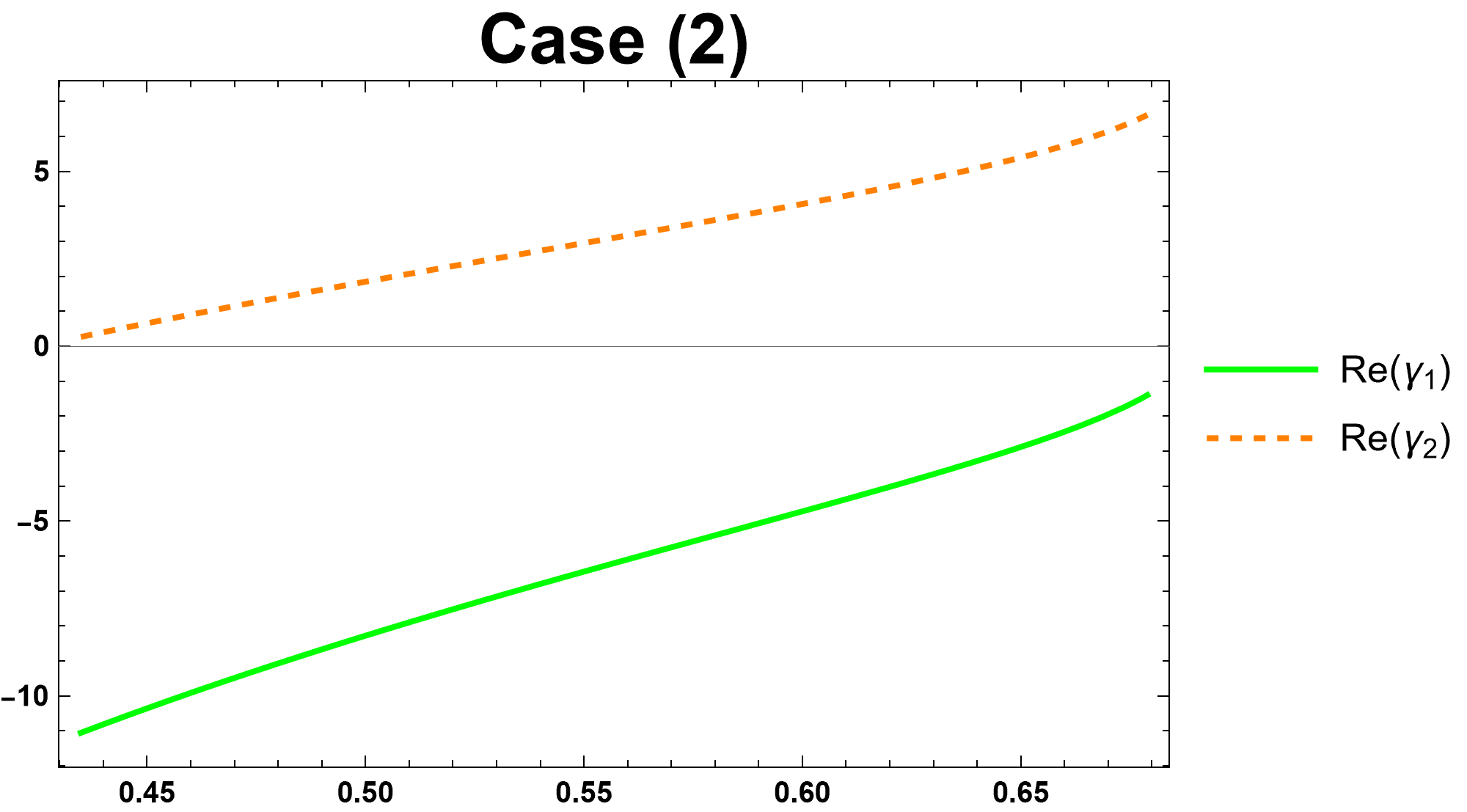}
\end{subfigure}\\
\begin{subfigure}{0.4\textwidth}
    \includegraphics[scale=0.45]{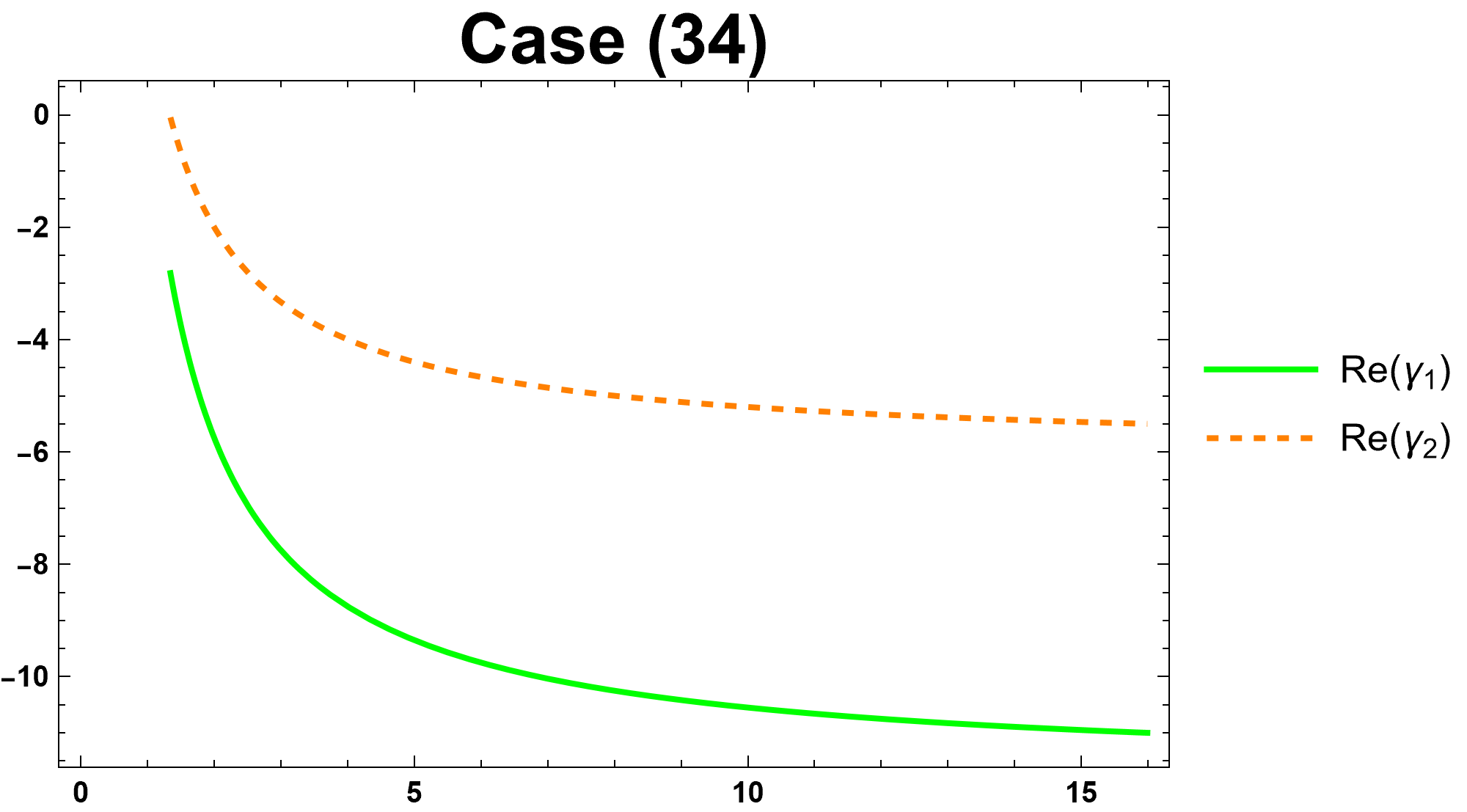}
\end{subfigure}
    \caption{Real part of the eigenvalues for $D$ and $E$ for a fixed value of one of the parameters according to the respective case. In cases (1) and (2), $\mu$ is fixed as $\mu=\frac{1}{2}\left(5-\sqrt{17}\right)$ while $m$ moves in two different intervals. In case (34), $\mu=\frac{35}{8}$ and $\frac{4}{5}<m<16.$ Both points behave as sinks or saddles whenever they exist and are hyperbolic.} 
    \label{fig:stability-of-DandE-2D-plots}
\end{figure}

\vspace{-9pt}
\begin{figure}[H]
   \begin{subfigure}{0.4\textwidth}
    \includegraphics[scale=0.45]{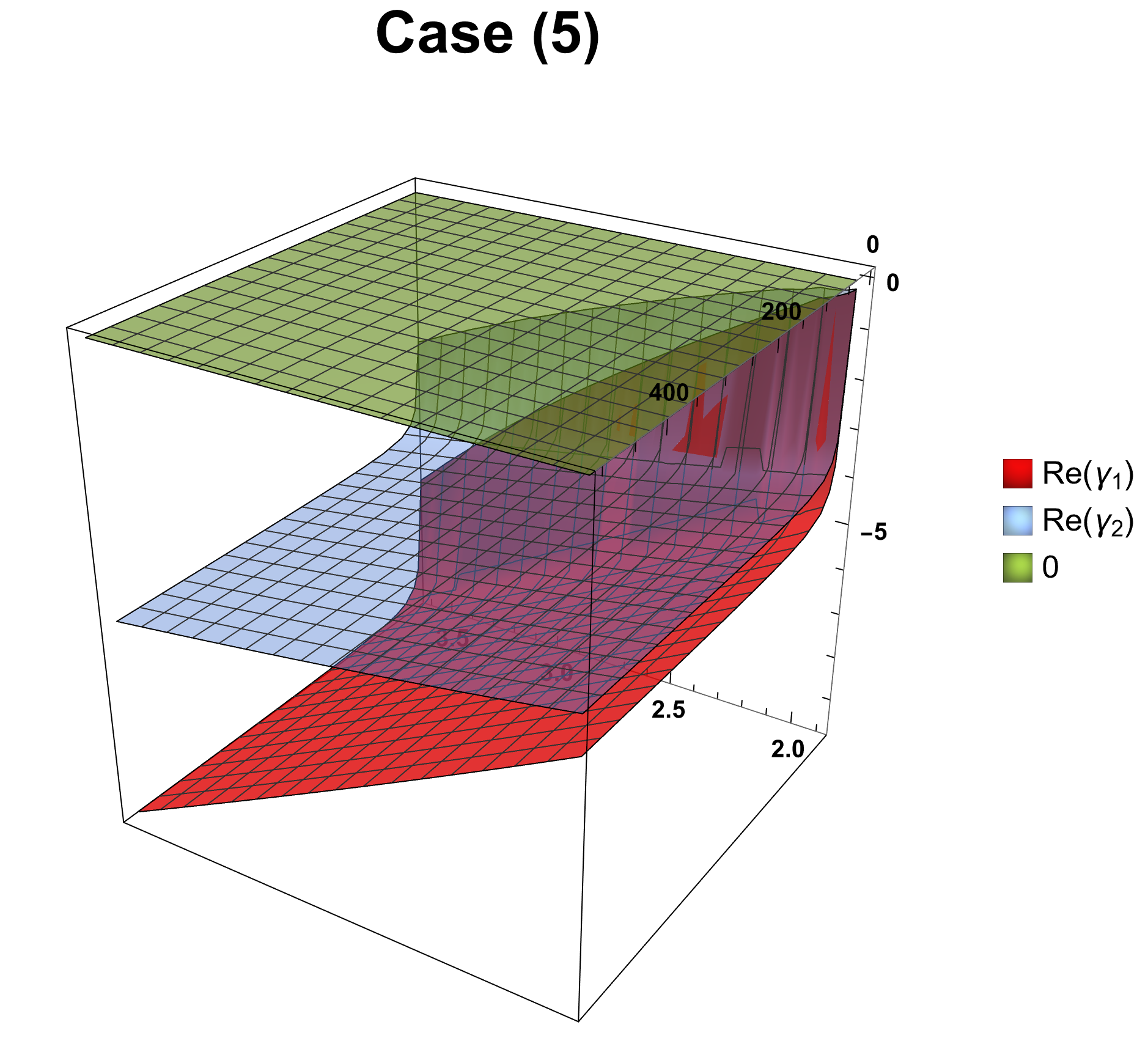}
\end{subfigure}\\
\begin{subfigure}{0.4\textwidth}
\includegraphics[scale=0.45]{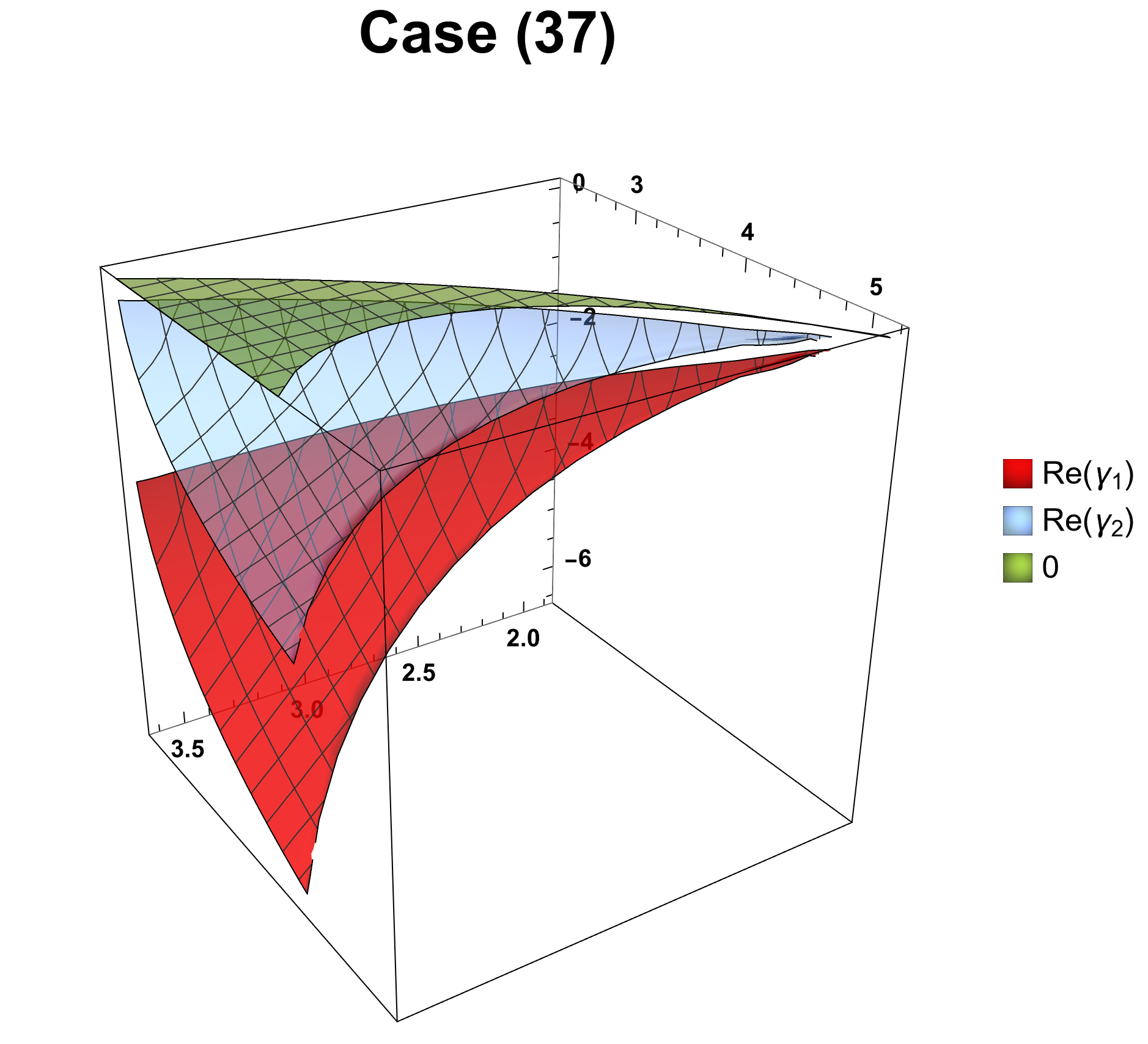}
\end{subfigure}\\
\begin{subfigure}{0.4\textwidth}
    \includegraphics[scale=0.455]{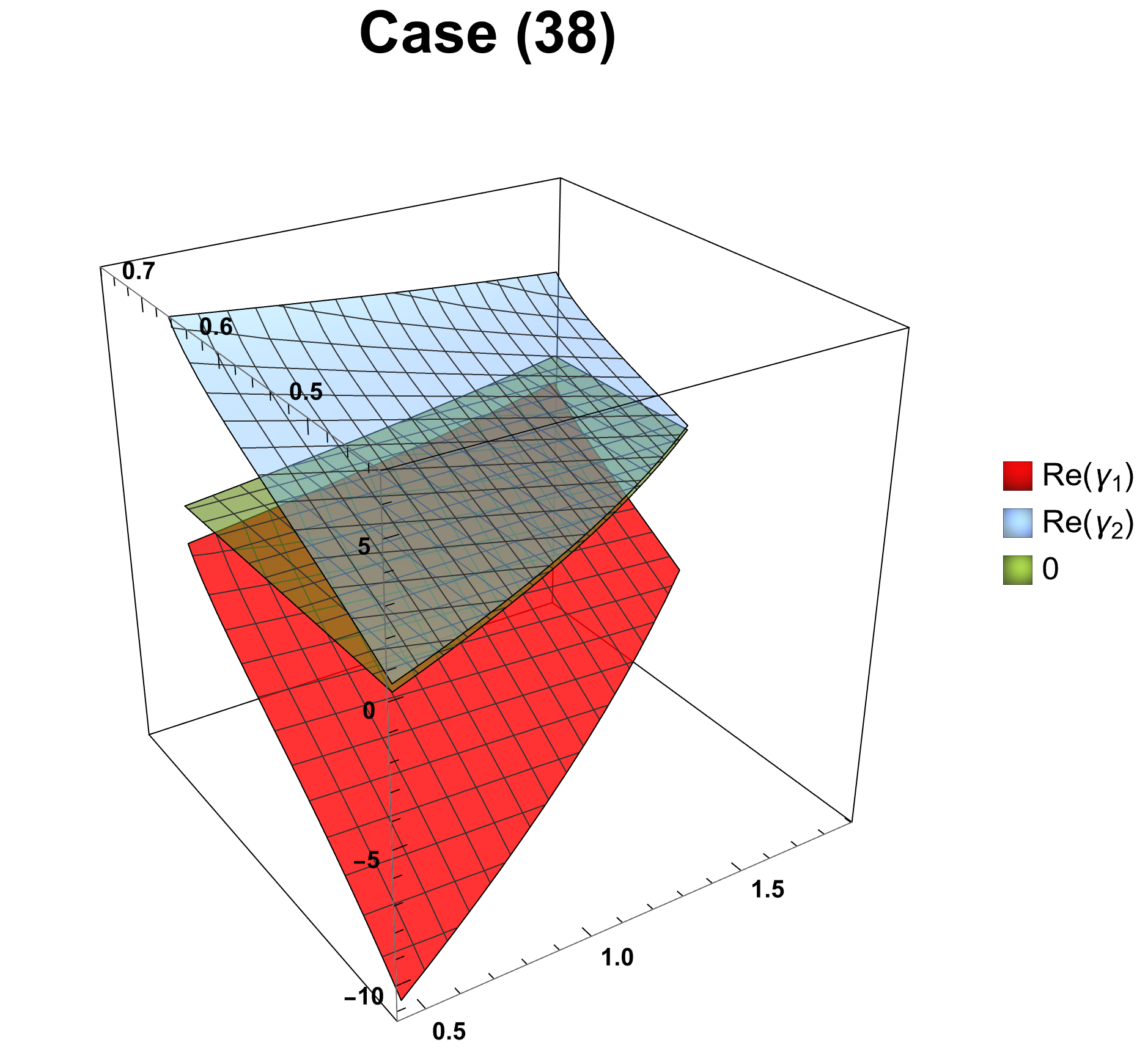}
\end{subfigure}
    \caption{{Real part} of the eigenvalues for $D$ and $E$ where both of the parameters $\mu, m$ remain free in some intervals. The cases depicted are cases (5), (37) and (38). Once again, as in  Figure \ref{fig:stability-of-DandE-2D-plots}, both points behave as sinks or saddles whenever they exist and are hyperbolic.} 
    \label{fig:stability-of-DandE-3D-plots}
\end{figure}

\begin{figure}[H]
    \includegraphics[scale=0.7]{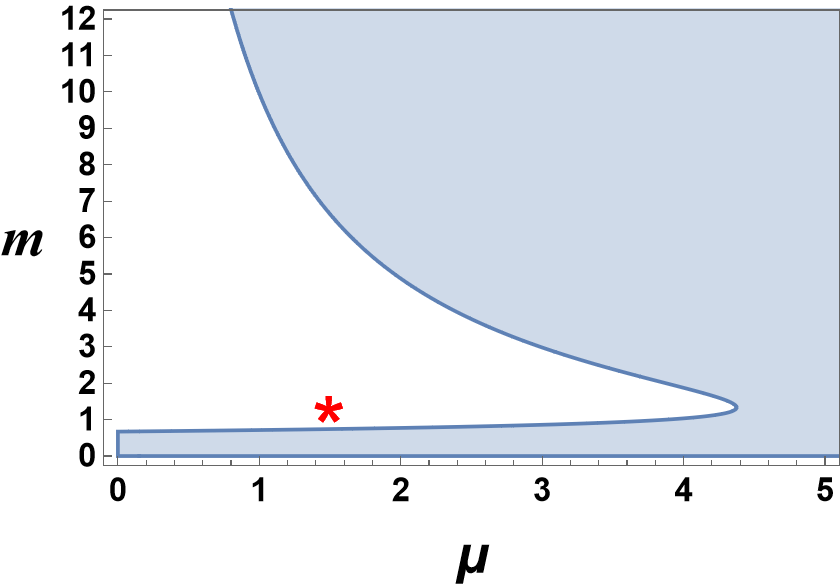}
    \caption{Existence region for points $B$, $C$, $D$ and $E$. The red star represents the point in the parameter space with the best-fit values of $\mu$ and $m$ according to the analysis in Section \ref{sec:fitting}.}
    \label{allowed-regions-2}
\end{figure}

From Figure  \ref{allowed-regions-2}, we have that  $B$, $C$, $D$, and $E$ do not exist for the best-fit values of $\mu$ and $m$ according to the analysis in Section \ref{sec:fitting}. 

Finally, we obtain the dynamics shown in Figure \ref{fig:Flux(epsilon,v)infty}. Here, we observe that for $(\mu=1, m=3)$ or $(\mu=1, m=4)$ or $(\mu=2, m=3)$,  point $A$ is an attractor. However, for $(\mu=2, m=4)$, there are two complex eigenvalues with zero real parts, namely $\left\{-\frac{i \sqrt{3}}{2},\frac{i \sqrt{3}}{2}\right\}$ which means $A$ is a center. Finally, for $(\mu=3, m=3)$, the point $A$ is a source, $B$ and $C$ are saddles and $D$ and $E$ are sinks. Finally for $(\mu=3, m=4)$, $A$ is a saddle point, $B$ and $C$ are sources and $D$ and $E$ are sinks. 

\begin{figure}[H]
   \begin{subfigure}{0.45\textwidth}\includegraphics[scale=0.4]{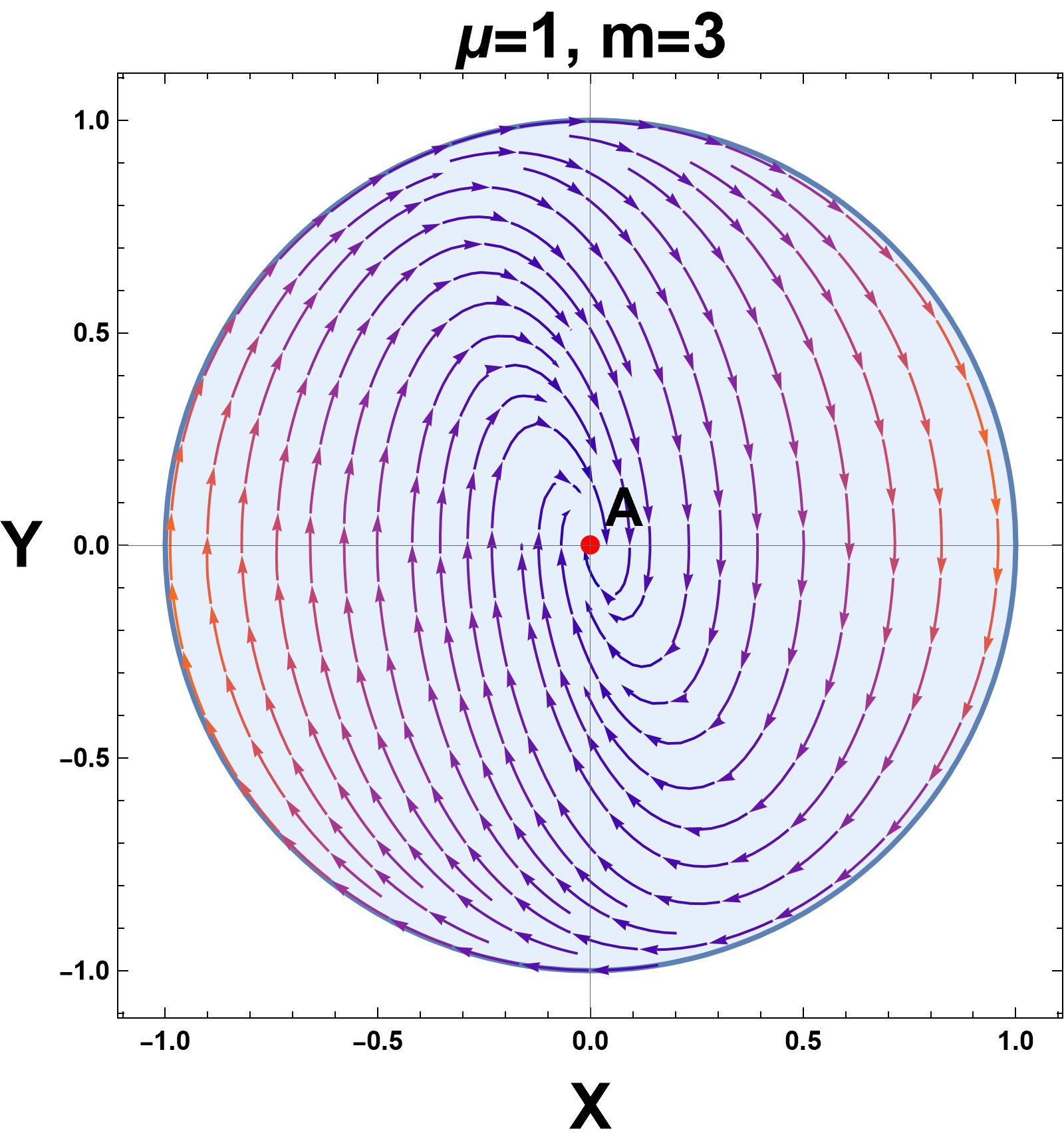}
\end{subfigure}
\begin{subfigure}{0.45\textwidth}
    \includegraphics[scale=0.4]{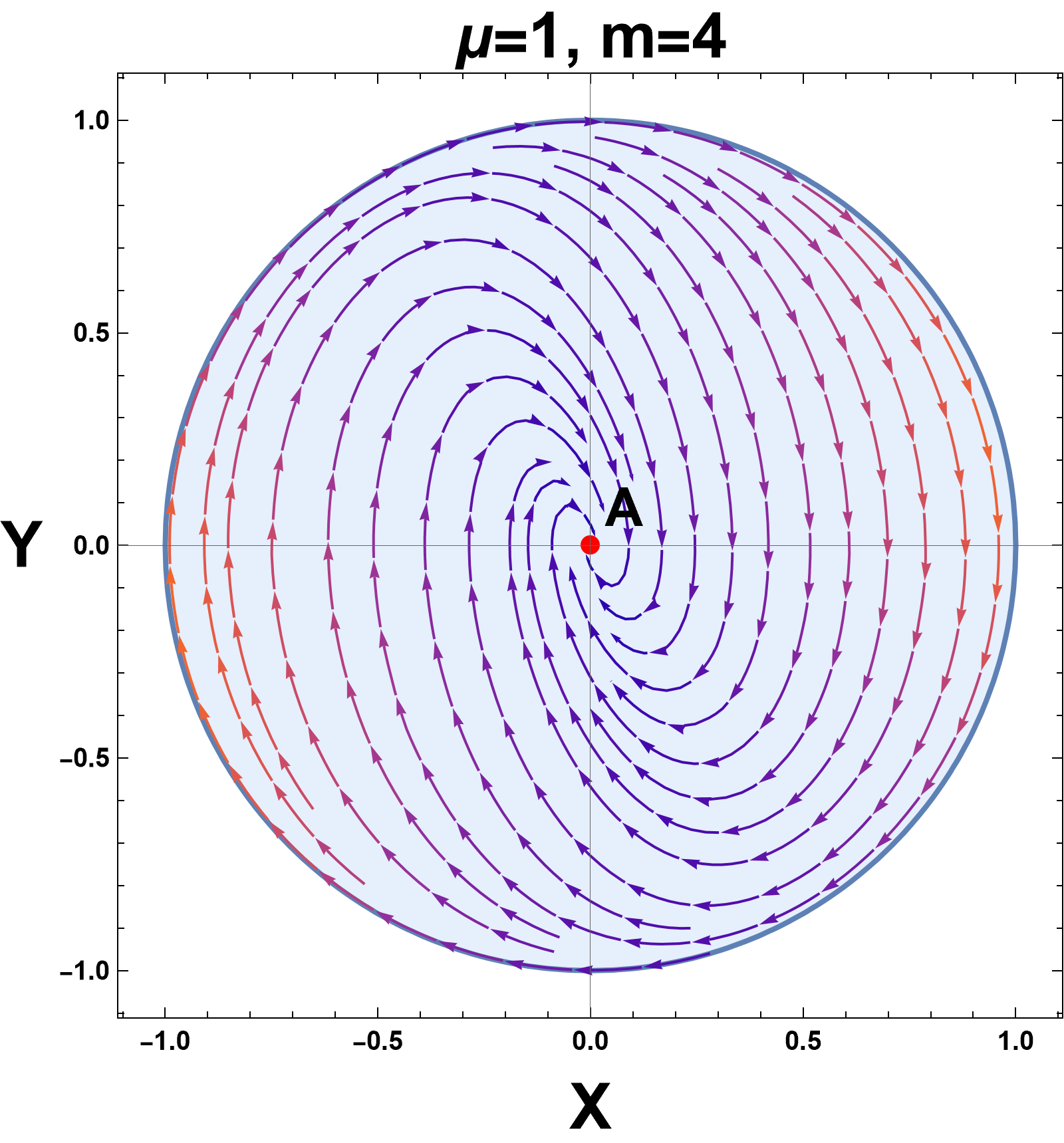}
\end{subfigure}\\
   \begin{subfigure}{0.45\textwidth}
    \includegraphics[scale=0.4]{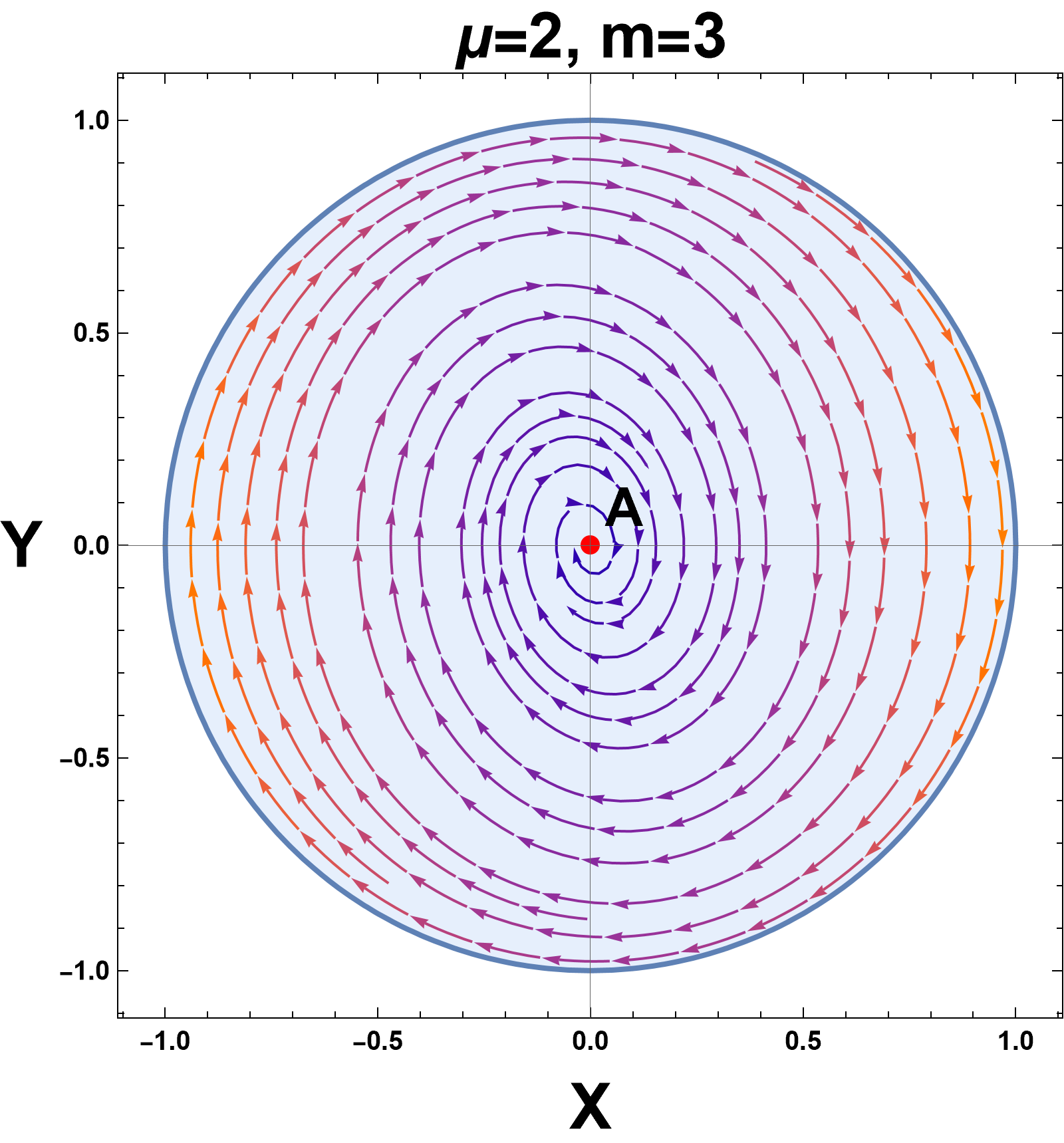}
\end{subfigure}
\begin{subfigure}{0.45\textwidth}
    \includegraphics[scale=0.4]{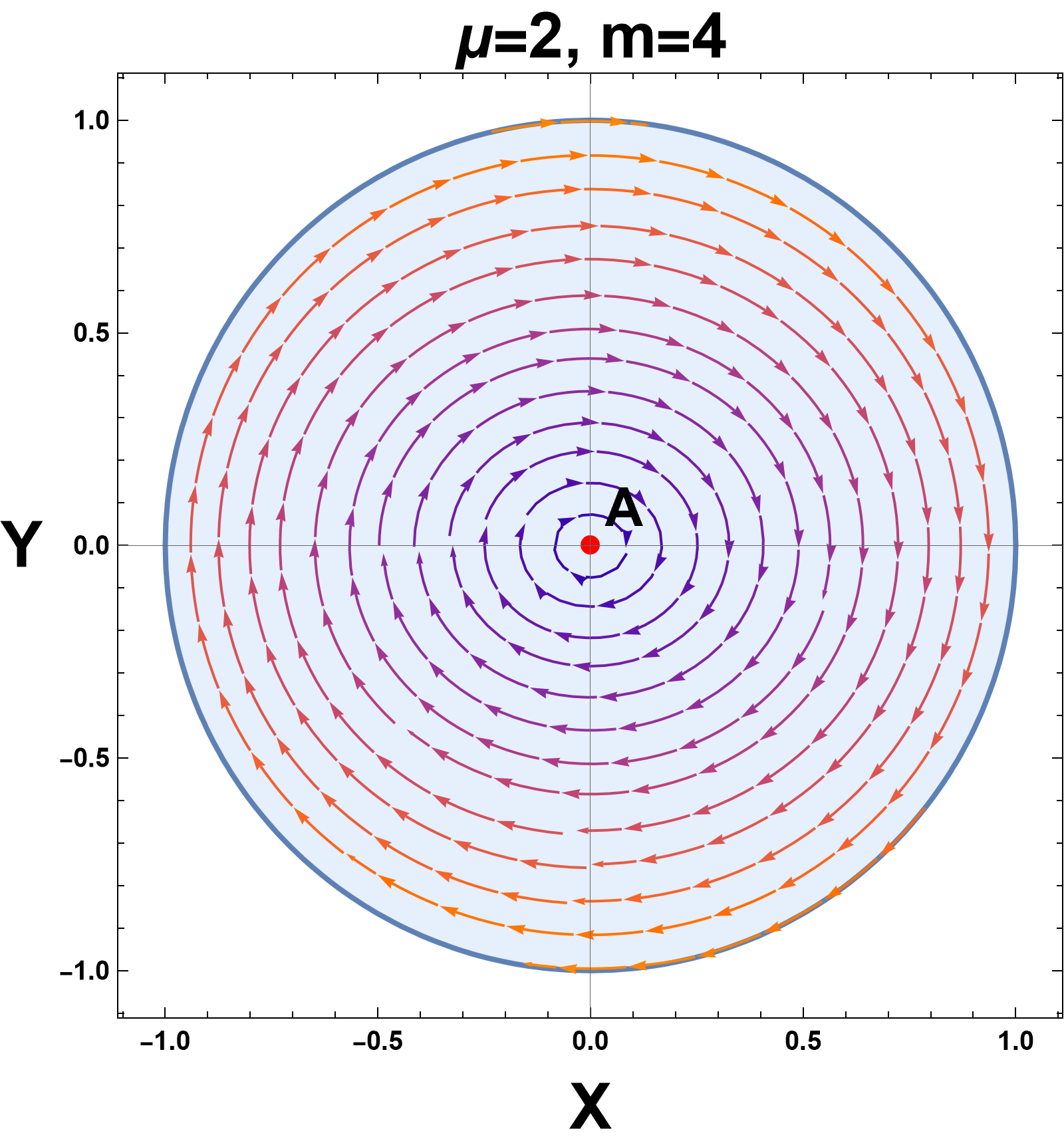}
\end{subfigure}\\
   \begin{subfigure}{0.45\textwidth}
    \includegraphics[scale=0.4]{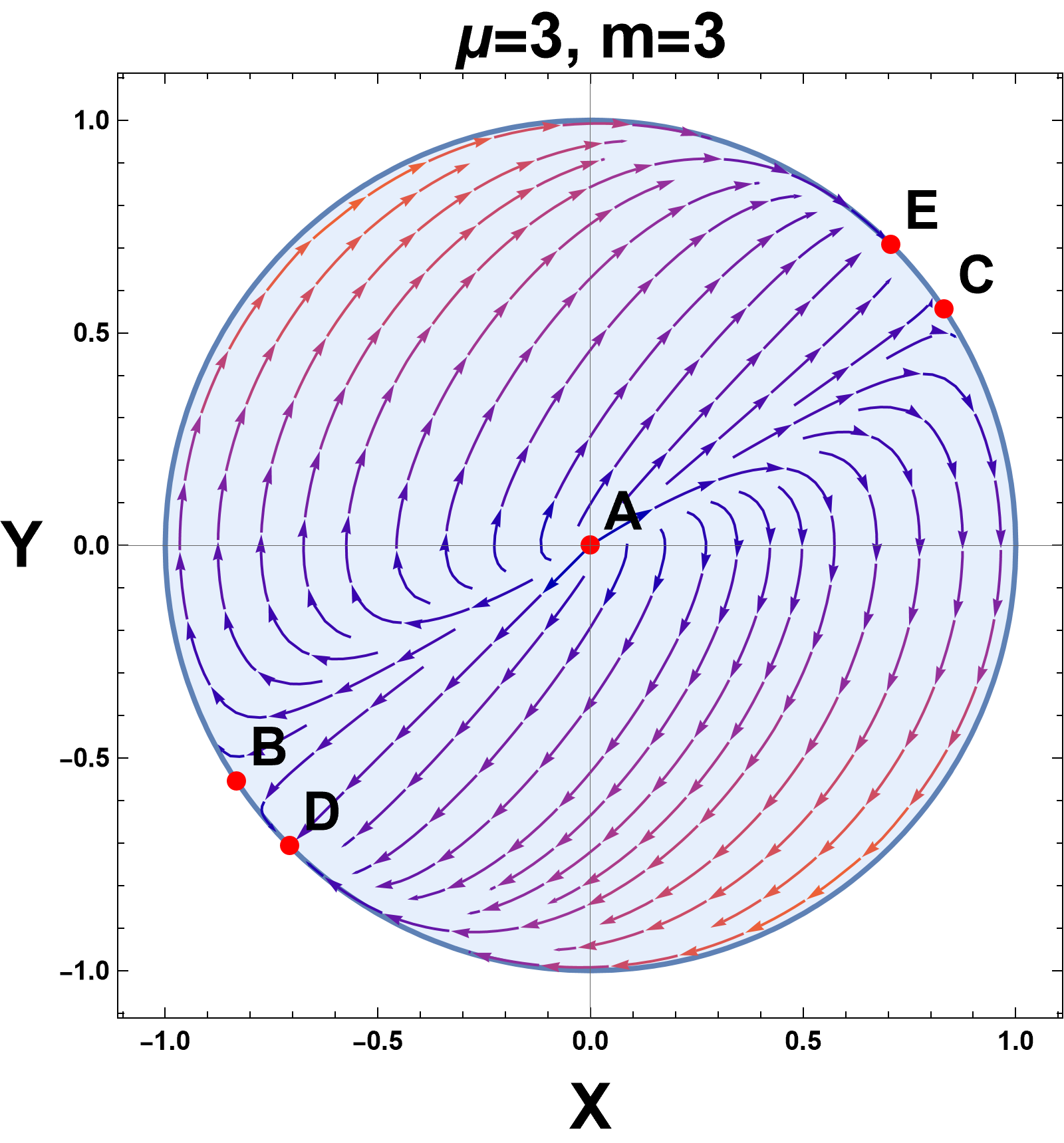}
\end{subfigure}
\begin{subfigure}{0.45\textwidth}
    \includegraphics[scale=0.4]{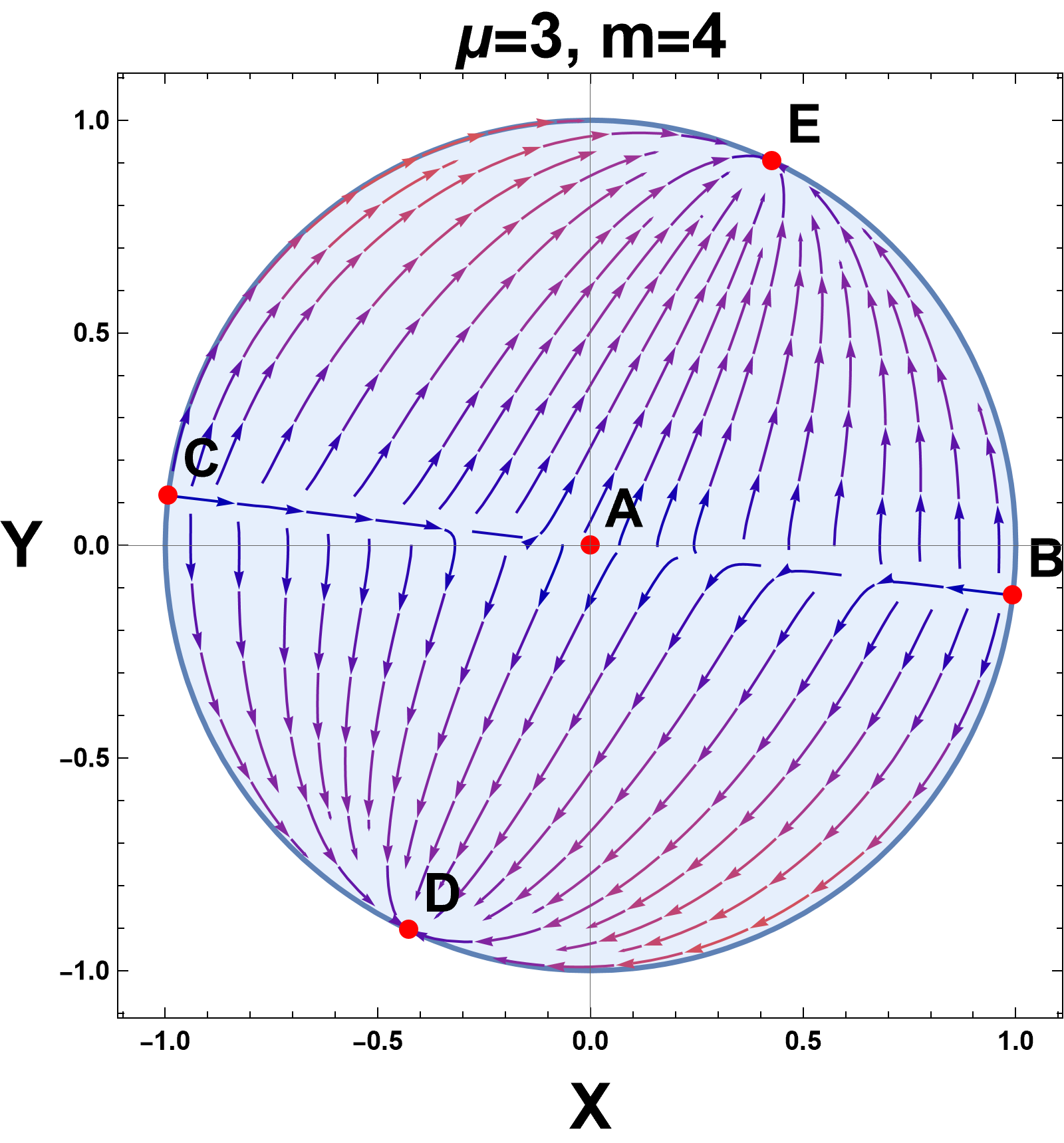}
\end{subfigure}
    \caption{{Dynamics of} system \eqref{infty-eq-1} and \eqref{infty-eq-2} in the compactified region $X^2+Y^2\leq 1$ for different values of the parameters $\mu$ and $m$. We observe that for $(\mu=1, m=3)$ or $(\mu=1, m=4)$ or $(\mu=2, m=3)$, point $A$ is an attractor. However, for $(\mu=2, m=4)$, there are two complex eigenvalues with zero real parts, namely $\left\{-\frac{i \sqrt{3}}{2},\frac{i \sqrt{3}}{2}\right\}$ which means $A$ is a center. For $(\mu=3, m=3)$, the point $A$ is a source, $B$ and $C$ are saddles and $D$ and $E$ are sinks. Finally for $(\mu=3, m=4)$, $A$ is a saddle point, $B$ and $C$ are sources and $D$ and $E$ are sinks. The streamlines are colored by default according to the magnitude of the vector field, with the arrow pointing in the time-increasing direction.} \label{fig:Flux(epsilon,v)infty}
\end{figure}

\section{\label{sec:fitting} Observational Constraints}
In this section, we constrain the free parameters of the exact scaling solution obtained for the exponential potential (see Section \ref{Scaling-Solutions}). To achieve this, we compute the best-fit parameters at the \(1\sigma\) confidence level (CL) using the affine-invariant Markov chain Monte Carlo (MCMC) method \cite{Goodman_Ensemble_2010}, implemented in the pure Python code emcee version 3.1.6 \cite{Foreman-Mackey:2012any}. This analysis incorporates data from SNe Ia, CC, GL, and BHS. 

In our procedure, we consider \(100\) chains, or ``walkers'', and use the autocorrelation time \(\tau_{\text{corr}}\) provided by the emcee module as a convergence test. At every \(50\) steps, we compute the value of \(\tau_{\text{corr}}\) for each free parameter. If the current step exceeds \(50\tau_{\text{corr}}\) and the value of \(\tau_{\text{corr}}\) changes by less than \(1\%\), we conclude that the chains have converged and stop the constraints. We discard the first \(5\tau_{\text{corr}}\) steps as ``burn-in'' steps, thin the chains by \(\tau_{\text{corr}}/2\), and flatten them. For this MCMC analysis, we consider the following Gaussian likelihood:
\begin{equation}\label{likelihood}
\mathcal{L}_{I}\propto\exp{\left(-\frac{\chi_{\text{I}}^{2}}{2}\right)},
\end{equation}
where $\chi_{\text{I}}^{2}$ is the merit function. In particular, we consider the combinations of data sets $\chi_{\text{SNe}}^{2}$, $\chi_{\text{SNe+CC}}^{2}=\chi_{\text{SNe}}^{2}+\chi_{\text{CC}}^{2}$, and $\chi_{\text{joint}}^{2}=\chi_{\text{SNe}}^{2}+\chi_{\text{CC}}^{2}+\chi_{\text{GL}}^{2}+\chi_{\text{BHS}}^{2}$. As initialization points, we consider a small vicinity around the values that maximize the likelihood. 

We begin by constructing the Hubble parameters that describe the exact scaling solution in the following subsections. Next, we provide a brief overview of the merit function for each data set included in the cosmological constraints. Finally, we present the results along with a discussion of their implications.

\subsection{Hubble Parameter for the Exponential Potential}
To derive the theoretical Hubble parameter for the scaling solution associated with the exponential potential, we substitute Equations \eqref{exactphi}, \eqref{exactpsi} and \eqref{exactV} into Equation \eqref{Hpf}. This yields
\begin{align}
    \dot{H}&=\frac{4 \left(12 c_4 \lambda ^2 t^3 H^3+1\right)}{3 \lambda ^2 t^2 \left(8 c_4 t^2H^2-1\right)}-\frac{(\mu -3) H}{t}-H^2,
\end{align}
where
\begin{align*}
    \lambda=\frac{1}{3} \sqrt{\frac{36-3 m \left[-4 \mu +((\mu -5) \mu +2) m+30\right]}{(\mu +2) m-6}}, \quad
    c_4= \frac{1}{96} m^2 \left[\frac{2 m
   \left(3 \lambda ^2+m\right)}{\lambda ^2 \left[(\mu -6) m+2\right]}+3\right].
\end{align*}
Hence, by introducing the logarithmic independent variable $s=-\ln(1+z)$, for which $s\rightarrow -\infty$ as $z\rightarrow \infty$, $s\rightarrow 0$ as $z\rightarrow 0$, $s\rightarrow\infty$ as $z\rightarrow -1$, and defining the age parameter as  $\alpha=tH$, we obtain the initial value problem
\begin{align}
   \alpha '(s)&= \frac{2 \left[16 c_4 \alpha (s)+3 \lambda ^2\right]}{3 \lambda ^2 \left[8 c_4 \alpha (s)^2-1\right]}-\mu-\frac{4}{3 \lambda ^2 \alpha (s)}-\alpha (s)+6, \label{eq(134)}\\
   t'(s)&=\frac{t(s)}{\alpha(s)}, \label{eq(135)}\\
   \alpha(0):&=\alpha_0=t_0 H_0, \quad t(0)=t_0. \label{eq(136)}
\end{align}
The above system has the exact solution 
\begin{equation}
    \alpha (s)=\frac{2}{m},\quad m\neq\frac{6}{\mu}, \label{eqA129}
\end{equation}
where $\alpha(0):=\alpha_0=\frac{2}{m}=t_0 H_0$ implies $m=\frac{2}{\alpha_0}$. So, integrating under the initial condition $t(0)= t_0$, we obtain $t(s)=t_0 e^{\frac{s}{\alpha_0}}$, 
and, therefore
\begin{equation}
    H(s)= \alpha(s) t(s)^{-1}=  H_0 e^{-\frac{s}{\alpha_0}} \quad\implies\quad H(z)= H_0  (1+z)^{\frac{1}{\alpha_0}}.
\label{eq:Ez}
\end{equation}
The deceleration parameter is  given by $q(z)=-1+\alpha_{0}^{-1}$. 
Therefore, the Universe is accelerated without invoking dark energy for $\alpha_{0}>1$. 

Finally, from Equation \eqref{Vf}, we have the effective matter sources:
\begin{align}
    \Omega_{\phi}:&= \frac{\frac{1}{2} \dot{\phi}^2 + V(\phi)}{3 H^2}= 1-\frac{2 \alpha_0 (3 \alpha_0-\mu -6)+4 (\mu-1)}{\alpha_0 \left[2 \alpha_0 \mu +3\alpha_0 (\alpha_0-5)-\mu ^2+5 \mu-2\right]}, \label{OmegaScaling1} 
\end{align}
and 
\begin{align}
    \Omega_{X} :&= \Omega_{\text{fracc}} + \Omega_{\text{GB}} + \Omega_{\text{GB,\,fracc}}  = \frac{-\frac{3 (\mu -1) H \left(8 H^2 \psi -1\right)}{t}+24 H^3 \dot{\psi}}{3 H^2} \nonumber \\
    & = \frac{2 \alpha_0 (3 \alpha_0-\mu -6)+4 (\mu-1)}{\alpha_0 \left[2 \mu \alpha_0+3\alpha_0 (\alpha_0-5)-\mu ^2+5 \mu-2\right]},
\end{align}
with 
\begin{equation}
    \Omega_{\phi} + \Omega_{X} =1,
\end{equation}
where, for this scaling solution, the scalar field behaves as dark matter, say 
\begin{equation}
      \Omega_{\phi}= \Omega_{DM}, \label{eqDM-phi}
\end{equation}
and the extra terms $\Omega_{\text{fracc}} + \Omega_{\text{GB}} + \Omega_{\text{GB,\,fracc}}$ mimic dark energy $\Omega_{X}$. 

To compare with the fractional cosmological model, we also constrain the $\Lambda$CDM model, for which the corresponding theoretical Hubble parameter is given by
\begin{equation}\label{LCDM}
    H=H_{0}\sqrt{\Omega_{m,0}(1+z)^{3}+1-\Omega_{m,0}},
\end{equation}
whose corresponding parameter space is $\boldsymbol{\theta}=(H_{0},\Omega_{m,0})$. 

For this scaling solution \eqref{eqA129}, we have that $\Omega_{\phi}$ and 
$\Omega_{X}$ are constants. Therefore, 
\begin{equation}
    1-\frac{2 \alpha_0 (3 \alpha_0-\mu -6)+4 (\mu-1)}{\alpha_0 \left[2 \alpha_0 \mu +3\alpha_0 (\alpha_0-5)-\mu ^2+5 \mu-2\right]}= \Omega_{m,0}.
\end{equation}
Hence, we have two solutions 
\begin{align}
\mu_\pm & = \frac{1}{2
   \alpha_0 (\Omega_{m,0}-1)} \Bigg\{\alpha_0 (2 \alpha_0
   (\Omega_{m,0}-1)+5 \Omega_{m,0}-7)+4  \nonumber \\
   & \pm \sqrt{\begin{array}{c}
    \alpha_0
   \Big[\alpha_0 \Big(16 \alpha_0^2
   (\Omega_{m,0}-1)^2-8 \alpha_0 (5 \Omega_{m,0}-7) (\Omega_{m,0}-1) \\ +\Omega_{m,0} (17 \Omega_{m,0}-86)+73\Big) 
   +8 (3 \Omega_{m,0}-5)\Big]+16  \end{array}}\Bigg\}.
\end{align}
Therefore, we use a Bayesian analysis to obtain the best-fit values of $\Omega_{m,0}$ and $\alpha_0$ and substitute in the previous expressions to obtain the physical values of $\mu$. 

\subsection{\label{Sec:CC} Cosmic Chronometers}
To constrain the model using cosmic chronometers, we utilize the data set from Ref.~\cite{Capozziello:2017nbu}, which comprises 31 data points spanning the redshift range \(0.0708 \leq z \leq 1.965\). These Hubble data points are derived using the differential age method, a model-independent approach \cite{Jimenez:2001gg}. Accordingly, the merit function for the CC data is {formulated as} \begin{equation}\label{meritCC}
    \chi_{CC}^{2}=\sum_{i=1}^{31}{\left[\frac{H_{i}-H_{th}(z_{i},\boldsymbol{\theta)}}{\sigma_{H,i}}\right]^{2}},
\end{equation}
where \( H_{i} \) represents the observational Hubble parameter at redshift \( z_{i} \), with an associated error \( \sigma_{H, i} \) provided by the cosmic chronometer (CC) sample. The term \( H_{\text{th}} \) denotes the theoretical Hubble parameter at the same redshift, while \( \boldsymbol{\theta} \) encompasses the free parameters of the model.

Note that the same theoretical Hubble parameter gives both analytical solutions of interest for the cosmological constraint, according to Equation \eqref{eq:Ez}, whose parameter space is $\boldsymbol{\theta}=(H_{0},\alpha_{0})$. Therefore, we consider for our MCMC analysis the flat priors $0.55<h<0.85$ and $0.5<\alpha_{0}<2.5$, where $h$ is the reduced Hubble constant according to the expression $H_{0}=100\frac{km/s}{Mpc}h$. We also constrain the $\Lambda$CDM model as a further comparison, whose respective theoretical Hubble parameter is given by \eqref{LCDM}, whose respective parameter space is $\boldsymbol{\theta}=(H_{0},\Omega_{m,0})$, for which we consider the same prior on $h$ as in the fractional cosmology, plus the flat prior $0<\Omega_{m,0}<1$.

\subsection{\label{sec:SNeIa} Type Ia Supernovae}
For the SNe Ia data, we consider the Pantheon+ sample \cite{Brout:2022vxf}, which consists of 1701 data points in the redshift range $0.001\leq z\leq 2.26$, whose respective merit function can be conveniently constructed in matrix notation (denoted by bold symbols) as
\begin{equation}\label{meritSNe}
\chi_{\text{SNe}}^{2}=\mathbf{\Delta D}(z,\boldsymbol{\theta},M)^{\dagger}\mathbf{C}^{-1}\mathbf{\Delta D}(z,\boldsymbol{\theta},M),
\end{equation}
where $\left[\mathbf{\Delta D}(z,\boldsymbol{\theta},M)\right]_{i}= m_{B,i}-M-\mu_{th}(z_{i},\boldsymbol{\theta})$ and $\mathbf{C}=\mathbf{C}_{\text{stat}}+\mathbf{C}_{\text{sys}}$, with $\mathbf{C}$ the total uncertainty covariance matrix. The matrices $\mathbf{C}_{\text{stat}}$ and $\mathbf{C}_{\text{sys}}$ account for the statistical and systematic uncertainties, respectively. The quantity $\mu_{i}=m_{B, i}-M$ corresponds to the observational distance modulus of the Pantheon+ sample, which is obtained by a modified version of Trip's formula \cite{Tripp:1997wt} and the BBC (BEAMS with Bias Corrections) approach \cite{Kessler:2016uwi}. In contrast, $m_{B, i}$ is the corrected apparent B-band magnitude of a fiducial SNe Ia at redshift $z_{i}$, and $M$ is the fiducial magnitude of an SNe Ia, which must be jointly estimated with the free parameters of the model under study.

The theoretical  distance modulus for a spatially flat FLRW spacetime is given by
\begin{equation}\label{theoreticaldistance}
\mu_{th}(z_{i},\boldsymbol{\theta})=5\log_{10}{\left[\frac{d_{L}(z_{i},\boldsymbol{\theta})}{\text{Mpc}}\right]}+25,
\end{equation}
with $d_{L}(z_{i},\boldsymbol{\theta})$ as the  luminosity distance given by
\begin{equation}\label{luminosity}
d_{L}(z_{i},\boldsymbol{\theta})=c(1+z_{i})\int_{0}^{z_{i}}{\frac{dz'}{H_{th}(z',\boldsymbol{\theta})}},
\end{equation}
where $c$ is the speed of light given in units of $\text{km/s}$.

In principle, there is a degeneration between $M$ and $H_{0}$. Hence, to constrain  $H_{0}$ using SNe Ia data alone, it is necessary to include the SH0ES (Supernovae and $H_{0}$ for Equation of State of the dark energy program) Cepheid host distance anchors, with a merit function of the form
\begin{equation}\label{Cepheidmerit}
\chi^{2}_{\text{Cepheid}}=\mathbf{\Delta D}_{\text{Cepheid}}\left(M\right)^{\dagger}\textbf{C}^{-1}\mathbf{\Delta D}_{\text{Cepheid}}\left(M\right),
\end{equation}
where 
$\left[\mathbf{\Delta D}_{\text{Cepheid}}\left(M\right)\right]_{i}=\mu_{i}\left(M\right)-\mu_{i}^{\text{Cepheid}}$,  where $\mu_{i}^{\text{Cepheid}}$ is the Cepheid calibrated host galaxy distance obtained by SH0ES \cite{Riess:2021jrx}. So, we use the Cepheid distances as the ``theory model'' to calibrate $M$, considering that the difference $\mu_{i}\left(M\right)-\mu_{i}^{\text{Cepheid}}$ is sensitive to $M$ and largely insensitive to other parameters of the cosmological model. Taking into account  that the total uncertainty covariance matrix for Cepheid is contained in the total uncertainty covariance matrix $\mathbf{C}$, we define the merit function for the SNe Ia data as
\begin{equation}\label{SNemeritfull}
\chi_{\text{SNe}}^{2}=\mathbf{\Delta D'}(z,\boldsymbol{\theta},M)^{\dagger}\mathbf{C}^{-1}\mathbf{\Delta D'}(z,\boldsymbol{\theta},M),
\end{equation}
where
\begin{equation}\label{SNeresidual}
\Delta\mathbf{D'}_{i}=\left\{\begin{array}{ll}
m_{B,i}-M-\mu_{i}^{\text{Cepheid}} & i\in\text{Cepheid host} \\
\\ m_{B,i}-M-\mu_{th}(z_{i},\boldsymbol{\theta}) & \text{otherwise}
\end{array}
\right..
\end{equation}
For the nuisance parameter $M$, we consider the flat prior $-20<M<-18$ in our MCMC analysis.

\subsection{\label{sec:lensing} Gravitational Lensing}
Multiple images are produced when a background object (the source) is lensed due to the gravitational force of a massive body (the lens). Therefore, the light rays emitted from the source will take different paths through spacetime at different image positions and arrive at the observer at different times. In this sense, the time delay of two different images $k$ and $l$ depends on the mass distribution along the line of sight of the lensing object, which can be calculated as follows:
\begin{equation}\label{lensing}
    \Delta t_{kl}=\frac{D_{\Delta t}}{c}\left[\frac{\left(\phi_{k}-\beta\right)^{2}}{2}-\psi(\phi_{k})-\frac{\left(\phi_{l}-\beta\right)^{2}}{2}+\psi(\phi_{l})\right],
\end{equation}
where $\phi_{k}$ and $\phi_{l}$ are the angular position of the images, $\beta$ is the angular position of the source, $\psi(\phi_{k})$ and $\psi(\phi_{l})$ are the lens potential at the image positions, and $D_{\Delta t}$ is the ``time-delay distance'', which is theoretically given by the expression \cite{Treu:2016ljm}
\begin{equation}\label{time-delay}
    D_{\Delta t}^{th}(\mathbf{z},\boldsymbol{\theta})=\left(1+z_{l}\right)\frac{d_{A,l}(z_{l},\boldsymbol{\theta})d_{A,s}(z_{s},\boldsymbol{\theta})}{d_{A,ls}(z_{ls},\boldsymbol{\theta})},
\end{equation}
where the subscripts $l$, $s$, and $ls$ stand for the lens, the source, and between the lens and the source, respectively; $\mathbf{z}=(z_{l},z_{s},z_{ls})$; and $d_{A,j}$ is the angular diameter distance, which can be written in terms of the luminosity distance \eqref{luminosity} as $d_{L}(z_{j},\boldsymbol{\theta})=d_{A,j}(1+z_{j})^{2}$ or

\begin{equation}\label{angulardistance}
    d_{A,j}(z_{j},\boldsymbol{\theta})=\frac{c}{(1+z_{j})}\int_{0}^{z_{j}}{\frac{dz'}{H_{th}(z',\boldsymbol{\theta})}}.
\end{equation}

In this paper, we consider the gravitational lensing compilation provided by the H0LiCOW collaboration \cite{Wong:2019kwg}, which consists of six lensed quasars:  {B1608+656} 
 \cite{Jee:2019hah}, SDSS 1206+4332 \cite{Birrer:2018vtm}, WFI2033-4723 \cite{Rusu:2019xrq}, RXJ1131-1231, HE 0435-1223, and PG 1115-080 \cite{Chen:2019ejq}, whose respective merit function can be written as
\begin{equation}\label{H0LiCOWmerit}
    \chi^{2}_{\text{GL}}=\sum_{i=1}^{6}{\left[\frac{D_{\Delta t,i}-D_{\Delta t}^{th}(\mathbf{z}_{i},\boldsymbol{\theta})}{\sigma_{D_{\Delta t},i}}\right]^{2}},
\end{equation}
where $D_{\Delta t, i}$ is the observational time-delay distance of the lensed quasar at redshift $\mathbf{z}_{i}=(z_{l, i};z_{s, i};z_{ls, i})$ with an associated error $\sigma_{D_{\Delta t}, i}$ (for more details, see Ref. \cite{Wong:2019kwg}). It is important to note that, for $z\to 0$, the angular diameter distance \eqref{angulardistance} tends to $d_{A}\to cz/H_{0}$ and, therefore, the gravitational lensing data of the H0LiCOW collaboration is sensitive to $H_{0}$, with a weak dependency on other cosmological parameters.

\subsection{\label{sec:BHS} Black Hole Shadows}
 Black Hole Shadow (BHS) data are of interest to study our local Universe since their dynamic is quite simple and can be seen as standard rulers if the angular size redshift $\alpha$, the relation between the size of the shadow, and the mass of the supermassive black hole that produces it are established \cite{Escamilla-Rivera:2022mkc}. In this paper, we are interested in two measures: the first one was made on the M87* supermassive black hole by The Event Horizon Telescope Collaboration \cite{EventHorizonTelescope:2019dse} (the first detection of a BHS) and the second one corresponds to the detection of Sagittarius A* (Sgr A*) \cite{EventHorizonTelescope:2022wkp}.

Light rays curve around its event horizon in a black hole (BH), creating a ring with a black spot center, the so-called shadow of the BH. The angular radius of the BHS for a Schwarzschild (SH) BH at redshift $z_{i}$ is given by
\begin{equation}\label{angularradius}
    \alpha_{SH}\left(z_{i},\boldsymbol{\theta}\right)=\frac{3\sqrt{3}m}{d_{A}(z_{i},\boldsymbol{\theta})},
\end{equation}
where $d_{A}(z_{i},\boldsymbol{\theta})$ is given by Equation \eqref{angulardistance} (note that the sub-index $j$ is not necessary in this case) and $m=GM_{BH}/c^{2}$ is the mass parameter of the BH, with $M_{BH}$ as the mass of the BH in solar masses units and $G$ as the gravitational constant. 

It is common to write Equation \eqref{angularradius} in terms of the shadow radius $\alpha_{SH}(z_{i},\boldsymbol{\theta})=R_{SH}/d_{A}(z_{i},\boldsymbol{\theta})$, where $R_{SH}=3\sqrt{3}GM_{BH}/c^{2}$ (the speed of light is given in units of $\text{m/s}$ in this case). Therefore, the merit function for the BHS data can be constructed as
\begin{equation}\label{BHSmerit}
    \chi^{2}_{BHS}=\sum_{i=1}^{2}{\left[\frac{\alpha_{i}-\alpha_{SH}(z_{i},\boldsymbol{\theta})}{\sigma_{\alpha,i}}\right]^{2}},
\end{equation}
where $\alpha_{i}$ is the observational angular radius of the BHS at redshift $z_{i}$ with an associated error $\sigma_{\alpha,i}$. It is important to note that for $z\to 0$, the angular radius \eqref{angularradius} tends to $\alpha_{SH}\to R_{SH}H_{0}/cz$, and, therefore, as well as in the gravitational lensing data, the BHS data are sensitive to $H_{0}$, with a weak dependency on other cosmological parameters. On the other hand, we divide Equation \eqref{angularradius} by a factor of $1.496\times 10^{11}$ to obtain $\alpha_{SH}$ in units of $\mu as$.

\subsection{\label{sec:Results} Results and Discussion}
\begin{figure}[H]
    \includegraphics[scale=0.45]{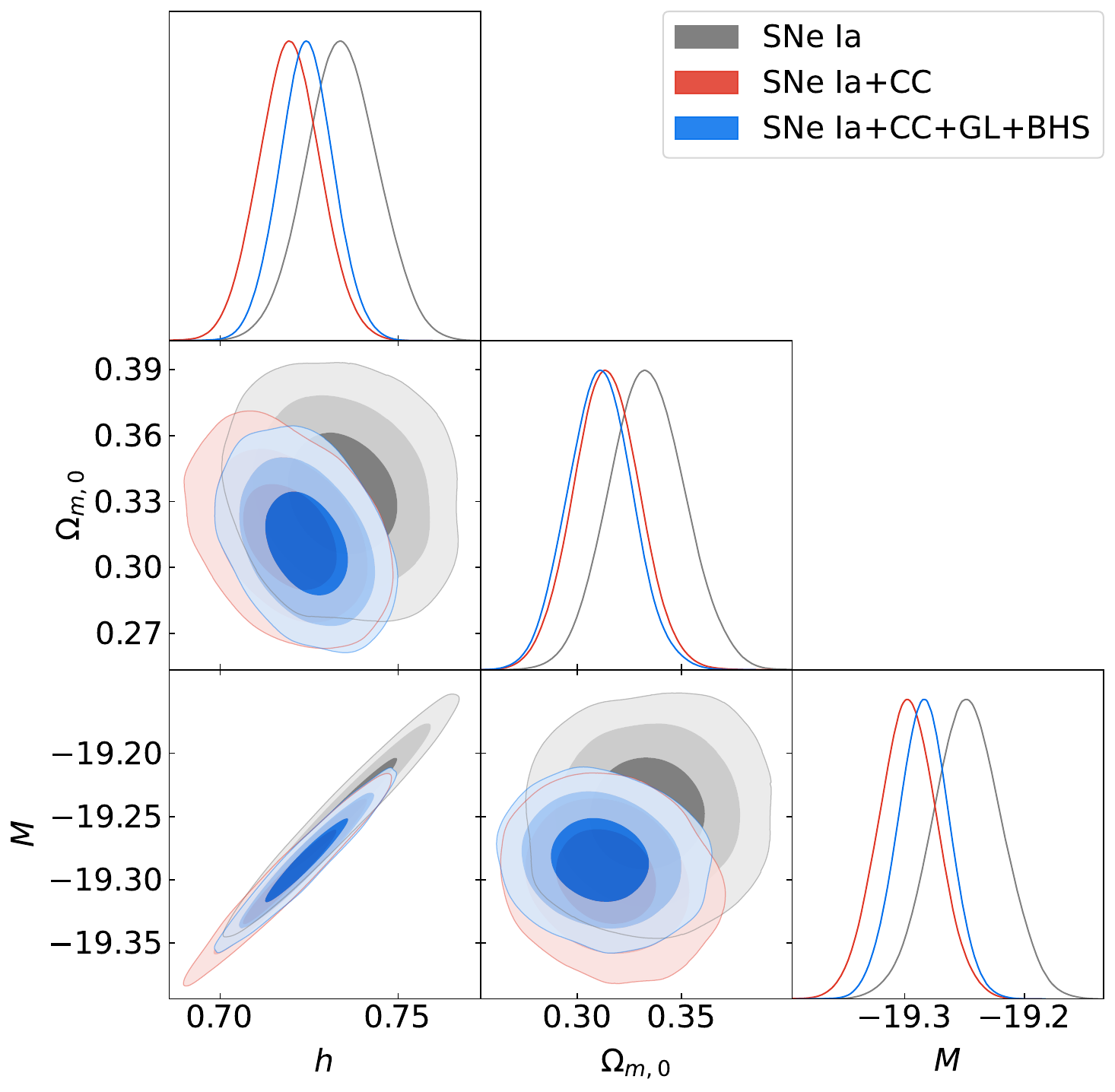}
    \caption{\label{fig: TriangleLCDMFit} Posterior 1D distribution and joint marginalized regions of the free parameter space of the $\Lambda$CDM cosmology for the SNe Ia, SNe Ia+CC, and SNe Ia+CC+GL+BHS (joint) data. The admissible joint regions correspond to $1\sigma$, $2\sigma$, and $3\sigma$ CL, respectively. The best-fit values for each model's free parameter are shown in Table \ref{tab:best-fit}.}
\end{figure}

\begin{figure}[H]
    \includegraphics[scale=0.45]{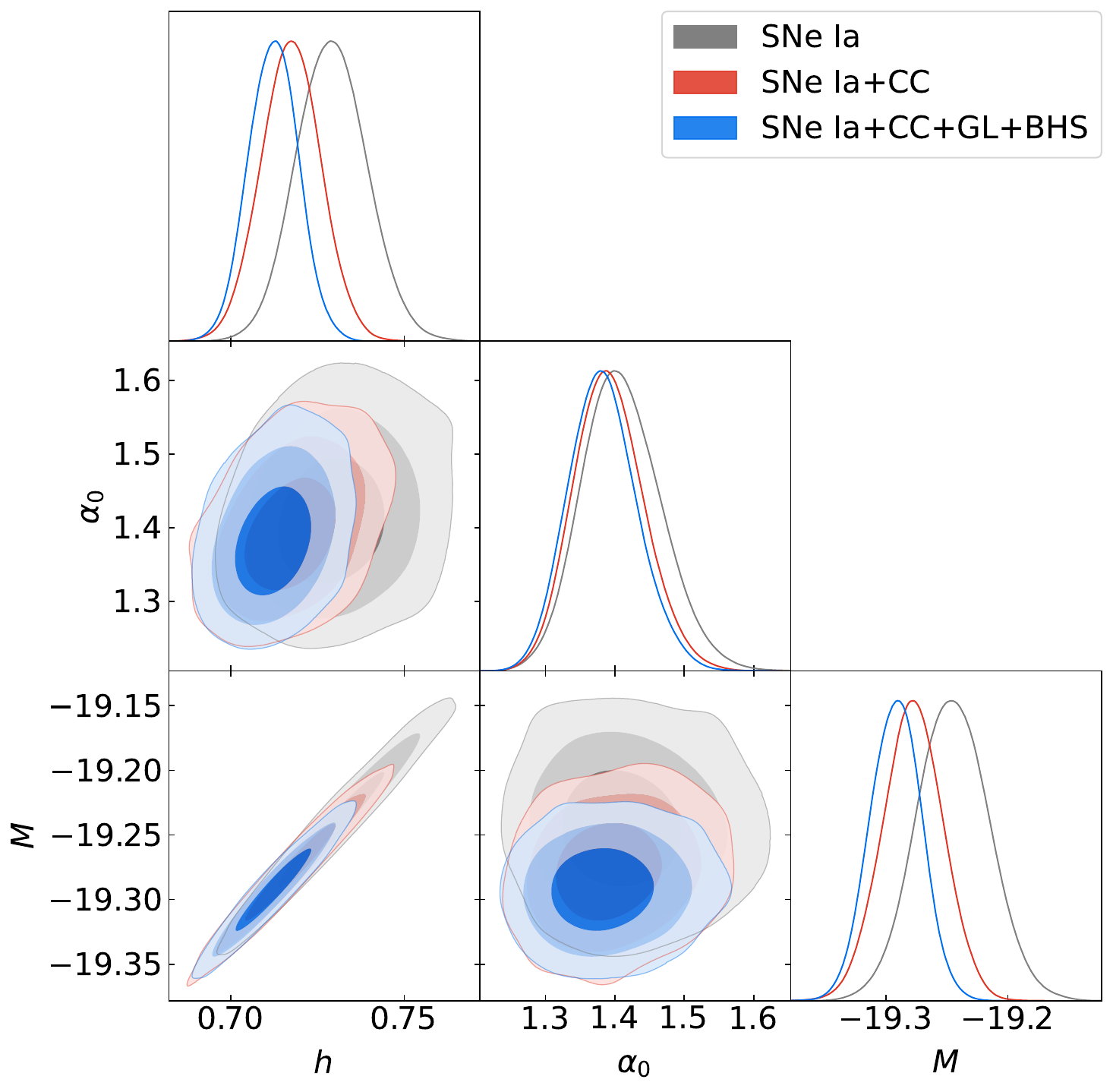}
    \caption{\label{fig: TriangleFractionalFit} Posterior 1D distribution and joint marginalized regions of the free parameter  space of the fractional cosmology for the SNe Ia, SNe Ia+CC, and SNe Ia+CC+GL+BHS (joint) data. The admissible joint regions correspond to $1\sigma$, $2\sigma$, and $3\sigma$ CL, respectively. The best-fit values for each model free parameter are shown in Table \ref{tab:best-fit}.}
\end{figure}

In Table \ref{tab:MCMCparameters}, we present the total number of steps and the correlation time for the free parameter  space of the $\Lambda$CDM and the fractional cosmology. In Table \ref{tab:best-fit}, we present their respective best-fit values at the $1\sigma$ CL and $\chi_{\text{min}}^{2}$ criteria. In Figures \ref{fig: TriangleLCDMFit} and \ref{fig: TriangleFractionalFit}, we depict the posterior 1D distribution and joint marginalized regions of the free parameter space of the $\Lambda$CDM and fractional cosmologies at $1\sigma \,(68.3\%)$, $2\sigma \,(95.5\%)$, and $3\sigma \,(99.7\%)$ CL, respectively. These results were obtained by the MCMC analysis described in Section \ref{sec:fitting} for the SNe Ia, SNe Ia+CC, and SNe Ia+CC+GL+BHS (joint) data.

\begin{table}[H]
    \centering\newcolumntype{C}{>{\centering\arraybackslash}X}
\begin{tabularx}{\textwidth}{ccCCCCC}
        \hline\hline
        \multirow{2}{*}{Data} & \multirow{2}{*}{Total steps} & \multicolumn{4}{c}{$\tau_{\rm{corr}}$} \\
        \cline{3-6}
         & & $h$ & $\Omega_{m,0}$ & $\alpha_{0}$ & $M$ \\
        \hline
        \multicolumn{6}{c}{$\Lambda$CDM cosmology} \\ 
        SNe Ia & $1250$ & $24.3$ & $22.7$ & $\cdots$ & $24.3$ \\
        SNe Ia+CC & $1300$ & $25.2$ & $24.3$ & $\cdots$ & $25.0$ \\
        joint & $1350$ & $23.4$ & $24.6$ & $\cdots$ & $23.6$ \\
        \hline
        \multicolumn{6}{c}{Fractional cosmology} \\
        SNe Ia & $1300$ & $24.5$ & $\cdots$ & $25.0$ & $24.6$ \\
        SNe Ia+CC & $1600$ & $25.0$ & $\cdots$ & $26.3$ & $25.5$ \\
        joint & $1350$ & $24.8$ & $\cdots$ & $23.0$ & $24.3$ \\
        \hline\hline
    \end{tabularx}
    \caption{\label{tab:MCMCparameters} Total number of steps and autocorrelation time $\tau_{\rm{corr}}$ for the free parameters space of the $\Lambda$CDM and the fractional cosmology, respectively. These values were obtained when the convergence test described in \S \ref{sec:fitting} is fulfilled for an MCMC analysis with 100 chains and the flat priors $0.55<h<0.85$, $0<\Omega_{m,0}<1$, $0.5<\alpha_{0}<2.5$, and $-20<M<-18$. This information is provided so that our results are replicable.}
\end{table}
\vspace{-6pt}

\begin{table}[H]
    \centering\newcolumntype{C}{>{\centering\arraybackslash}X}
\begin{tabularx}{\textwidth}{ccCCCCC}
    \hline\hline
         \multirow{2}{*}{Data} & \multicolumn{4}{c}{Best-fit values} & \multirow{2}{*}{$\chi_{\text{min}}^{2}$} \\
         \cline{2-5}
          & $h$ & $\Omega_{m,0}$ & $\alpha_{0}$ & $M$ & \\
         \hline
         \multicolumn{6}{c}{$\Lambda$CDM cosmology} \\
         SNe Ia & $0.734\pm 0.010$ & $0.333\pm 0.018$ & $\cdots$ & $-19.25\pm 0.03$ & $1523$ \\
         SNe Ia+CC & $0.719\pm 0.009$ & $0.314\pm 0.016$ & $\cdots$ & $-19.30\pm 0.03$ & 1547  \\
         Joint & $0.724\pm 0.008$ & $0.311\pm 0.016$ & $\cdots$ & $-19.28\pm 0.02$ & $1681$ \\
         \hline
         \multicolumn{6}{c}{Fractional cosmology} \\
         SNe Ia & $0.729\pm 0.010$ & $\cdots$ & $1.41\pm 0.06$ & $-19.25\pm 0.03$ & $1532$ \\
         SNe Ia+CC & $0.717\pm 0.009$ & $\cdots$ & $1.39\pm 0.05$ & $-19.28\pm 0.02$ & $1564$ \\
         Joint & $0.712\pm 0.007$ & $\cdots$ & $1.38\pm 0.05$ & $-19.29\pm 0.02$ & $1706$ \\
         \hline\hline
    \end{tabularx}
    \caption{\label{tab:best-fit}  Best-fit values and $\chi_{\text{min}}^{2}$ criteria of the $\Lambda$CDM and the fractional cosmology respectively, for the SNe Ia, SNe Ia+CC, and SNe Ia+CC+GL+BHS (joint) data. The uncertainties presented correspond to $1\sigma$ CL.}
\end{table}

\vspace{-6pt}

\begin{figure}[H]
    \includegraphics[scale = 0.63]{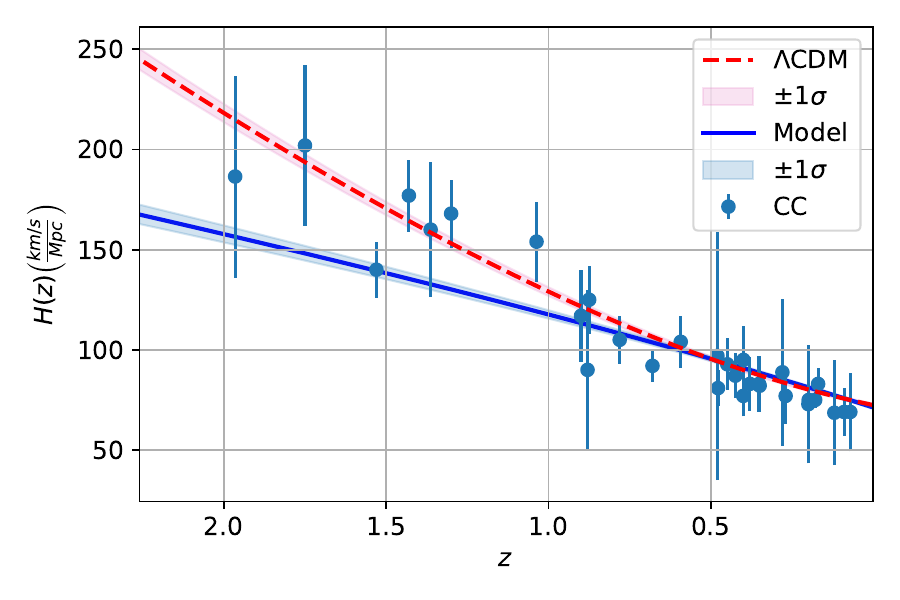}
    \caption{Theoretical Hubble parameter for the $\Lambda$CDM cosmology (red dashed line) and the fractional cosmology for the exponential potential (solid blue line) as a function of the redshift $z$, contrasted with the CC sample. The shaded curve represents the confidence regions of the Hubble parameter at a $1\sigma$ CL. The figure was obtained using the chains of the MCMC procedure described in Section \ref{sec:fitting} for the joint analysis.}
    \label{fig:CCFractional}
\end{figure}

In Table \ref{tab:best-fit}, we can see that the $\Lambda$CDM model exhibits lower values of the $\chi_{\text{min}}^{2}$ criteria compared to the scaling solution explored in this paper. However, it is essential to note that this solution does not fully represent the behavior of the fractional cosmology studied in this paper. It only provides an approximate solution to the complete picture. Therefore, the constraint presented in the table offers insights into the ability of this fractional cosmology to describe the observed Universe. An explanation of this behavior can be seen in Figure~\ref{fig:CCFractional}, where we depict the Hubble parameter for the $\Lambda$CDM cosmology and the fractional cosmology for the exponential potential as a function of the redshift $z$. The figure shows that the solution can only mimic the $\Lambda$CDM model at a redshift $z<0.5$. The solutions are moving away from this point, with the differences being greater when the redshift increases.

To reconstruct the other parameters that characterized the solution, we use the chains obtained in our MCMC analysis described in Section \ref{sec:fitting} for the fractional cosmology in the case of the joint analysis. Following this line, considering that $\alpha=2/m$, we obtain at $1\sigma$ CL the value $m=1.44\pm 0.05$. Also, considering that $\alpha_{0}=t_{0}H_{0}$, we obtain at $1\sigma$ CL the value $t_{0}=19.0\pm 0.7$ [Gyr]. On the other hand, to infer the value of $\mu$ for the scaling solution obtained for the exponential potential, we also consider the chains obtained in our MCMC analysis described in Section \ref{sec:fitting} for the $\Lambda$CDM model in the case of the joint analysis. Therefore, from Equation \eqref{OmegaScaling1}, we obtain the approximated values $\mu=1.48 \pm 0.17$ and $\mu=4.974 \pm 0.10$.

\begin{figure}[H]
    \includegraphics[scale = 0.63]{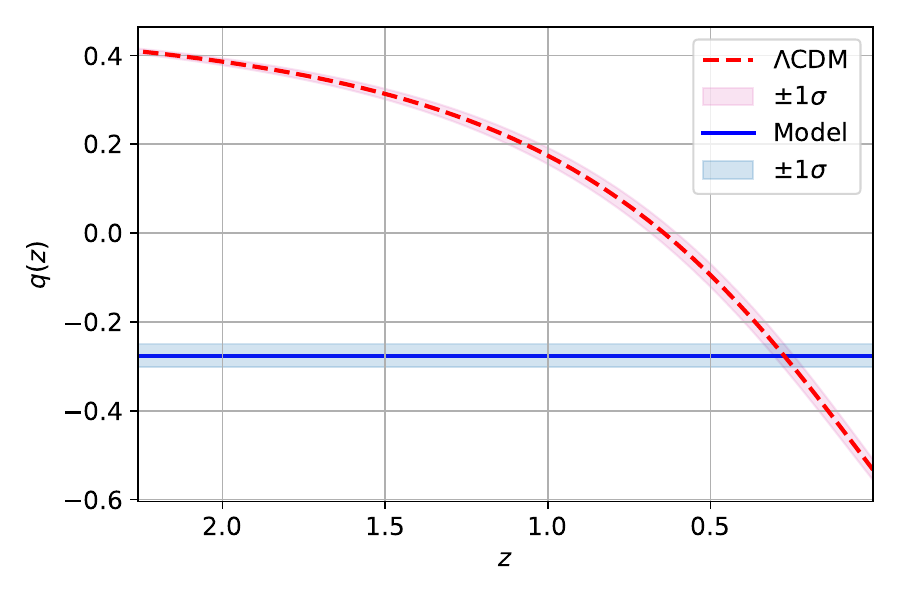}
    \caption{Deceleration parameters for the $\Lambda$CDM cosmology (red dashed line) and the fractional cosmology for the exponential potential (solid blue line) as a function of the redshift $z$. The shaded curves represent the confidence region of the deceleration parameter at a $1\sigma$ CL. The figure was obtained using the chains of the MCMC procedure described in Section \ref{sec:fitting} for the joint analysis.}
    \label{fig:qFractional}
\end{figure}

Focusing on the observational suitable solution for the joint analysis, in Figure \ref{fig:qFractional}, we depict the deceleration parameter for the fractional cosmology obtained for the exponential potential as a function of the redshift $z$, with an error band at $1\sigma$ CL. We also depict the deceleration parameter for the $\Lambda$CDM cosmology as a reference model. From this figure, we can conclude that this solution only represents an always expanding solution and does not exhibit a transition between a decelerated solution and an accelerated one such as the $\Lambda$CDM model. The deceleration parameter for this fractional solution is constant and given by $q=-1+\alpha_{0}^{-1}$, with a value at $1\sigma$ CL of $q_{0}=-0.28\pm 0.03$ at the current time; i.e., this solution also represents a less accelerated solution than the $\Lambda$CDM cosmology at the current time. Nevertheless, we must again emphasize that this solution is only a particular solution and does not represent the complete picture of fractional cosmology. This solution sheds some light on the capability of fractional cosmology in describing the cosmological background, where, as we can see, we can obtain an accelerated solution at the current time without invoking any dark energy.

\begin{figure}[H]
    \includegraphics[scale = 0.63]{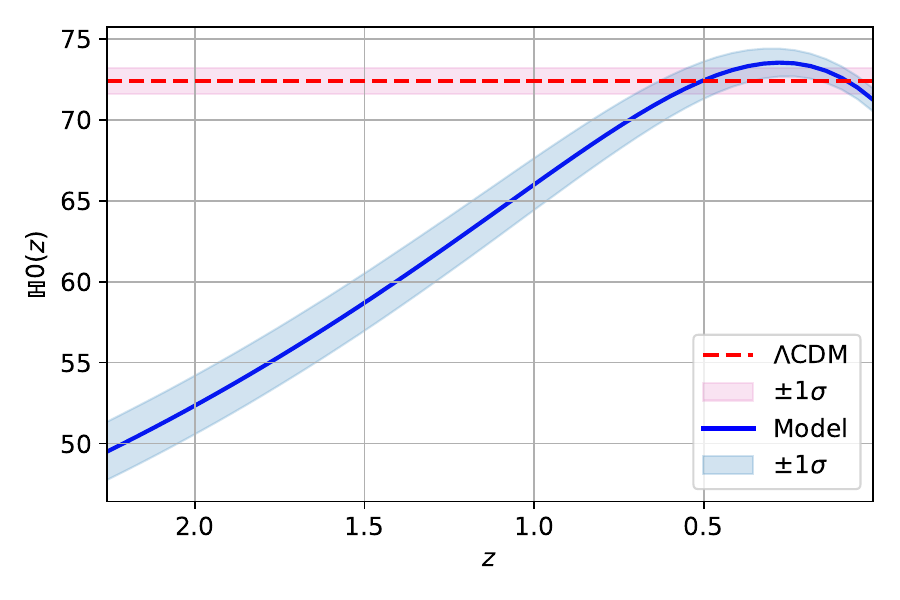}
    \caption{$\mathbf{\mathbb{H}}0$ diagnostics for the $\Lambda$CDM cosmology (red dashed line) and the fractional cosmology for the exponential potential (solid blue line) as a function of the redshift $z$. The shaded curve represents the confidence regions of the $\mathbf{\mathbb{H}}0$ diagnostic at a $1\sigma$ CL. The figure was obtained using the chains of the MCMC procedure described in Section \ref{sec:fitting} for the joint analysis.}
    \label{fig:H0diagnostic}
\end{figure}

Ultimately, we construct the $\mathbf{\mathbb{H}}0(z)$ diagnostic \citep{H0diagnostic:2021} for the observational suitable fractional solution, which can give us insights into the possibility of alleviating the $H_{0}$ tension. Following this line, the $\mathbf{\mathbb{H}}0(z)$ diagnostic is defined by
\begin{align}
    \mathbf{\mathbb{H}}0(z)= {H(z)}/{\sqrt{\Omega_{m,0}(1+z)^{3}+1-\Omega_{m,0}}}, \label{HH00diagnostic}
\end{align}
where the Hubble parameter is obtained numerically by $H(z)=H_{th}(z)$. So, in Figure~\ref{fig:H0diagnostic}, we illustrate the $\mathbf{\mathbb{H}}0$ diagnostic for the $\Lambda$CDM cosmology and the fractional solution as a function of the redshift $z$, with an error band at $1\sigma$ CL. This figure shows that at redshift $z>0.5$ (the same redshift where both models are moving away according to Figure~\ref{fig:CCFractional}), the value of $H_0$ for the fractional cosmology presents a run to values lower than $H_{0}=74.03\pm 1.42 \frac{km/s}{Mpc}$, obtained by model-independent measurements of cepheid \cite{Riess:2019cxk}. Therefore, fractional cosmology can be a suitable framework to alleviate the $H_{0}$ tension.

\vspace{-6pt}

\vspace{-9pt}

\section{Concluding Remarks}
\label{CONclusion}
We have developed a new theoretical framework called the Fractional Einstein--Gauss--Bonnet scalar field cosmology, which has important physical implications. We used fractional calculus to modify the gravitational force with an arbitrary constant fractional-order derivative. This allowed us to derive a modified Friedmann equation and a modified Klein--Gordon equation using the Fractional Action-like Variational Approach.

One critical issue is whether the obtained numerical results will be the same if the model is formulated with another fractional-order derivative. Caputo's definition of fractional and integral derivatives has several advantages compared to Riemann–Louville or Grünwald–Letnikov. First, it considers the values of the function and its derivatives at the origin (or, in general, at any lower or upper bound $a$), making it suitable for solving fractional-order initial value problems using Laplace transforms. 
Additionally, the Caputo fractional derivative of a constant is $0$, making it more consistent with classical calculation. However, after applying different definitions of the fractional derivative to the same function, we found that they give different results, but all these definitions are equally well--founded. Physicists can test different definitions of the fractional derivative by comparing theoretical results with experimental data. Depending on the specific problem, one of the definitions will show the most significant agreement with the experiment.

In our research, we utilized fractional calculus to derive modified Friedmann and Klein--Gordon equations, resulting in changes to the gravitational action integral described in Section \ref{Einstein-G}. We discovered non-trivial solutions related to exponential potential, exponential couplings to the GB term, and a logarithmic scalar field. Our findings show that we achieved late-time accelerating power-law solutions for the scale factor and Hubble parameter with inverse time power-law expansion, as discussed in Section \ref{Scaling-Solutions}. We also analyzed the phase space structure using linear stability theory and examined the dynamical effects of the Gauss--Bonnet couplings. The stability analysis of the solutions, which is a critical aspect of our research, was conducted in Section \ref{sect4.1}, where we presented a reconstruction procedure for the potential and the coupling functions. To reconstruct the other parameters characterizing the solution, we utilized the chains obtained in our MCMC analysis described in Section \ref{sec:fitting} for the fractional cosmology in the case of the joint analysis. Additionally, by considering that $\alpha=2/m$, we derived specific approximate values for cosmological quantities, assuming they are definable, as the fractional modification of the derivative concept can yield results that closely align with observational cosmological measurements. Notably, the parameters $m$ and $\alpha_{0}$ were estimated using data from cosmic chronometers, type Ia supernovae, supermassive Black Hole Shadows, and strong gravitational lensing, indicating that the solution is consistent with an accelerated expansion ($q(z)<0$) at late times with the values $\alpha_0=1.38\pm 0.05$, and $m=1.44\pm 0.05$. The Universe's age in the literature was reported as $26.7$ [Gyr] for hybrid covarying coupling constants together with tired light models in \cite{Gupta:2023mgg} using Pantheon + data fit parameters with $5.8$ [Gyr]
at $z = 10$ and $3.5$ [Gyr] at $z = 20$, giving enough time to form massive galaxies. However, in previous works within the fractional cosmology context, the authors reported other Universe ages that are larger than expected under the standard paradigm; for example, $33.633^{+14.745}_{-15.095}$ [Gyr] (CC), $33.837^{+27.833}_{-10.788}$ [Gyr] (SNe Ia) and $33.617^{+3.411}_{-4.511}$ [Gyr] (joint) were reported in \cite{Garcia-Aspeitia:2022uxz}. Moreover, the value $25.62_{-4.46}^{+6.89}$ [Gyr]  (SNe Ia+OHD data) at 3$\sigma$ CL was obtained in \cite{Gonzalez:2023who}. Our research reports a value of $t_{0}=19.0\pm 0.7$ [Gyr] (SNe Ia+CC+GL+BHS) at 1$\sigma$ CL. Still, it could be larger than the age of the Universe expected under the standard paradigm. However, these values do not contradict the minimum bound expected for the Universe's age imposed by globular clusters $t_0=13.5^{+0.16}_{-0.14}\pm 0.23$ \citep{Valcin:2021}, and, as far as we know, the maximum bound does not exist and is model-dependent.

We then used Bayesian analysis to determine the value for $\mu$ for the scaling solution found for the exponential potential; we also examined the chains generated in our MCMC analysis outlined in Section \ref{sec:fitting} for the $\Lambda$CDM model when performing joint analysis. Thus, using Equation \eqref{OmegaScaling1}, we derived the approximate values of $\mu=1.48 \pm 0.17$ and $\mu=4.974 \pm 0.10$ (consistent with $\Omega_{m,0}=0.311\pm 0.016$ and $h=0.712\pm 0.007$). 
We focused on finding an appropriate observational solution for joint analysis (with $\mu=1.48 \pm 0.17$). Hence, we depicted the deceleration parameter for fractional cosmology with an exponential potential as a function of the redshift ($z$). We also included an error band at a confidence level of $1\sigma$. We used the $\Lambda$CDM cosmology as a reference model. This solution shows a constant deceleration parameter and does not transition from a decelerated phase to an accelerated one like the $\Lambda$CDM model. The deceleration parameter for this fractional solution is constant and given by $q=-1+\alpha_{0}^{-1}$, with a value at a confidence level of $1\sigma$ of $q_{0}=-0.28\pm 0.03$ at the current time. This means this solution represents a less accelerated expansion than the current $\Lambda$CDM cosmology. However, it is essential to note that this solution is only a specific case and does not represent the complete picture of fractional cosmology. This solution provides insight into the potential of fractional cosmology in describing the cosmological background, showing that we can achieve an accelerated expansion at the current time without invoking dark energy.

We have constructed the $\mathbf{\mathbb{H}}0(z)$ diagnostic \citep{H0diagnostic:2021} to analyze the fractional solution for observational suitability. This diagnostic provided insights into the potential to alleviate the $H_{0}$ tension. Following this approach, the $\mathbf{\mathbb{H}}0(z)$ diagnostic \eqref{HH00diagnostic} involves obtaining the Hubble parameter numerically as $H(z)=H_{th}(z)$. We have plotted the $\mathbf{\mathbb{H}}0$ diagnostic for both the $\Lambda$CDM cosmology and the fractional solution as a function of the redshift $z$, along with a 1$\sigma$ CL error band. Our analysis reveals that for redshifts $z>0.5$, the value of $H_{0}$ for the fractional cosmology consistently trended lower than $H_{0}=74.03\pm 1.42 \frac{km/s}{Mpc}$, which was obtained from model-independent measurements of cepheids \cite{Riess:2019cxk}. Therefore, fractional cosmology shows promise as a framework for addressing the $H_{0}$ tension. 

Our model identifies various equilibrium points corresponding to cosmological scenarios, including accelerated and scaling solutions. The scaling behavior at specific equilibrium points suggests that the geometric corrections in the coupling to the GB scalar can replicate the behavior of the dark sector in modified gravity. 
Moreover, attracting scaling solutions solves the Coincidence Problem. 

This cosmological model, derived from fractional calculus, is consistent with recent cosmological observations and provides an alternative explanation for cosmic acceleration, distinct from the $\Lambda$CDM model. Our results build upon and significantly enhance previous work in the field, emphasizing the practical implications of fractional calculus in cosmology. Additionally, our model shows promise in addressing the persistent tension in the Hubble constant, offering hope for future research \cite{LeonTorres:2023ehd}.

\vspace{6pt} 


\section*{Author contributions}

Conceptualization, B.M.-R., B.D., A.D.M. and G.L.; methodology, G.L.; software, B.M.-R., B.D., A.D.M., G.L., E.G. and J.M.; validation, B.M.-R., B.D., A.D.M., G.L., E.G. and J.M.; formal analysis, B.M.-R., B.D., A.D.M., G.L., E.G. and J.M.; investigation, B.M.-R., B.D., A.D.M., G.L., E.G. and J.M.; resources, G.L. and J.M.; writing—original draft preparation, G.L.; writing—review and editing, G.L.; visualization, B.M.-R., B.D., A.D.M., G.L., E.G. and J.M.; supervision, G.L., B.D. and A.D.M.; project administration, G.L., J.M., B.M.-R. and B.D.; funding acquisition, B.M.-R., B.D., A.D.M., G.L., E.G. and J.M. All authors have read and agreed to the published version of the manuscript.

\section*{Funding}

B.M.-R., G.L., J.M., and A.D.M. acknowledge the financial support of Agencia Nacional de Investigación y Desarrollo (ANID), Chile, through Proyecto Fondecyt Regular 2024, Folio 1240514, Etapa 2024. G.L. was also funded by Vicerrectoría de Investigación y Desarrollo Tecnológico (VRIDT) at Universidad Católica del Norte (UCN) through Resolución VRIDT N°026/2023, Resolución VRIDT N°027/2023, Proyecto de Investigación Pro Fondecyt 2023 (Resolución VRIDT N°076/2023) and Resolución VRIDT N°09/2024. A.D.M. was also supported by Agencia Nacional de Investigación y Desarrollo - ANID Subdirección de Capital Humano/Doctorado Nacional/año 2020 folio 21200837, Gastos operacionales proyecto de Tesis/2022 folio 242220121 and VRIDT-UCN. E.G. was funded by Vicerrectoría de Investigación y Desarrollo Tecnológico (VRIDT) at Universidad Católica del Norte (UCN) through Proyecto de Investigación Pro Fondecyt 2023, Resolución VRIDT N°076/2023.

\section*{Data availability}

The data supporting this article can be found in Section~\ref{sec:fitting}.

\acknowledgments{G.L. is thankful for the support of Núcleo de Investigación Geometría Diferencial y Aplicaciones, Resolución VRIDT N°096/2022. E.G. acknowledges the scientific support of Núcleo de Investigación No. 7 UCN-VRIDT 076/2020, Núcleo de Modelación y Simulación Científica (NMSC).}

\section*{Conflicts of interest}
We declare no conflict of interest. The funders had no role in the design of the study; in the collection, analyses, or interpretation of data; in the writing of the manuscript; or in the decision to publish the results.

\appendix

\section[\appendixname~\thesection]{Variational equations}
\label{app0}

We start with an action defined as \cite{10.5555/1466940.1466942}
\begin{equation}
    S=\frac{1}{\Gamma(\mu)}\int_0^\tau \mathcal{L}\left(\theta,q_i(\theta),\Dot{q}_i(\theta),\Ddot {q}_i(\theta)\right)(\tau-\theta)^{\mu-1}\,d\theta, \label{Action_1}
\end{equation}
where $\Dot{q}_i(\theta)$, $\Ddot {q}_i(\theta)$ means integer order derivatives with respect to the argument $\theta$. 

To obtain the equations of motion of the fields, we do the infinitesimal transformation of the fields
\begin{equation}
    q_i \mapsto q_i + \delta q_i,\quad  \delta \mathcal{L} = \mathcal{L}(q_i + \delta q_i)-  \mathcal{L}(q_i).
\end{equation}
Then, we write the variation of the action:
\begin{equation}
\begin{split}
    \delta S&=\frac{1}{\Gamma(\mu)}\int_0^\tau \left[(\delta\mathcal{L})(\tau-\theta)^{\mu-1}+\mathcal{L}\delta ((\tau-\theta)^{\mu-1})\right]\,d\theta\\
    &=\frac{1}{\Gamma(\mu)}\int_0^\tau \left[\left(\frac{\partial \mathcal{L}}{\partial q_i}\delta q_i+\frac{\partial \mathcal{L}}{\partial \Dot{q}_i}\delta \Dot{q}_i+\frac{\partial \mathcal{L}}{\partial\Ddot{q}_i}\delta \Ddot{q}_i\right)(\tau-\theta)^{\mu-1}\right]\,d\theta.
\end{split}
\end{equation}

We can integrate by parts:
\begin{equation}
\begin{split}
    \int_0^\tau \frac{\partial \mathcal{L}}{\partial\Dot{q}_i}(\tau-\theta)^{\mu-1} \delta \Dot{q}_i\,d\theta=&\frac{\partial \mathcal{L}}{\partial \Dot{q}_i}(\tau-\theta)^{\mu-1} \delta q_i\Big|_0^\tau -\int_0^\tau  \frac{d}{d\theta}\left[\frac{\partial \mathcal{L}}{\partial \Dot{q}_i} (\tau-\theta)^{\mu-1}\right] \delta q_i\,d\theta,
\end{split}
\end{equation}
\begin{equation}
\begin{split}
    \int_0^\tau \frac{\partial \mathcal{L}}{\partial \Ddot{q}_i}(\tau-\theta)^{\mu-1} \delta \Ddot{q}_i\,d\theta&= \frac{\partial \mathcal{L}}{\partial \Ddot{q}_i}(\tau-\theta)^{\mu-1} \delta \Dot{q}_i\Big|_0^\tau-\int_0^\tau \frac{d}{d \theta}\left[\frac{\partial \mathcal{L}}{\partial \Ddot{q}_i} (\tau-\theta)^{\mu-1}\right] \delta \Dot{q}_i  \,d\theta\\
    &= -\frac{d}{d \theta}\left[\frac{\partial \mathcal{L}}{\partial \Ddot{q}} (\tau-\theta)^{\mu-1}\right]\delta q_i \Big|_0^\tau + \frac{\partial \mathcal{L}}{\partial \Ddot{q}_i}(\tau-\theta)^{\mu-1} \delta \Dot{q}_i\Big|_0^\tau\\
    & + \int_0^\tau \frac{d^2}{d \theta^2}\left[\frac{\partial \mathcal{L}}{\partial \Ddot{q}_i} (\tau-\theta)^{\mu-1}\right] \delta q_i\,d\theta.
\end{split}
\end{equation}
Therefore, the variation of the action can be written as
\begin{equation}
\begin{split}
    \delta S&= \frac{1}{\Gamma(\mu)}\int_0^\tau   \left\{ \frac{\partial \mathcal{L}}{\partial q_i}
(\tau-\theta)^{\mu-1}-\frac{d}{d\theta}\left[\frac{\partial \mathcal{L}}{\partial \Dot{q}_i} (\tau-\theta)^{\mu-1}\right] + \frac{d^2}{d \theta^2}\left[\frac{\partial \mathcal{L}}{\partial \Ddot{q}_i} (\tau-\theta)^{\mu-1}\right]\right\} \delta q_i\,d\theta\\
    &+  \left\{\frac{\partial \mathcal{L}}{\partial \Dot{q}_i}(\tau-\theta)^{\mu-1} -\frac{d}{d \theta}\left[\frac{\partial \mathcal{L}}{\partial \Ddot{q}} (\tau-\theta)^{\mu-1}\right]\right\}\delta q_i \Bigg|_0^\tau + \frac{\partial \mathcal{L}}{\partial \Ddot{q}_i}(\tau-\theta)^{\mu-1} \delta \Dot{q}_i \Bigg|_0^\tau.
\end{split}
\end{equation}
Then, assuming that $\delta q_i(0)=0$ at the lower end, we have the equations of motion:
\begin{equation}
 \frac{\partial \mathcal{L}}{\partial q_i}
(\tau-\theta)^{\mu-1}-\frac{d}{d\theta}\left[\frac{\partial \mathcal{L}}{\partial \Dot{q}_i} (\tau-\theta)^{\mu-1}\right] + \frac{d^2}{d \theta^2}\left[\frac{\partial \mathcal{L}}{\partial \Ddot{q}_i} (\tau-\theta)^{\mu-1}\right] =0. \label{Variational-Eqs}
\end{equation}

For the more general case where the Lagrangian depends on derivatives of up to order $n>2$ of the generalized coordinates $q_i$, the same procedure can be used to show that, through successive integration by parts, alternating signs emerge, and the equations of motion are
\begin{equation}
    \sum_{k=0}^n(-1)^k\frac{d^k}{d\theta^k}\left[\frac{\partial \mathcal{L}}{\partial \left( q_i^{(k)}\right)}\left(\tau-\theta\right)^{\mu-1}\right]=0.
\end{equation}

\section[\appendixname~\thesection]{Stability analysis of power-law solutions}
\label{app1}

In general, when we have an ordinary differential equation
\begin{equation}
\mathcal{F}\left(
t,\psi\left(  t\right), \dot{\psi}\left(  t\right), \ddot{\psi}\left(  t\right)
,...\right)  \equiv 0, \label{bc}
\end{equation}  
where $t$ is the independent variable and $\psi\left(
t\right)  $ is the dependent variable; we can provide the analysis of the stability of the solution $\psi_{s}(t)$ in the
interval $0< t <\infty$ using similar methods as in \cite{Ratra:1987rm,Liddle:1998xm} and  \cite{Uzan:1999ch}. Defining the new time variable 
\begin{equation}
t= e^{\tau},\quad   -\infty <\tau <\infty,
\end{equation}
such that $t\rightarrow 0$ as $\tau \rightarrow -\infty$ and  $t\rightarrow \infty$ as $\tau \rightarrow \infty$, as well as the 
ratio 
\begin{equation}
u(\tau)= \frac{\psi(\tau)}{\psi_s(\tau)},
\end{equation}
where $\psi(\tau)= \psi(e^{\tau})$ and $\psi_s(\tau)=\psi_s(e^{\tau})$ given.
Notice that evaluated  at $\psi_s(\tau)$ we have $u=1$, therefore, defining $\varepsilon= \frac{\psi(\tau)}{\psi_s(\tau)}-1$ the solution is shifted to $\varepsilon=0$. 
Let 
\begin{equation}
\psi^{\prime} \equiv \frac{d \psi}{d \tau},
\end{equation}
then, 
\begin{equation}
\dot \psi = \frac{d \tau}{d t} \psi^{\prime} = e^{-\tau} \psi^{\prime},\quad \ddot \psi= e^{-2 \tau} \left(\psi^{\prime \prime} - \psi^{\prime} \right).
\end{equation}

Therefore, the equation (\ref{bc}) becomes 
\begin{equation}
\mathcal{G}\left(
\tau,u\left( \tau\right), u^{\prime}\left(\tau\right), u^{\prime \prime}\left(\tau\right)
,...\right)  \equiv 0.
\end{equation}

According to \eqref{system2}, $\psi(t)$ satisfies
\begin{equation}\label{psi pp}
    \ddot{\psi}=c_1+c_2\frac{\psi}{t^2}+c_3\frac{\dot{\psi}}{t},
\end{equation}
where $c_1$, $c_2$, and $c_3$ are the constants. 
Additionally, given the solution found for the exponential potential, \eqref{exactpsi}, the critical solution \eqref{Nexactpsi} can be expressed as:
\begin{equation}\label{psicritic}
    \psi_c(t)=c_4 t^2,
\end{equation}
where $c_4$ is also a constant. 

Thus, replacing $\ddot\psi$, $\dot{\psi}$, and $t$, \eqref{psi pp} becomes
\begin{equation}
    e^{-2\tau}\left[\psi''(\tau)-\psi'(\tau)\right]=c_1+c_2e^{-2\tau}\psi(\tau)+c_3e^{-2\tau}\psi'(\tau),
\end{equation}
That is,
\begin{equation}
    \psi''(\tau)=c_1 e^{2\tau}+c_2\psi(\tau)+\left(c_3+1\right)\psi'(\tau). \label{eq129}
\end{equation}
Using the definition of $\varepsilon$, we see that
\begin{equation}
\begin{split}
    \psi(\tau)&=(\varepsilon(\tau)+1)\psi_c(\tau),\\
    \psi'(\tau)&=\varepsilon'(\tau)\psi_c(\tau)+(\varepsilon(\tau)+1)\psi'_c(\tau),\\
    \psi''(\tau)&=\varepsilon''(\tau)\psi_c(\tau)+2\varepsilon'(\tau)\psi'_c(\tau)+(\varepsilon(\tau)+1)\psi''_c(\tau).
\end{split}
\end{equation}
Therefore,
\begin{equation}
\varepsilon''\psi_c+2\varepsilon'\psi'_c+(\varepsilon+1)\psi''_c=c_1e^{2\tau}+c_2(\varepsilon+1)\psi_c+(c_3+1)\left[\varepsilon'\psi_c+(\varepsilon+1)\psi'_c\right]. \label{Eq131}
    \end{equation}
Equation \eqref{Eq131}  can be written as 
\begin{align}
    \varepsilon''\psi_c+2\varepsilon'\psi'_c+\varepsilon \psi''_c + \underbrace{\left[\psi''_c - c_1e^{2\tau} - c_2\psi_c - (c_3+1) \psi'_c\right]}_{=0, \; \psi_c(\tau)\;\text{satisfies}\; \eqref{eq129}.}= c_2 \varepsilon \psi_c+(c_3+1)\left(\varepsilon'\psi_c+ \varepsilon \psi'_c\right) \label{eq132}
\end{align}
Due to $\psi_c(\tau)$ being a non-trivial particular solution of equation \eqref{eq129}, the term inside squared brackets on the left-hand side of \eqref{eq132} is zero. 
After simplification, we acquire 
\begin{equation}
    \varepsilon''+2\varepsilon'\frac{\psi'_c}{\psi_c}+\varepsilon\frac{\psi''_c}{\psi_c} =c_2\varepsilon+(c_3+1)\left(\varepsilon'+\varepsilon\frac{\psi'_c}{\psi_c}\right).
\end{equation}
Note that
\begin{equation}
    \psi_c(\tau)=c_4e^{2\tau},\quad 
    \psi'_c(\tau)=2c_4e^{2\tau},\quad
    \psi''_c(\tau)=4c_4e^{2\tau}.
\end{equation}
Therefore,
\begin{equation}
    \frac{\psi'_c}{\psi_c}=2,\quad\frac{\psi''_c}{\psi_c}=4.
\end{equation}
Hence, by reducing terms, we have 
\begin{equation}
    \varepsilon''=\varepsilon'(c_3-3)+\varepsilon(c_2+2c_3-2).
\end{equation}
Defining $v=\varepsilon'$, we have the linear system 
\begin{equation}
    \varepsilon'=v, \quad 
    v'=\varepsilon(c_2+2c_3-2) + v(c_3-3),
\end{equation}
that was investigated using dynamical systems tools in \S \ref{sect4.1} for the choice of the constants:
\begin{align*}
    c_1 & = \frac{1}{32} \left[m (-4 \mu +(\mu -2) (\mu -1) m+18)-12\right],\quad 
    c_2 = \frac{m \left[\mu  (-\mu  m+m+4)-22\right]+12}{m^2},\\
    c_3 & = 2 \left(\mu -\frac{2}{m}\right),\quad 
    c_4= \frac{1}{96} m^2 \left(\frac{2 m
   \left(3 \lambda ^2+m\right)}{\lambda ^2 \left[(\mu -6) m+2\right]}+3\right).
\end{align*}

\bibliography{main}

\end{document}